# The spinor-tensor gravity of the classical Dirac field

Piero Chiarelli

*National Council of Research of Italy, Area of Pisa, 56124 Pisa, Moruzzi 1, Italy*

*Interdepartmental Center "E. Piaggio" University of Pisa*
Phone: +39-050-315-2359
Fax : +39-050-315-2166

Email: pchiare@ifc.cnr.it.

**Abstract**: In this work, with the help of the quantum hydrodynamic formalism, the gravitational equation associated to the Dirac field is derived. The gravitational coupling of the fermion field is obtained through three steps:

1. By deriving the quantum hydrodynamic representation of the Dirac equation in the Lagrangean form where the bi-spinor fields $\Psi = \begin{pmatrix} Œ_1 \\ Œ_2 \end{pmatrix}$ is described by the evolution of four mass densities $/Œ_{\pm 1}/^2, /Œ_{\pm 2}/^2$ subject to the theory-defined quantum potential, where $Œ_{\pm i} = \begin{pmatrix} Œ_{\pm 1} \\ Œ_{\pm 2} \end{pmatrix} = \begin{pmatrix} /Œ_{\pm 1}/ \, exp[\, i \frac{S_{\pm 1}}{\hbar}\,] \\ /Œ_{\pm 2}/ \, exp[\, i \frac{S_{\pm 2}}{\hbar}\,] \end{pmatrix} \equiv Œ_{\pm} = \frac{Œ_1 \pm Œ_2}{\sqrt{2}}$, that move with momenta $p_{\pm \sim} = \begin{pmatrix} p_{\pm 1 \sim} \\ p_{\pm 2 \sim} \end{pmatrix} = -\begin{pmatrix} \partial_{\sim} S_{\pm 1} \\ \partial_{\sim} S_{\pm 2} \end{pmatrix}$ ;
2. By generalizing the quantum equations to the curved space-time in the covariant form;
3. By defining the metric of the space-time through the Einstein-like equation derived by using the minimum action principle applied to the overall action of both the motion of the mass densities $/Œ_{\pm 1}/^2, /Œ_{\pm 2}/^2$ and the gravitational field.

The derived gravity for the Dirac field shows an energy-impulse tensor density (EITD) that is a function of the fermion field with the spontaneous emergence of the "cosmological" gravity tensor (CGT) that in the classical limit can include the cosmological constant (CC). Even if the classical cosmological constant is set to zero, the gravity equation (GE) shows the emergence of a theory-derived non-zero CGT of quantum-mechanical origin that, in principle, allows to have a stable quantum vacuum (out of the *collapsed branched polymer phase*) without the classical CC. In the classical macroscopic limit, the gravity equation of the Dirac field leads to the general relativity equation.

In the perturbative approach, the work shows that the CGT leads to a second order correction to the Newtonian gravity. The outputs of the theory show that the expectation value of the CGT takes contribution from the space where the mass is localized (and the space-time is curvilinear) while tends to zero as the space-time approaches to the flat vacuum, regardless the zero-point vacuum energy density, leading to an overall mean cosmological effect on the motion of the galaxies that may possibly be compatible with the astronomical observations.

The spinor-tensor form of the Dirac field gravity shows analogies with the modified Brans-Dick gravity where each spinor term brings an effective gravity constant G divided by its field squared.

Finally, the work shows that, in order to obtain the gravitational equation associated to the Dirac field by the minimum action principle, the Copenhagen interpretation of quantum mechanics, the Madelung-De Broglie-Bohm pilot-wave and the decoherence approaches must be contemporarily utilized.

## 1. Introduction

One of the problems of nowadays physics is to describe how the quantum mechanical properties of the space-time (ST) and the second quantization affect the gravity.

Even if the general relativity (GR) has opened some understanding about the cosmological dynamics [1-6] the complete explanation of generation of matter and its distribution in the universe need the integration of the cosmological physics with the quantum one. To this end the quantum gravity (QG) represents the goal of the century for the theoretical research [7-9]. Nevertheless, difficulties arise when one attempts to apply the quantum field theory (QFT) to the force of gravity [10-11].

The difficulties about the integration of QFT and the GR become really evident in the so called cosmological constant (CC) that Einstein introduced into its equation to give stability to the solution of universe evolution but that then he refused [6].

Actually, the renewed interest for the CC is basically stated by semi-empirical arguments to explain the motion of the galaxies [10-11]. The great difficulty of the QFT to give a correct value of the CC, relies in the fact that the energy-impulse tensor density (EITD) in the Einstein equation [12] (for classical bodies) owns a point-dependence by the mass density, but does not have any analytical connection with the fields of matter described by the Dirac or by the Klein-Gordon equation (KGE).

If we place matter into a classic ST, each infinitesimal element of mass can be put freely in an infinitesimal volume without any interference with the neighboring ones. In the quantum mechanical ST, the things are different and it is more and more difficult to add more and more matter in the same place (e.g., forbidding punctual mass distribution).

Obviously, these two kinds of space-times are different and leads to distinct types of gravity. The incompleteness of general relativity in describing the gravity of the physical quantum-ST is recognized by all and modified gravity models (on semi-empirical basis) have been proposed. One example is the scalar-tensor gravity about which the Brans-Dicke [13-15 one is the most known example. If, on one hand, the conceptual generalization about the possible sources of gravity has been fruitful showing how problems such as the cosmological constant, the universe inflation and the dark matter can be solved [16-18], on the other hand, it has brought to the appearance of a scalar field of obscure physical reading.

The derivation of the quantum-mechanical gravity of a mass distribution obeying to the Klein-Gordon equation [19], by using the classical-like hydrodynamic quantum representation, that leads to terms that are identical to those contained into the Brans-Dicke model, shows that the scalar field of the BD theory, by fact, reproduces the effects of the quantum-mechanical property of the ST upon the gravity.

Taking in account the contribution of the quantum potential energy (i.e., the energy connected to the pilot wave) in the definition of the space-time curvature, the author has shown that the quantization of a scalar uncharged field may lead to a *mean* cosmological constant value (on the universe scale) compatible with the astronomical observation with the energy cut-off defined by the Planck wave length [20].

The objective of this work is to generalize the gravitational equation (GE) of the scalar KGE field [17] to the field of half integer spin charged particles obeying to the Dirac-Fock-Weyl equation (DFWE).

The work is carried out by utilizing the quantum hydrodynamic representation where the bi-spinor fields $\Psi = \begin{pmatrix} \psi_1 \\ \psi_2 \end{pmatrix}$ is described by the evolution of four mass densities $|\psi_{\pm 1}|^2, |\psi_{\pm 2}|^2$, where $\psi_{\pm} = \frac{\psi_1 \pm \psi_2}{\sqrt{2}} = \begin{pmatrix} \psi_{\pm 1} \\ \psi_{\pm 2} \end{pmatrix} = \begin{pmatrix} |\psi_{\pm 1}| \, exp[i \frac{S_{\pm 1}}{\hbar}] \\ |\psi_{\pm 2}| \, exp[i \frac{S_{\pm 2}}{\hbar}] \end{pmatrix}$, that move with momenta $p_{\pm 1 \mu} = -\partial_{\mu} S_{\pm 1}$ $p_{\pm 2 \mu} = -\partial_{\mu} S_{\pm 2}$ subject to the non-local interaction of the theory-defined quantum potential [20-24] whose energy, following the basic principle of the general relativity, contributes to the definition of the space-time curvature. The gravity generated by the bi-spinor fields $\Psi$ is derived by using the minimum action principle applied to the Lagrangean form of the quantum hydrodynamic motion equations [21]. Thence, the perturbative approach to the gravity-DFWE coupled system of equations is developed in curved space-time.

In section five, the expectation value of the cosmological constant generated by the quantum Dirac field, is derived for flat and near-Minkowskian space-time.

Finally, in section six, some aspects of the gravity of the Dirac field are discussed:

i. The Dirac field gravity and the cosmological constant in the pure quantum gravity;
ii. Analogies with the Brans-Dicke model;
iii. Quantum mechanical gravity and the foundations of the quantum theory;
iv. Experimental tests.

## 2. The hydrodynamic form of the DFWE

### 2.0. The hydrodynamic form of the DFWE in the Minkowskian space-time

In reference [21] the Dirac equation

$$\left( i\hbar \gamma^{\mu} \left( \partial_{\mu} + \frac{ie}{\hbar} A_{\mu} \right) - mc \right) \Psi = 0 \tag{2.0.1}$$

where

$$\gamma^{\mu} = \left( \gamma^0, \gamma^i \right) = \left( \begin{pmatrix} I & 0 \\ 0 & -I \end{pmatrix}, \begin{pmatrix} 0 & \sigma_i \\ -\sigma_i & 0 \end{pmatrix} \right), \tag{2. 0.2}$$

where $\sigma^{\mu} = (\sigma_0, \sigma_i)$ are the 4-D extended Pauli matrices [22], where

$$\Psi = \begin{pmatrix} \psi_1 \\ \psi_2 \end{pmatrix} \qquad (2.\ 0.3)$$

$$\overline{\Psi} = \Psi^{\dagger}\gamma^{0}, \qquad (2.\ 0.4)$$

and where $A_{\mu} = (\frac{\phi}{c},\ -A_i)$ is the electromagnetic potential, has been expressed as a function of the components $\psi_k$ of the bi-spinor $\Psi = \begin{pmatrix} \psi_1 \\ \psi_2 \end{pmatrix}$, in the form

$$\frac{i\hbar}{mc}\sigma^{\mu}\left(\partial_{\mu} + \frac{ie}{\hbar}A_{\mu}\right)\psi_{+} = \psi_{-} \qquad (2.\ 0.5)$$

$$\frac{i\hbar}{mc}\bar{\sigma}^{\mu}\left(\partial_{\mu} + \frac{ie}{\hbar}A_{\mu}\right)\psi_{-} = \psi_{+} \qquad (2.\ 0.6)$$

where $\psi_{\pm} = \frac{\psi_1 \pm \psi_2}{\sqrt{2}}$. Moreover, by posing

$$\begin{aligned} \bar{\sigma}^{\mu}\sigma^{\nu} &= g^{\mu\nu} + \rho_{+}^{\mu\nu} \\ \sigma^{\mu}\bar{\sigma}^{\nu} &= g^{\mu\nu} + \rho_{-}^{\mu\nu} \end{aligned} \qquad (2.\ 0.7)$$

where $\bar{\sigma}^{\mu} = (\bar{\sigma}_0, \bar{\sigma}_i) = (\sigma_0, -\sigma_i) = \sigma_{\mu}$, substituting (2.0.5) in (2.0.6) it follows that

$$\left[\left(\partial_{\mu} + \frac{ie}{\hbar}A_{\mu}\right)\left(\partial^{\mu} + \frac{ie}{\hbar}A^{\mu}\right) + \left(\frac{m^2c^2}{\hbar^2} + \frac{ie}{2\hbar}\rho_{\pm}^{\mu\nu}F_{\mu\nu}\right)\right]\psi_{\pm} = 0 \qquad (2.\ 0.8)$$

where $F_{\mu\nu}$ is the electromagnetic (EM) tensor [22].
Furthermore, by using the hydrodynamic notation

$$\psi_{\pm i} = \begin{pmatrix} \psi_{\pm 1} \\ \psi_{\pm 2} \end{pmatrix} = \begin{pmatrix} |\psi_{\pm 1}|\, exp\left[i\frac{S_{\pm 1}}{\hbar}\right] \\ |\psi_{\pm 2}|\, exp\left[i\frac{S_{\pm 2}}{\hbar}\right] \end{pmatrix} \qquad (2.\ 0.9)$$

and equating the real and imaginary part of equation (2.0.8), both the quantum hydrodynamic Hamilton-Jacobi equation (HHJE) of motion and the current conservation equation [21] of the spinors $\psi_{\pm i}$ follow.
The HHJE reads (see ref. [21])

$$\left(\partial_{\mu}S_{\pm i} + eA_{\mu}\right)\left(\partial^{\mu}S_{\pm i} + eA^{\mu}\right) = \left(m^2c^2 - mV_{qu\pm i}\right), \qquad (2.\ 0.10)$$

where $S_{\pm i} = \begin{pmatrix} S_{\pm 1} \\ S_{\pm 2} \end{pmatrix}$, where i=1,2 is the spinor italic index and where the quantum potential reads

$$V_{qu\pm i} = \begin{pmatrix} V_{qu\pm 1} \\ V_{qu\pm 2} \end{pmatrix} = -\frac{\hbar^2}{m}\begin{pmatrix} \dfrac{\partial_{_}\partial^{\tilde{}}\ /\text{Œ}_{\pm 1}/}{/\text{Œ}_{\pm 1}/} \\ \dfrac{\partial_{_}\partial^{\tilde{}}\ /\text{Œ}_{\pm 2}/}{/\text{Œ}_{\pm 2}/} \end{pmatrix} - Re\{\frac{ie\hbar}{2m}\begin{pmatrix} \text{r}^{\}|}_{\pm\ 1j}\dfrac{/\text{Œ}_{\pm j}/}{/\text{Œ}_{\pm 1}/} \\ {}^{\}|}_{\pm\ 2j}\dfrac{/\text{Œ}_{\pm j}/}{/\text{Œ}_{\pm 2}/} \end{pmatrix} F_{\}|}\ \}. \qquad (2.0.11)$$

The current equation for the spinors $\text{Œ}_{\pm i}$ reads [21]

$$-\left(g^{\tilde{}\text{€}} + \text{r}_{\pm}^{\tilde{}\text{€}}\right)\partial_{_}\left(\frac{/\text{Œ}_{\pm i}/^{2}}{m}\left(\partial_{\text{€}} S + eA_{\text{€}}\right)\right) = \partial_{_} J^{(\pm i)\tilde{}}\ , \qquad (2.0.12)$$

that, being $\text{Œ}_{+}$ and $\text{Œ}_{-}$ coupled each other through the EM potential in equations (2.0.8), leads to the overall conservation of current $J^{\tilde{}}$

$$\frac{\partial\left(\left(J^{(+1)\tilde{}} + J^{(+2)\tilde{}}\right) + \left(J^{(-1)\tilde{}} + J^{(-2)\tilde{}}\right)\right)}{\partial q^{\tilde{}}} = \frac{\partial J^{(+)\tilde{}} + J^{(-)\tilde{}}}{\partial q^{\tilde{}}} = \frac{\partial J^{\tilde{}}}{\partial q^{\tilde{}}} = 0 \qquad (2.0.13)$$

where the current $J^{\tilde{}}$ reads [21]

$$\begin{aligned} J^{\tilde{}} &= J^{(+)\tilde{}} + J^{(-)\tilde{}} \\ &= \frac{\hbar}{2im}\begin{bmatrix} \text{Œ}_{+}{}^{\dagger}\text{ŕ}^{\tilde{}}\dagger^{\text{€}}\left(\partial_{\text{€}} - \dfrac{e}{i\hbar}A_{\text{€}}\right)\text{Œ}_{+} - \text{Œ}_{+}\text{ŕ}^{\tilde{}}\dagger^{\text{€}}\left(\partial_{\text{€}} + \dfrac{e}{i\hbar}A_{\text{€}}\right)\text{Œ}_{+}{}^{\dagger} \\ +(\text{Œ}_{-}{}^{\dagger}\dagger^{\tilde{}}\text{ŕ}^{\text{€}}\left(\partial_{\text{€}} - \dfrac{e}{i\hbar}A_{\text{€}}\right)\text{Œ}_{-} - \text{Œ}_{-}\dagger^{\tilde{}}\text{ŕ}^{\text{€}}\left(\partial_{\text{€}} + \dfrac{e}{i\hbar}A_{\text{€}}\right)\text{Œ}_{-}{}^{\dagger}\ ) \end{bmatrix} \end{aligned} \qquad (2.0.14)$$

that by using (2.0.5-6), leads to [21] the known relation

$$\begin{aligned} J^{\tilde{}} &= -\frac{\hbar}{2im}\begin{bmatrix} \text{Œ}_{+}{}^{\dagger}\text{ŕ}^{\tilde{}}\text{Œ}_{-} - \text{Œ}_{+}\text{ŕ}^{\tilde{}}\text{Œ}_{-}{}^{\dagger} \\ +\text{Œ}_{-}{}^{\dagger}\text{ŕ}^{\tilde{}}\text{Œ}_{+} + \text{Œ}_{-}\text{ŕ}^{\tilde{}}\text{Œ}_{+}{}^{\dagger} \end{bmatrix} \\ &= -\frac{\hbar}{4im}\begin{bmatrix} \left(\text{Œ}^{1} + \text{Œ}^{2}\right)^{\dagger}\text{ŕ}^{\tilde{}}\left(\text{Œ}^{1} - \text{Œ}^{2}\right) + \left(\text{Œ}^{1} + \text{Œ}^{2}\right)\text{ŕ}^{\tilde{}}\left(\text{Œ}^{1} - \text{Œ}^{2}\right)^{\dagger} \\ +\left(\text{Œ}^{1} - \text{Œ}^{2}\right)^{\dagger}\text{ŕ}^{\tilde{}}\left(\text{Œ}^{1} + \text{Œ}^{2}\right) + \left(\text{Œ}^{1} - \text{Œ}^{2}\right)\text{ŕ}^{\tilde{}}\left(\text{Œ}^{1} + \text{Œ}^{2}\right)^{\dagger} \end{bmatrix} \\ &= \frac{\hbar}{4im}\begin{bmatrix} -\text{Œ}^{1\dagger}\text{ŕ}^{\tilde{}}\text{Œ}^{1} + \text{Œ}^{2\dagger}\text{ŕ}^{\tilde{}}\text{Œ}^{2} - \text{Œ}^{1}\text{ŕ}^{\tilde{}}\text{Œ}^{1\dagger} + \text{Œ}^{2}\text{ŕ}^{\tilde{}}\text{Œ}^{2\dagger} \\ -\text{Œ}^{1\dagger}\text{ŕ}^{\tilde{}}\text{Œ}^{1} + \text{Œ}^{2\dagger}\text{ŕ}^{\tilde{}}\text{Œ}^{2} - \text{Œ}^{1}\text{ŕ}^{\tilde{}}\text{Œ}^{1\dagger} + \text{Œ}^{2}\text{ŕ}^{\tilde{}}\text{Œ}^{2\dagger} \end{bmatrix} \\ &= \frac{\hbar}{2im}\left[-\text{Œ}^{1\dagger}\text{ŕ}^{\tilde{}}\text{Œ}^{1} + \text{Œ}^{2\dagger}\text{ŕ}^{\tilde{}}\text{Œ}^{2} - \text{Œ}^{1}\text{ŕ}^{\tilde{}}\text{Œ}^{1\dagger} + \text{Œ}^{2}\text{ŕ}^{\tilde{}}\text{Œ}^{2\dagger}\right] \\ &= -\frac{\hbar}{2im}\left[\overline{\Psi}\text{x}^{\tilde{}}\Psi + \overline{\Psi^{*}}\text{x}^{\tilde{}}\Psi^{*}\right] = J^{(+)\tilde{}} + J^{(-)\tilde{}} = -\frac{\hbar}{im}\overline{\Psi}\text{x}^{\tilde{}}\Psi \end{aligned} \qquad (2.0.15)$$

**2.1. The solution of the hydrodynamic Hamilton-Jacobi equation for eigenstates**

Bu using the hydrodynamic notations

$$\partial_\mu S_{\pm i} = \partial_\mu \begin{pmatrix} S_{\pm 1} \\ S_{\pm 2} \end{pmatrix} = -\begin{pmatrix} p_{\pm 1\mu} \\ p_{\pm 2\mu} \end{pmatrix} = -p_{\pm i\mu} \tag{2.1.1}$$

the Hamilton-Jacobi equation (2.0.10), as a function of the spinors components,

$$\left(\partial_\mu \begin{pmatrix} S_{\pm 1} \\ S_{\pm 2} \end{pmatrix} + eA_\mu\right)\left(\partial^\mu \begin{pmatrix} S_{\pm 1} \\ S_{\pm 2} \end{pmatrix} + eA^\mu\right) = m^2c^2\left(\mathbf{1} - \frac{1}{mc^2}\begin{pmatrix} V_{qu\pm 1} \\ V_{qu\pm 2} \end{pmatrix}\right) \tag{2.1.2}$$

for the $k$-th eigenstate reads

$$\begin{aligned} &\left(\partial_\mu \begin{pmatrix} S_{\pm 1(k)} \\ S_{\pm 2(k)} \end{pmatrix} + eA_\mu\right)\left(\partial^\mu \begin{pmatrix} S_{\pm 1(k)} \\ S_{\pm 2(k)} \end{pmatrix} + eA^\mu\right) = -\left(\begin{pmatrix} p_{\pm 1(k)\mu} \\ p_{\pm 2(k)\mu} \end{pmatrix} - eA_\mu\right)\left(\begin{pmatrix} p_{\pm 1(k)}{}^{\mu} \\ p_{\pm 2(k)}{}^{\mu} \end{pmatrix} - eA^\mu\right) \\ &= m^2c^2\left(\mathbf{1} - \frac{1}{mc^2}\begin{pmatrix} V_{qu\pm 1(k)} \\ V_{qu\pm 2(k)} \end{pmatrix}\right) \end{aligned} \tag{2.1.3}$$

Moreover, by posing

$$\gamma_{\pm i} = \frac{1}{\sqrt{1 - \frac{\dot{q}_{\pm i}{}^{2}}{c^2}}} = \frac{1}{\sqrt{\frac{g_{\mu\nu}\dot{q}^{\nu}_{\pm i}\dot{q}^{\mu}_{\pm i}}{c^2}}}, \tag{2.1.4}$$

by using the following identity

$$\begin{aligned} &\left(\partial_\mu S_{\pm i} + eA_\mu\right)\left(\partial^\mu S_{\pm i} + eA^\mu\right) = m^2\gamma^2_{\pm i}\dot{q}_{\mu\pm i}\dot{q}^{\mu}{}_{\pm i}\left(\mathbf{1} - \frac{V_{qu\pm i}}{mc^2}\right) \\ &= m^2\gamma^2_{\pm i}\, c^2\left(\mathbf{1} - \frac{V_{qu\pm i}}{mc^2}\right) - m^2\gamma^2_{\pm i}\, \dot{q}^2_{\pm i} - \left(\mathbf{1} - \frac{V_{qu\pm i}}{mc^2}\right) \\ &= \left(p_{\pm i\mu} + eA_\mu\right)\left(p_{\pm i}{}^{\mu} + eA^\mu\right) = m^2\gamma^2_{\pm i}\dot{q}_{\mu\pm i}\dot{q}^{\mu}{}_{\pm i}\left(1 - \frac{V_{qu\pm i}}{mc^2}\right) \end{aligned} \tag{2.1.5}$$

and by using the notation

$$p_\mu{}^{(\pm)}{}_{\pm i(k)} = \pm p_\mu{}_{\pm i(k)}, \tag{2.1.6}$$

where the superscript $(\pm)$ stands for positive and negative solutions of (2.1.5), respectively, for the $k$ eigenstate it follows that

$$p_{\mu\pm i(k)} = \left(m\gamma_{\pm i}\, \dot{q}_{\mu\pm i(k)}\sqrt{1 - \frac{V_{qu\pm i(k)}}{mc^2}} + eA_\mu\right) \tag{2.1.7}$$

$$E_{\pm i(k)} - eW = mX_{\pm i}c^2\sqrt{1-\frac{V_{qu\pm i(k)}}{mc^2}} \qquad (2.1.8)$$

From (2.1.1, 2.1.7) it follows that the hydrodynamic Lagrangean for the eigenstates reads

$$L_{\pm i(k)} = \frac{dS_{\pm i(k)}}{dt} = -p_{\pm i(k)\sim}\,\dot{q}_{\pm i(k)}{}^{\sim} = -\frac{mc^2}{\gamma}\sqrt{1-\frac{V_{qu\pm i(k)}}{mc^2}} + eA_{\sim}\dot{q}_{\pm i(k)}{}^{\sim} \qquad (2.1.9)$$

where $V_{qu\pm i(k)}$ is given by (2.0.11) and where the Lagrangean for negative-energy states reads

$$L^{(-)}_{\pm i(k)} = -p^{(-)}{}_{\pm i(k)\sim}\,\dot{q}_{\pm i(k)}{}^{\sim} = p_{\pm i(k)\sim}\,\dot{q}_{\pm i(k)}{}^{\sim} = -L_{\pm i(k)}\ . \qquad (2.1.10)$$

Moreover, it is useful to recast (2.1.7, 2.1.9) as follows, respectively,

$$p_{\sim\pm i(k)} = p_{\sim\pm i(k)class} + p_{\sim\pm i(k)Q} \qquad (2.1.11)$$

$$p_{\sim\pm i(k)class} = \left(m\gamma_{\pm i}\ \dot{q}_{\sim\pm i(k)} + eA_{\sim}\right) = \frac{p_{\sim\pm i(k)} - eA_{\sim}\left(1-\sqrt{1-\frac{V_{qu\pm i(k)}}{mc^2}}\right)}{\sqrt{1-\frac{V_{qu\pm i(k)}}{mc^2}}}$$
$$= -\frac{\partial_{\sim}S_{\pm i(k)} + eA_{\sim}a_{\left(V_{qu\pm i(k)}\right)}}{1-a_{\left(V_{qu\pm i(k)}\right)}} \qquad (2.1.12)$$

$$p_{\sim\pm i(k)Q} = -a_{\left(V_{qu\pm i(k)}\right)}m\gamma_{\pm i}\ \dot{q}_{\sim\pm i(k)}$$
$$= -a_{\left(V_{qu\pm i(k)}\right)}\left(p_{\sim\pm i(k)class} - eA_{\sim}\right) \qquad (2.1.13)$$

and

$$L_{\pm i(k)} = -\frac{mc^2}{\gamma}\sqrt{1-\frac{V_{qu\pm i(k)}}{mc^2}} + eA_{\sim}\dot{q}_{\pm i(k)}{}^{\sim} = L_{\pm i(k)class} + L_{\pm i(k)Q}\,, \qquad (2.1.14)$$

where

$$L_{\pm i(k)class} = -\frac{mc^2}{\gamma} + eA_{\sim}\dot{q}_{\pm i(k)}{}^{\sim}\,, \qquad (2.1.15)$$

$$L_{\pm i(k)Q} = a_{\left(V_{qu\pm i(k)}\right)}\frac{mc^2}{\gamma}$$
$$= -\left(1-\sqrt{1-\frac{V_{qu\pm i(k)}}{mc^2}}\right)\left(L_{\pm i(k)class} - eA_{\sim}\dot{q}_{\pm i(k)}{}^{\sim}\right) \qquad (2.1.16)$$

where

$$a_{\left(V_{qu\pm i(k)}\right)} = \left( 1 - \sqrt{1 - \frac{V_{qu\pm i(k)}}{mc^2}} \right), \tag{2.1.17}$$

**2.2. The Lagrangean motion equations for the Dirac field**

Since the quantum hydrodynamic description submitted to irrotational condition is equivalent to the Dirac-Fock-Weil equation (DFWE) [22], given the generic fermion field

$$\text{Œ}_{\pm i} = \begin{pmatrix} \text{Œ}_{\pm 1} \\ \text{Œ}_{\pm 2} \end{pmatrix} = \begin{pmatrix} /\text{Œ}_{\pm 1}/\, exp\left[ i \frac{S_{\pm 1}}{\hbar} \right] \\ /\text{Œ}_{\pm 2}/\, exp\left[ i \frac{S_{\pm 2}}{\hbar} \right] \end{pmatrix} = \begin{pmatrix} \sum\limits_{k'=-\infty}^{\infty} a_{\pm 1k}\, exp[\frac{iS_{\pm 1(k')}}{\hbar}] \\ \sum\limits_{k'=-\infty}^{\infty} a_{\pm 2k}\, exp[\frac{iS_{\pm 2(k')}}{\hbar}] \end{pmatrix}$$

$$= \begin{pmatrix} \sum\limits_{k=0}^{\infty} \text{r}_{\pm 1k}\, exp[\frac{iS_{\pm 1(k)}}{\hbar}] + \text{s}^{\dagger}_{\pm 1k}\, exp[\frac{iS_{\pm 1(-k)}}{\hbar}] \\ \sum\limits_{k=0}^{\infty} \text{r}_{\pm 2k}\, exp[\frac{iS_{\pm 2(k)}}{\hbar}] + \text{s}^{\dagger}_{\pm 2k}\, exp[\frac{iS_{\pm 2(-k)}}{\hbar}] \end{pmatrix} \tag{2.2.1}$$

$$\equiv \sum_{k=0}^{\infty} \text{r}_{\pm ik}\, exp[\frac{iS_{\pm i(k)}}{\hbar}] + \text{s}^{\dagger}_{\pm ik}\, exp[\frac{iS_{\pm i(-k)}}{\hbar}]$$

the action reads

$$S_{\pm i} = \frac{\hbar}{2i} ln[\frac{\text{Œ}_{\pm i}}{\text{Œ}_{\pm i}{}^{*}}] = \frac{\hbar}{2i} \begin{pmatrix} ln[\sum\limits_{k'=-\infty}^{\infty} a_{\pm ik}\, exp(\frac{iS_{\pm i(k')}}{\hbar})] \\ -ln[\sum\limits_{k'=-\infty}^{\infty} a^{*}{}_{\pm ik}\, exp(-\frac{iS_{\pm i(k')}}{\hbar})] \end{pmatrix}$$

$$= \frac{\hbar}{2i} \begin{pmatrix} ln[\sum\limits_{k=0}^{\infty} \text{r}_{\pm ik}\, exp(\frac{iS_{\pm i(k)}}{\hbar}) + \text{s}^{*}_{\pm ik}\, exp(\frac{iS_{\pm i(-k)}}{\hbar})] \\ -ln[\sum\limits_{k=0}^{\infty} \text{r}^{*}_{\pm ik}\, exp(-\frac{iS_{\pm i(k)}}{\hbar}) + \text{s}_{\pm ik}\, exp(-\frac{iS_{\pm i(-k)}}{\hbar})] \end{pmatrix} \tag{2.2.2}$$

where for the $k$ -eigenstates $S_{\pm i(k)}$ reads

$$S_{\pm i(k)} = \begin{pmatrix} S_{\pm 1(k)} \\ S_{\pm 2(k)} \end{pmatrix} = \frac{\hbar}{2i} \begin{pmatrix} ln \frac{\text{Œ}_{\pm 1(k)}}{\text{Œ}_{\pm 1(k)}{}^{*}} \\ ln \frac{\text{Œ}_{\pm 2(k)}}{\text{Œ}_{\pm 2(k)}{}^{*}} \end{pmatrix}. \tag{2.2.3}$$

Moreover, from (2.2.1), the hydrodynamic momentum $p_{\pm i \sim}$ and the Lagrangean function $L_{\pm}$ read, respectively,

$$p_{\pm i\sim} = -\partial_{\sim} S_{\pm i} = -\frac{1}{2}\frac{\sum_{k=0}\left(\begin{array}{l}\mathsf{r}_{\pm ik}\; exp(\frac{iS_{\pm i(k)}}{\hbar})\left(\frac{\hbar}{i}\partial_{\sim}\, ln\,\mathsf{r}_{\pm ik} - p_{\sim_{\pm i(k)}}\right)\\ +\mathsf{s}^{*}_{\pm ik}\; exp(\frac{iS_{\pm i(-k)}}{\hbar})\left(\frac{\hbar}{i}\partial_{\sim}\, ln\,\mathsf{s}^{*}_{\pm ik} - p_{\sim_{\pm i(-k)}}\right)\end{array}\right)}{\sum_{k=0}\mathsf{r}_{\pm ik}\; exp(\frac{iS_{\pm i(k)}}{\hbar}) + \mathsf{s}^{*}_{\pm ik}\; exp(\frac{iS_{\pm i(-k)}}{\hbar})}$$

$$+\frac{1}{2}\frac{\sum_{k=0}\left(\begin{array}{l}\mathsf{r}^{*}_{\pm ik}\; exp(-\frac{iS_{\pm i(k)}}{\hbar})\left(\frac{\hbar}{i}\partial_{\sim}\, ln\,\mathsf{r}^{*}_{\pm ik} + p_{\sim_{\pm i(k)}}\right)\\ +\mathsf{s}_{\pm ik}\; exp(-\frac{iS_{\pm i(-k)}}{\hbar})\left(\frac{\hbar}{i}\partial_{\sim}\, ln\,\mathsf{s}^{*}_{\pm ik} + p_{\sim_{\pm i(-k)}}\right)\end{array}\right)}{\sum_{k=0}\mathsf{r}^{*}_{\pm ik}\; exp(-\frac{iS_{\pm i(k)}}{\hbar}) + \mathsf{s}_{\pm ik}\; exp(-\frac{iS_{\pm i(-k)}}{\hbar})\, )]} = Tr\left(\boldsymbol{p}_{\pm i\sim}\right) \qquad (2.2.4)$$

where the matrix $\boldsymbol{p}_{\pm i\sim}$ reads

$$\boldsymbol{p}_{\pm i\sim} = \left(\begin{array}{c} -\frac{1}{2}\frac{\left(\begin{array}{l}\mathsf{r}_{\pm ik}\; exp(\frac{iS_{\pm i(k)}}{\hbar})\left(\frac{\hbar}{i}\partial_{\sim}\, ln\,\mathsf{r}_{\pm ik} - p_{\sim_{\pm i(k)}}\right)\\ +\mathsf{s}^{*}_{\pm ik}\; exp(\frac{iS_{\pm i(-k)}}{\hbar})\left(\frac{\hbar}{i}\partial_{\sim}\, ln\,\mathsf{s}^{*}_{\pm ik} - p_{\sim_{\pm i(-k)}}\right)\end{array}\right)}{\sum_{k_j=0}\mathsf{r}_{\pm ik_j}\; exp(\frac{iS_{\pm i(k_j)}}{\hbar}) + \mathsf{s}^{*}_{\pm ik_j}\; exp(\frac{iS_{\pm i(-k_j)}}{\hbar})\, )} \\ +\frac{1}{2}\frac{\left(\begin{array}{l}\mathsf{r}^{*}_{\pm ik}\; exp(-\frac{iS_{\pm i(k)}}{\hbar})\left(\frac{\hbar}{i}\partial_{\sim}\, ln\,\mathsf{r}^{*}_{\pm ik} + p_{\sim_{\pm i(k)}}\right)\\ +\mathsf{s}_{\pm ik}\; exp(-\frac{iS_{\pm i(-k)}}{\hbar})\left(\frac{\hbar}{i}\partial_{\sim}\, ln\,\mathsf{s}_{\pm ik} + p_{\sim_{\pm i(-k)}}\right)\end{array}\right)}{\sum_{k_j=0}\mathsf{r}^{*}_{\pm ik_j}\; exp(-\frac{iS_{\pm i(k_j)}}{\hbar}) + \mathsf{s}_{\pm ik_j}\; exp(-\frac{iS_{\pm i(-k_j)}}{\hbar})\, )]} \end{array}\right)\mathsf{u}_{mk} = \tilde{p}_{(k)\pm i\sim}\,\mathsf{u}_{mk}\,, \qquad (2.2.5)$$

$$L_{\pm i}=\frac{dS_{\pm i}}{dt}=\frac{\partial S_{\pm i}}{\partial t}+\frac{\partial S_{\pm i}}{\partial q_i}\dot{q}_i=-p_{\sim\pm i}\dot{q}^{\sim}{}_{\pm i}=Tr\left(\boldsymbol{L}_{\pm i}\right)$$

$$=\frac{1}{2}\frac{\sum\limits_{k=0}\left(\begin{array}{l}\mathrm{r}_{\pm ik}\,exp(\frac{iS_{\pm i(k)}}{\hbar})\left(\frac{\hbar}{i}\dot{q}^{\sim}{}_{\pm i}\partial_{\sim}\,ln\,\mathrm{r}_{\pm ik}-p_{\sim_{\pm i(k)}}\dot{q}^{\sim}{}_{\pm i}\right)\\+\mathrm{s}^{*}{}_{\pm ik}\,exp(\frac{iS_{\pm i(-k)}}{\hbar})\left(\frac{\hbar}{i}\dot{q}^{\sim}{}_{\pm i}\partial_{\sim}\,ln\,\mathrm{s}^{*}{}_{\pm ik}-p_{\sim_{\pm i(-k)}}\dot{q}^{\sim}{}_{\pm i}\right)\end{array}\right)}{\sum\limits_{k=0}\left(\mathrm{r}_{\pm ik}\,exp(\frac{iS_{\pm i(k)}}{\hbar})+\mathrm{s}\,{*}_{\pm ik}\,/\text{Œ}_{\pm 1(-k)}/\,exp(\frac{iS_{\pm i(-k)}}{\hbar})\right)}\,.$$
$$-\frac{1}{2}\frac{\sum\limits_{k=0}\left(\begin{array}{l}\mathrm{r}^{*}{}_{\pm ik}\,exp(-\frac{iS_{\pm i(k)}}{\hbar})\left(\frac{\hbar}{i}\dot{q}^{\sim}{}_{\pm i}\partial_{\sim}\,ln\,\mathrm{r}^{*}{}_{\pm ik}+p_{\sim_{\pm i(k)}}\dot{q}^{\sim}{}_{\pm i}\right)\\+\mathrm{s}_{\pm ik}\,exp(-\frac{iS_{\pm i(-k)}}{\hbar})\left(\frac{\hbar}{i}\dot{q}^{\sim}{}_{\pm i}\partial_{\sim}\,ln\,\mathrm{s}_{\pm ik}+p_{\sim_{\pm i(-k)}}\dot{q}^{\sim}{}_{\pm i}\right)\end{array}\right)}{\sum\limits_{k=0}\left(\mathrm{r}^{*}{}_{\pm ik}\,exp(-\frac{iS_{\pm i(k)}}{\hbar})+\mathrm{s}_{\pm ik}\,exp(-\frac{iS_{\pm i(-k)}}{\hbar})\right)}\qquad(2.2.6)$$

where

$$\boldsymbol{L}_{\pm i}=-\boldsymbol{p}_{\sim_{\pm i}}\dot{q}^{\sim}{}_{\pm i}=-\tilde{p}_{\sim_{\pm i(k)}}\dot{q}^{\sim}{}_{\pm i}\mathrm{u}_{m\tilde{k}}=\tilde{L}_{\pm i(k)}\mathrm{u}_{m\tilde{k}}\qquad(2.2.7)$$

where

$$\dot{q}_{\sim\pm i}=\frac{\left(p_{\sim\pm i}+eA_{\sim}\right)}{m\mathrm{x}_{\pm i}\sqrt{1-\frac{V_{qu\pm i}}{mc^{2}}}}=c^{2}\frac{\left(p_{\sim\pm i}+eA_{\sim}\right)}{E_{\pm i}}=-c^{2}\frac{\left(p_{\sim\pm i}+eA_{\sim}\right)}{\frac{\partial S_{\pm i}}{\partial t}+e\mathrm{W}}\qquad(2.2.8)$$

where

$$\tilde{L}_{\pm i(k)}=\frac{1}{2}\frac{\left(\begin{array}{l}\mathrm{r}_{\pm ik}\,exp(\frac{iS_{\pm i(k)}}{\hbar})\left(\frac{\hbar}{i}\dot{q}^{\tilde{}}_{\pm i}\partial_{\sim}\,ln\,\mathrm{r}_{\pm ik}-p_{\sim_{\pm i(k)}}\dot{q}^{\sim}{}_{\pm i}\right)\\+\mathrm{s}^{*}{}_{\pm ik}\,exp(\frac{iS_{\pm i(-k)}}{\hbar})\left(\frac{\hbar}{i}\dot{q}^{\tilde{}}_{\pm i}\partial_{\sim}\,ln\,\mathrm{s}^{*}{}_{\pm ik}-p_{\sim_{\pm i(-k)}}\dot{q}^{\sim}{}_{\pm i}\right)\end{array}\right)}{\sum\limits_{k_j=0}\left(\mathrm{r}_{\pm ik_j}\,exp(\frac{iS_{\pm i(k_j)}}{\hbar})+\mathrm{s}\,{*}_{\pm ik_j}\,exp(\frac{iS_{\pm i(-k_j)}}{\hbar})\right)}\,.$$
$$-\frac{1}{2}\frac{\left(\begin{array}{l}\mathrm{r}^{*}{}_{\pm ik}\,exp(-\frac{iS_{\pm i(k)}}{\hbar})\left(\frac{\hbar}{i}\dot{q}^{\tilde{}}_{\pm i}\partial_{\sim}\,ln\,\mathrm{r}^{*}{}_{\pm ik}+p_{\sim_{\pm i(k)}}\dot{q}^{\sim}{}_{\pm i}\right)\\+\mathrm{s}_{\pm i\tilde{k}}\,exp(-\frac{iS_{\pm i(-\tilde{k})}}{\hbar})\left(\frac{\hbar}{i}\dot{q}^{\tilde{}}_{\pm i}\partial_{\sim}\,ln\,\mathrm{s}_{\pm i\tilde{k}}+p_{\sim_{\pm i(-\tilde{k})}}\dot{q}^{\sim}{}_{\pm i}\right)\end{array}\right)}{\sum\limits_{k_j=0}\left(\mathrm{r}\,{*}_{\pm ik_j}\,exp(-\frac{iS_{\pm i(k_j)}}{\hbar})+\mathrm{s}_{\pm ik_j}\,exp(-\frac{iS_{\pm i(-k_j)}}{\hbar})\right)}\qquad(2.2.9)$$

Thence, by using (2.2.5, 2.2.9) it follows that

$$-\frac{\partial \tilde{L}_{\pm i(k)}}{\partial \dot{q}^{\sim}} = -\frac{1}{2}\frac{\left(\begin{array}{l}\mathrm{r}_{\pm i\tilde{k}}\, exp(\frac{iS_{\pm i(k)}}{\hbar})\left(\frac{\hbar}{i}\partial_{\sim}\, ln\,\mathrm{r}_{\pm ik} + \frac{\partial p_{\sim\pm i(k)}\dot{q}^{\sim}{}_{\pm i}}{\partial \dot{q}^{\sim}}\right)\\ +\mathrm{s}^{*}{}_{\pm i\tilde{k}}\, exp(\frac{iS_{\pm i(-k)}}{\hbar})\left(\frac{\hbar}{i}\partial_{\sim}\, ln\,\mathrm{s}^{*}{}_{\pm i\tilde{k}} + \frac{\partial p_{\sim\pm i(-k)}\dot{q}^{\sim}{}_{\pm i}}{\partial \dot{q}^{\sim}}\right)\end{array}\right)}{\sum_{k_j=0}\left(\mathrm{r}_{\pm ik_j}\, exp(\frac{iS_{\pm i(k_j)}}{\hbar}) + \mathrm{s}^{*}{}_{\pm ik_j}\, exp(\frac{iS_{\pm i(-k_j)}}{\hbar})\right)}$$

$$+\frac{1}{2}\frac{\left(\begin{array}{l}\mathrm{r}^{*}{}_{\pm ik}\, exp(-\frac{iS_{\pm i(k)}}{\hbar})\left(\frac{\hbar}{i}\partial_{\sim}\, ln\,\mathrm{r}^{*}{}_{\pm ik} - \frac{\partial p_{\sim\pm i(k)}\dot{q}^{\sim}{}_{\pm i}}{\partial \dot{q}^{\sim}}\right)\\ +\mathrm{s}_{\pm ik}\, exp(-\frac{iS_{\pm i(-k)}}{\hbar})\left(\frac{\hbar}{i}\partial_{\sim}\, ln\,\mathrm{s}_{\pm ik} - \frac{\partial p_{\sim\pm i(-k)}\dot{q}^{\sim}{}_{\pm i}}{\partial \dot{q}^{\sim}}\right)\end{array}\right)}{\sum_{j=0}\left(\mathrm{r}^{*}{}_{\pm ik_j}\, exp(-\frac{iS_{\pm i(k_j)}}{\hbar}) + \mathrm{s}_{\pm ik_j}\, exp(-\frac{iS_{\pm i(-k_j)}}{\hbar})\right)}$$

$$= \tilde{p}_{(k)\pm i\sim} \quad , \qquad (2.2.10)$$

Moreover, given that, for the $\tilde{k}$-eigenstates

$$\tilde{L}_{(\tilde{k})\pm i} = L_{(\tilde{k})\pm i} \qquad (2.2.11)$$

$$\tilde{p}_{(\tilde{k})\pm i_{\sim}} = p_{(\tilde{k})\pm i_{\sim}} \qquad (2.2.12)$$

and that the hydrodynamic Lagrangean function of th $\tilde{k}$–eigenstate $L_{\pm i(\tilde{k})}$ (2.1.9) of time-independent systems does not explicitly depend by time, it follows that

$$-\frac{\partial L_{\pm i(\tilde{k})}}{\partial q^{\sim}} = -\frac{\partial}{\partial q^{\sim}}\frac{dS_{\pm i(\tilde{k})}}{dt} = \left(0, \frac{\partial}{\partial q_i}\frac{dS_{\pm i(\tilde{k})}}{dt}\right)$$

$$= \left(0, \frac{d}{dt}\frac{\partial S_{\pm i(\tilde{k})}}{\partial q_i}\right) = \frac{d}{dt}\left(-\frac{\partial S_{\pm i(\tilde{k})}}{\partial q^{\sim}}\right) = \dot{p}_{\pm i(\tilde{k})_{\sim}} \quad , \qquad (2.2.13)$$

and that, for the eigenstates, the Lagrangean hydrodynamic motion equations read

$$p_{(\tilde{k})\pm i_{\sim}} = -\frac{\partial L_{(\tilde{k})\pm i}}{\partial \dot{q}^{\sim}_{\pm i(\tilde{k})}} , \qquad (2.2.14)$$

$$\dot{p}_{(\tilde{k})\pm i_{\sim}} = -\frac{\partial L_{(\tilde{k})\pm i}}{\partial q^{\sim}} \qquad (2.2.15)$$

leading to the motion equation of spinors

$$\frac{d}{dt}\left(-\frac{\partial L_{\pm i(\tilde{k})}}{\partial \dot{q}_{\pm i(\tilde{k})}{}^{\sim}}\right) = -\frac{\partial L_{\pm i(\tilde{k})}}{\partial q^{\sim}} \quad . \qquad (2.2.16)$$

It must be observed that, in principle, the solutions of (2.2.16) have to be submitted to the irrotational condition [22]) and to the current conservation condition (2.0.12).
On the other hand, as shown in [23] the stationary states (i.e., the eigenstates), of (2.2.16) obey to the current conservation (2.0.12) and are irrotational solutions of the DFWE.
Equation (2.2.16) (whose non-relativistic limit (i.e., the Pauli equation) is given in ref. [22]) is not solved here since we are not interested in the quantum hydrodynamic description but in deriving the Lagrangean quantum hydrodynamic equation of motion from a minimum action principle.
Moreover, generally speaking, by using the identities (2.2.5, 2.2.9) the motion equation

$$
\begin{aligned}
-\frac{\partial \tilde{L}_{(k)\pm i}}{\partial q^{\sim}} &= -\frac{\partial}{\partial q^{\sim}}\left( \tilde{p}_{(k)\pm i\text{€}}\, \dot{q}_{(k)\pm i}{}^{\text{€}} \right) = -\left( \frac{\partial}{\partial q^{\sim}}\, \tilde{p}_{(k)\pm i\text{€}} \right) \dot{q}_{(k)\pm i}{}^{\text{€}} - \frac{\partial \tilde{S}_{(k)\pm i}}{\partial q^{\text{€}}} \frac{\partial \dot{q}_{(k)\pm i}{}^{\text{€}}}{\partial q^{\sim}} \\
&= \dot{\tilde{p}}_{(k)\pm i_{\sim}} + \tilde{p}_{(k)\pm i\text{€}} \frac{\partial \dot{q}_{(k)\pm i\text{€}}}{\partial q^{\sim}} = \frac{d}{dt}\left( -\frac{\partial \tilde{L}_{(k)\pm i}}{\partial \dot{q}^{\sim}_{\pm i(k)}} \right) + \tilde{p}_{(k)\pm i\text{€}} \frac{\partial \dot{q}_{(k)\pm i\text{€}}}{\partial q^{\sim}}
\end{aligned} \quad . \quad (2.2.17)
$$

of the generic DFWE state (2.2.1) follows.
Moreover, by making the summation over $k$ equation (2.2.17) leads to

$$
Tr(\, \dot{\boldsymbol{p}}_{\pm i_{\sim}} \,) = \frac{d}{dt} Tr(\, \boldsymbol{p}_{\pm i_{\sim}} \,) = -\frac{\partial Tr\left( \boldsymbol{L}_{\pm i} \right)}{\partial q^{\sim}} - Tr\left( \boldsymbol{p}_{\pm i\text{€}}\, \partial_{\sim} \dot{\boldsymbol{q}}_{\pm i\text{€}} \right) \qquad (2.2.18)
$$

or to

$$
\dot{p}_{\pm i_{\sim}} = -\frac{\partial L_{\pm i}}{\partial q^{\sim}} - \sum_{k} \tilde{p}_{(k)\pm i\text{€}} \frac{\partial \dot{q}_{(k)\pm i\text{€}}}{\partial q^{\sim}} \qquad (2.2.19)
$$

where, in the following, it is useful to pose

$$
Tr\left( \boldsymbol{L}_{\pm i} \right) = L_{\pm i} = L_{\pm i_{Class}} + L_{\pm i_{Q}} + L_{\pm i_{mix}} \, , \qquad (2.2.20)
$$

where

$$
\begin{aligned}
L_{\pm i_{class}} = &-\frac{1}{2} \frac{\sum_{k=0} \left( \mathsf{r}_{\pm ik}\, e^{\frac{iS_{\pm i(k)}}{\hbar}}\, p_{\sim_{\pm i(k)class}}\, \dot{q}^{\sim}{}_{\pm i} + \mathsf{s}^{*}{}_{\pm ik}\, e^{\frac{iS_{\pm i(-k)}}{\hbar}}\, p_{\sim_{\pm i(-k)class}}\, \dot{q}^{\sim}{}_{\pm i} \right)}{\sum_{k=0} \left( \mathsf{r}_{\pm ik}\, e^{\frac{iS_{\pm i(k)}}{\hbar}} + \mathsf{s}^{*}{}_{\pm ik}\, e^{\frac{iS_{\pm i(-k)}}{\hbar}} \right)} \\
&-\frac{1}{2} \frac{\sum_{k=0} \left( \mathsf{r}^{*}{}_{\pm ik}\, e^{-\frac{iS_{\pm i(k)}}{\hbar}}\, p_{\sim_{\pm i(k)class}}\, \dot{q}^{\sim}{}_{\pm i} + \mathsf{s}_{\pm ik}\, e^{-\frac{iS_{\pm i(-k)}}{\hbar}}\, p_{\sim_{\pm i(-k)class}}\, \dot{q}^{\sim}{}_{\pm i} \right)}{\sum_{k=0} \left( \mathsf{r}^{*}{}_{\pm ik}\, e^{-\frac{iS_{\pm i(k)}}{\hbar}} + \mathsf{s}_{\pm ik}\, e^{-\frac{iS_{\pm i(-k)}}{\hbar}} \right)}
\end{aligned} \, , \quad (2.2.21)
$$

$$\tilde{L}_{\pm iQ} = -\frac{1}{2}\frac{\sum_{k=0}\left(\begin{array}{l} r_{\pm ik}\, e^{\frac{iS_{\pm i(k)}}{\hbar}}\, a_{\left(V_{qu_{\pm i}(k)}\right)}\left(p_{\sim \pm i(k)class}\,\dot{q}^{\sim}{}_{\pm i} + eA_{\sim}\dot{q}_{\pm i}{}^{\sim}\right)\\ +s^{*}{}_{\pm ik}\, e^{\frac{iS_{\pm i(-k)}}{\hbar}}\, a_{\left(V_{qu_{\pm i}(-k)}\right)}\left(+p_{\sim \pm i(-k)class}\,\dot{q}^{\sim}{}_{\pm i} + eA_{\sim}\dot{q}_{\pm i}{}^{\sim}\right)\end{array}\right)}{\sum_{k=0}\left(r_{\pm ik}\, e^{\frac{iS_{\pm i(k)}}{\hbar}} + s^{*}{}_{\pm ik}\, e^{\frac{iS_{\pm i(-k)}}{\hbar}}\right)}$$

$$-\frac{1}{2}\frac{\sum_{k=0}\left(\begin{array}{l} r^{*}{}_{\pm ik}\, e^{-\frac{iS_{\pm i(k)}}{\hbar}}\, a_{\left(V_{qu_{\pm i}(k)}\right)}\left(p_{\sim \pm i(k)class}\,\dot{q}^{\sim}{}_{\pm i} + eA_{\sim}\dot{q}_{\pm i}{}^{\sim}\right)\\ +s_{\pm ik}\, e^{-\frac{iS_{\pm i(-k)}}{\hbar}}\, a_{\left(V_{qu_{\pm i}(-k)}\right)}\left(p_{\sim \pm i(-k)class}\,\dot{q}^{\sim}{}_{\pm i} + eA_{\sim}\dot{q}_{\pm i}{}^{\sim}\right)\end{array}\right)}{\sum_{k=0}\left(r^{*}{}_{\pm ik}\, e^{-\frac{iS_{\pm i(k)}}{\hbar}} + s_{\pm ik}\, e^{-\frac{iS_{\pm i(-k)}}{\hbar}}\right)} \qquad (2.2.22)$$

(2.2.23)

## 2.3. The hydrodynamic energy-impulse tensor of the Dirac field

In this section we derive the expression of the hydrodynamic energy-impulse tensor as a function of the Dirac field.

Given the EITD $T_{\pm i(k)\sim}{}^{€}{}_{F} = /Œ_{\pm i(k)}/^{2}\, T_{\pm i(\pm k)\sim}{}^{€}$ [21] where

$$T_{\pm i(k)\sim}{}^{\text{€}} = -\left(\frac{\partial L_{\pm i(\tilde{k})}}{\partial \dot{q}^{\sim}}\dot{q}^{\text{€}}_{\pm i(k)} - L_{\pm i(k)}u_{\sim}{}^{\text{€}}\right) = -\left(-p_{\sim \pm i(k)}\dot{q}^{\text{€}}_{\pm i(k)} + p_{r \pm i(k)}\dot{q}^{r}_{\pm i(k)}u_{\sim}{}^{\text{€}}\right)$$

$$= c^2\left[\frac{\partial S_{\pm i(k)}}{\partial t} + eW\right]^{-1}\left(p_{\sim \pm i(k)}\left(p^{\text{€}}{}_{\pm i(k)} - eA^{\text{€}}\right) - p_{r \pm i(k)}\left(p^{r}{}_{\pm i(k)} - eA^{r}\right)_{\pm i(k)}u_{\sim}{}^{\text{€}}\right)$$

$$= \frac{mc^2}{X}\left(\begin{array}{l}\left(u_{\sim \pm i(k)}\sqrt{1-\frac{V_{qu\pm i(k)}}{mc^2}} + \frac{eA_{\sim}}{mc}\right)u^{\text{€}}{}_{\pm i(k)} \\ -\left(u_{r \pm i(k)}\sqrt{1-\frac{V_{qu\pm i(k)}}{mc^2}} + \frac{eA_{r}}{mc}\right)u^{r}{}_{\pm i(k)}u_{\sim}{}^{\text{€}}\end{array}\right),$$

$$= \frac{mc^2}{X}\left(\begin{array}{l}\left(u_{\sim \pm i(k)}\sqrt{1-\frac{V_{qu\pm i(k)}}{mc^2}}u^{\text{€}}{}_{\pm i(k)} - u_{r \pm i(k)}\sqrt{1-\frac{V_{qu\pm i(k)}}{mc^2}}u^{r}{}_{\pm i(k)}\right) \\ +\frac{eA_{\sim}}{mc}u^{\text{€}}{}_{\pm i(k)} - \frac{eA_{r}}{mc}u^{r}{}_{\pm i(k)}\;u_{\sim}{}^{\text{€}}\end{array}\right)$$

(2.3.1)

that it is useful to recast in the form

$$T_{\pm i(k)\sim}{}^{\text{€}} = T_{\pm i(k)Class\sim}{}^{\text{€}} + L_{\pm i(k)Class}u_{\sim}{}^{\text{€}} + T_{\pm i(k)Q\sim}{}^{\text{€}} \quad , \qquad (2.3.2)$$

where

$$T_{\pm i(k)Class\sim}{}^{\text{€}} = \frac{mc^2}{\gamma}\left(u_{\sim \pm i(k)} + \frac{eA^{\text{€}}}{mc}\right)u^{\text{€}}_{\pm i(k)}$$

$$= -c^2 p_{\sim \pm i(k)class}\frac{\left(p^{\text{€}}{}_{\pm i(k)} - eA^{\text{€}}\right)}{E_{\pm i(k)} - eW}, \qquad (2.3.3)$$

$$T_{\pm i(k)Q\sim}{}^{\text{€}} = -a_{\left(V_{qu\pm i(k)}\right)}\frac{mc^2}{\gamma_{\pm i(k)}}\left(u_{\sim \pm i(k)}u^{\text{€}}{}_{\pm i(k)} - u_{r \pm i(k)}u^{r}{}_{\pm i(k)}u_{\sim}{}^{\text{€}}\right), \quad (2.3.4)$$

with the help of the identities (see (2.1.1, 2.1.7, 2.2.1) and the following one

$$S_{\pm i} = \frac{\hbar}{2i}ln\left[\frac{\text{Œ}_{\pm i}}{\text{Œ}_{\pm i}*}\right] = \begin{pmatrix}S_{\pm 1} \\ S_{\pm 2}\end{pmatrix} = \frac{\hbar}{2i}\begin{pmatrix}ln\left[\frac{\text{Œ}_{\pm 1}}{\text{Œ}_{\pm 1}*}\right] \\ ln\left[\frac{\text{Œ}_{\pm 2}}{\text{Œ}_{\pm 2}*}\right]\end{pmatrix} \qquad (2.3.5)$$

it is possible to obtain the expression of the EITD (2.3.1) as a function of the Dirac field, independently by the hydrodynamic formalism, as follows

$$T_{\pm i(k)\sim}{}^{\text{€}}{}_{F} = |\text{Œ}_{\pm i(k)}|^2 \frac{i\hbar c^2}{2}\left(\frac{\partial}{\partial t} ln[\frac{\text{Œ}_{\pm i(k)}}{\text{Œ}_{\pm i(k)}{}^*}] + \frac{2ie}{\hbar}\text{W}\right)^{-1}$$

$$\left( \frac{\partial ln[\frac{\text{Œ}_{\pm i(k)}}{\text{Œ}_{\pm i(k)}{}^*}]}{\partial q^{\sim}} \left(\frac{\partial ln[\frac{\text{Œ}_{\pm i(\tilde{k})}}{\text{Œ}_{\pm i(k)}{}^*}]}{\partial q_{\text{€}}} - \frac{2ie}{\hbar}A^{\text{€}}\right) - \frac{\partial ln[\frac{\text{Œ}_{\pm i(k)}}{\text{Œ}_{\pm i(k)}{}^*}]}{\partial q^{\text{r}}} \left(\frac{\partial ln[\frac{\text{Œ}_{\pm i(k)}}{\text{Œ}_{\pm i(k)}{}^*}]}{\partial q_{\text{r}}} - \frac{2ie}{\hbar}A^{\text{r}}\right) \text{u}_{\sim}{}^{\text{€}} \right) \qquad (2.3.6)$$

It is useful to note that expression (2.3.6) excludes the hydrodynamic solutions that do not satisfy the irrotational condition.
Moreover, since for

$$k' > 0,\ k = k',\ \text{S}_{\pm ik} = 0 \qquad (2.3.7)$$

$$k' < 0,\ k = -k',\ \text{r}_{\pm ik} = 0 \qquad (2.3.8)$$

the positive ($k' > 0$, superscript$^{(+)}$) and negative *energy* ($k' < 0$, superscript$^{(-)}$) eigenstates, (2.3.6) reads, respectively,

$$T^{(+)}_{\pm i(k)\sim}{}^{\text{€}}{}_{F} = T_{\pm i(k'>0)\sim}{}^{\text{€}}{}_{F} = |\text{r}_{\pm i(k)}|^2 \frac{i\hbar c^2}{2}\left(\frac{\partial}{\partial t}\left( ln[\frac{\text{r}_{\pm i(k)}}{\text{r}^*_{\pm i(k)}}] + \frac{2iS_{\pm 1(k)}}{\hbar}\right) + \frac{2ie}{\hbar}\text{W}\right)^{-1}$$

$$\left( \frac{\partial ln[\frac{\text{r}_{\pm i(k)}}{\text{r}^*_{\pm i(k)}}] + \frac{2iS_{\pm 1(k)}}{\hbar}}{\partial q^{\sim}} \left(\frac{\partial ln[\frac{\text{r}_{\pm i(k)}}{\text{r}^*_{\pm i(k)}}] + \frac{2iS_{\pm 1(k)}}{\hbar}}{\partial q_{\text{€}}} - \frac{2ie}{\hbar}A^{\text{€}}\right) - \frac{\partial ln[\frac{\text{r}_{\pm i(k)}}{\text{r}^*_{\pm i(k)}}] + \frac{2iS_{\pm 1(k)}}{\hbar}}{\partial q^{\text{r}}} \left(\frac{\partial ln[\frac{\text{r}_{\pm i(k)}}{\text{r}^*_{\pm i(k)}}] + \frac{2iS_{\pm 1(k)}}{\hbar}}{\partial q_{\text{r}}} - \frac{2ie}{\hbar}A^{\text{r}}\right) \text{u}_{\sim}{}^{\text{€}} \right) \qquad (2.3.9)$$

$$T^{(-)\ \ \text{€}}_{\pm i(k)\tilde{}\ \ F} = T^{\ \ \text{€}}_{\pm i(k'<0)\tilde{}\ \ F} = |\mathrm{S}_{\pm i(k)}|^2 \frac{i\hbar c^2}{2}\left(\frac{\partial}{\partial t}\left(ln[\frac{\mathrm{S}_{\pm i(k)}}{\mathrm{S}^*_{\pm i(k)}}]+\frac{2iS_{\pm 1(-k)}}{\hbar}\right)+\frac{2ie}{\hbar}\mathrm{W}\right)^{-1}$$

$$\left(\begin{array}{l}\dfrac{\partial\, ln[\frac{\mathrm{S}_{\pm i(k)}}{\mathrm{S}^*_{\pm i(k)}}]+\frac{2iS_{\pm 1(-k)}}{\hbar}}{\partial q^{\tilde{}}}\left(\dfrac{\partial\, ln[\frac{\mathrm{S}_{\pm i(k)}}{\mathrm{S}^*_{\pm i(k)}}]+\frac{2iS_{\pm 1(-k)}}{\hbar}}{\partial q_{\text{€}}}-\dfrac{2ie}{\hbar}A^{\text{€}}\right)\\ -\dfrac{\partial\, ln[\frac{\mathrm{S}_{\pm i(k)}}{\mathrm{S}^*_{\pm i}}]+\frac{2iS_{\pm 1(-k)}}{\hbar}}{\partial q^{\mathrm{r}}}\left(\dfrac{\partial\, ln[\frac{\mathrm{S}_{\pm i(k)}}{\mathrm{S}^*_{\pm i(k)}}]+\frac{2iS_{\pm 1(-k)}}{\hbar}}{\partial q_{\mathrm{r}}}-\dfrac{2ie}{\hbar}A^{\mathrm{r}}\right)\mathrm{u}_{\tilde{}}^{\ \text{€}}\end{array}\right) \qquad (2.3.10)$$

So far, we have not introduced the explicit form of the Minkowskian DFWE field since we are interested in covariantly generalizing the theory to the curvilinear space-time

**2.4. The macroscopic limit**

Since in the macroscopic classical limit (i.e., when the action values are much bigger than the Planck constant and when the physical length is much bigger than the De Broglie length [24]) the superposition of states (2.2.1) undergo collapse to an eigenstate (due to the quantum decoherence produced by fluctuations [23-26]), the macroscopic GE of the general relativity cannot be obtained just by applying the limiting condition $lim_{\hbar\to 0}$ but also by imposing the decoherence condition (terming it $lim_{dec}$ ) as follows

$$lim_{macro} \equiv lim_{dec}\, lim_{\hbar\to 0} = lim_{\hbar\to 0}\, lim_{dec}\,. \qquad (2.4.1)$$

where the *decoherence* effect on the quantum state reads

$$lim_{dec}\text{Œ}_{\pm i} = lim_{dec}\left(\begin{array}{l}\sum_{k=0}\mathrm{r}_{\pm 1k}\, exp[\frac{iS_{\pm 1(k)}}{\hbar}]+\mathrm{s}^{\dagger}_{\pm 1k}\, exp[\frac{iS_{\pm 1(-k)}}{\hbar}]\\ \sum_{k=0}\mathrm{r}_{\pm 2k}\, exp[\frac{iS_{\pm 2(k)}}{\hbar}]+\mathrm{s}^{\dagger}_{\pm 2k}\, exp[\frac{iS_{\pm 2(-k)}}{\hbar}]\end{array}\right)$$

$$= \begin{array}{l}\text{for } k'>0\left(\begin{array}{l}\mathrm{r}_{\pm 1\tilde{k}}\, exp[\frac{iS_{\pm 1(k)}}{\hbar}]\\ \mathrm{r}_{\pm 2\tilde{k}}\, exp[\frac{iS_{\pm 2(k)}}{\hbar}]\end{array}\right)\\ \text{for } k'<0\left(\begin{array}{l}\mathrm{s}^{\dagger}_{\pm 1\tilde{k}}\, exp[\frac{iS_{\pm 1(-k)}}{\hbar}]\\ \mathrm{s}^{\dagger}_{\pm 2\tilde{k}}\, exp[\frac{iS_{\pm 2(-k)}}{\hbar}]\end{array}\right)\end{array} \qquad (2.4.2)$$

leading to

$$lim_{dec}\, lim_{\hbar\to 0}\text{Œ}_{\pm i} = lim_{dec}\text{Œ}_{0} = lim_{dec}\sum_{k=0} \text{r}_{0k}\, exp[\frac{iS_{0(k)}}{\hbar}] + \text{s}_{0}{}^{\dagger}{}_{k}\, exp[\frac{iS_{0(-k)}}{\hbar}]$$

$$= \begin{matrix} \text{for } k'>0 = \text{r}_{0\tilde{k}}\, exp[\frac{iS_{0(k)}}{\hbar}] \\ \text{for } k'<0 = \text{s}_{0}{}^{\dagger}{}_{\tilde{k}}\, exp[\frac{iS_{0(-k)}}{\hbar}] \end{matrix} \qquad (2.4.3)$$

where $\tilde{k}$ is the eigenstates to which the wave-function collapses and where

$$lim_{\hbar\to 0}\text{Œ}_{\pm i} = \text{Œ}_{0} = \sum_{k=0} \text{r}_{0k}\, exp[\frac{iS_{0(k)}}{\hbar}] + \text{s}_{0}{}^{\dagger}{}_{k}\, exp[\frac{iS_{0(-k)}}{\hbar}] \qquad (2.4.4)$$

is the solution of the following equation

$$lim_{\hbar\to 0}\left[\hbar^{2}\left(\partial_{\sim} + \frac{ie}{\hbar}A_{\sim}\right)\left(\partial^{\sim} + \frac{ie}{\hbar}A^{\sim}\right) + \left(m^{2}c^{2} + \frac{ie\hbar}{2}\text{r}^{\sim€}F_{\sim€}\right)\right]\text{Œ}_{\pm}$$

$$\equiv \left[\left(\partial_{\sim} + \frac{ie}{\hbar}A_{\sim}\right)\left(\partial^{\sim} + \frac{ie}{\hbar}A^{\sim}\right) + m^{2}c^{2}\right]\text{Œ}_{0} = 0 \qquad (2.4.5)$$

and $\text{r}_{k}$, $\text{s}^{\dagger}{}_{k}$ obey to the relations

$$lim_{\hbar\to 0}\text{r}_{\pm 1\tilde{k}} = lim_{\hbar\to 0}\text{r}_{\pm 2\tilde{k}} = \text{r}_{0\tilde{k}} \qquad (2.4.6)$$

$$lim_{\hbar\to 0}\text{s}^{\dagger}{}_{\pm 1\tilde{k}} = lim_{\hbar\to 0}\text{s}^{\dagger}{}_{\pm 2\tilde{k}} = \text{s}_{0}{}^{\dagger}{}_{\tilde{k}}\,. \qquad (2.4.7)$$

Moreover, by introducing (2.4.6-7) into (2.4.2) the reversibility of the two limits

$$lim_{dec}\, lim_{\hbar\to 0}\text{Œ}_{\pm i} = lim_{\hbar\to 0}\, lim_{dec}\text{Œ}_{\pm i} \qquad (2.4.8)$$

follows.

The detailed stochastic hydrodynamic derivation of the macroscopic limit (2.4.1) is given in ref. [24-26], but it is suffice to say here that the quantum superposition of state can be maintained beyond the De Broglie length in the case of strong interactions [23-26] such as in the strong gravitational field of a black hole. Therefore, we assume the existence of the macroscopic limit in the curved space-time taking care that the macroscopic scale may go much beyond the De Broglie length in very high curvature space-time.

**2.5. The minimum action in the hydrodynamic formalism**

Since the hydrodynamic Lagrangean depends also by the quantum potential and hence by $|\text{Œ}_{\pm i}| = f_{1}(|\text{Œ}_{\pm i(k)}|)$ and $\partial_{\sim}|\text{Œ}_{\pm i}| = f_{2}(|\text{Œ}_{\pm i(k)}|, \partial_{\sim}|\text{Œ}_{\pm i(k)}|)$, the problem of defining the equation of motion can be generally carried out by using the set of variables : $x_{(k)\sim} = (q_{\sim}, \dot{q}_{\sim\pm i(k)}, |\text{Œ}_{\pm i(k)}|, \partial_{\sim}|\text{Œ}_{\pm i(k)}|)$.

Thence, the variation of the hydrodynamic action $\mathcal{S} = \int\iiint \mathcal{L}dVdt$ (between the fixed starting and ending points, $q_{\sim a}$ $q_{\sim b}$, respectively) reads [19]

$$\delta \mathcal{S}_{\pm i} = \int\!\!\iiint |\psi_{\pm i}|^2 \left(\frac{\partial Tr\mathbf{L}}{\partial x_{(k)}^{\nu}}\right)\delta x_{(k)}^{\nu}\, dVdt = \int\!\!\iiint |\psi_{\pm i}|^2 \left(\sum_k \frac{\partial \tilde{L}_{\pm i(k)}}{\partial x_{(k)}^{\nu}}\right)\delta x_{(k)}^{\nu}\, dVdt$$

$$= \int\!\!\iiint |\psi_{\pm i}|^2 \sum_k \begin{pmatrix} \dfrac{\partial \tilde L_{\pm i(k)}}{\partial q^\nu}\delta q^\nu + \dfrac{\partial \tilde L_{\pm i(k)}}{\partial \dot q^\nu_{(k)}}\delta \dot q^\nu_{\pm i(k)} \\ + \dfrac{\partial \tilde L_{\pm i(k)}}{\partial |\psi_k|}\delta|\psi_{\pm i(k)}| + \dfrac{\partial \tilde L_{\pm i(k)}}{\partial \partial^\nu |\psi_k|}\delta \partial^\nu |\psi_{\pm i(k)}|\end{pmatrix} dVdt \quad . (2.5.1)$$

$$= \frac{1}{c}\int\!\!\iiint |\psi_{\pm i}|^2 \sum_k \begin{pmatrix} \left(\dfrac{\partial \tilde L_{\pm i(k)}}{\partial q^\nu} - \dfrac{d}{dt}\dfrac{\partial \tilde L_{\pm i(k)}}{\partial \dot q^\nu_{\pm i(k)}}\right)\delta q^\nu \\ + \left(\dfrac{\partial \tilde L_{\pm i(k)}}{\partial |\psi_{\pm i(k)}|} - \partial^\nu \dfrac{\partial \tilde L_{\pm i(k)}}{\partial\partial^\nu |\psi_{\pm i(k)}|}\right)\delta|\psi_{\pm i(k)}|\end{pmatrix} d\Omega$$

Given that the quantum motion equations for eigenstates satisfy the condition

$$\left(\frac{\partial \tilde L_{\pm i(k)}}{\partial q^\nu} - \frac{d}{dt}\frac{\partial \tilde L_{\pm i(k)}}{\partial \dot q^\nu_{\pm i(k)}}\right) = \left(\frac{\partial L_{\pm i(k)}}{\partial q^\nu} - \frac{d}{dt}\frac{\partial L_{\pm i(k)}}{\partial \dot q^\nu_{\pm i(k)}}\right) = 0 \ , \qquad (2.5.2)$$

the variation of the action $\delta \mathcal{S}_{\pm i(k)}$ for the $k$-th eigenstates reads

$$\delta \mathcal{S}_{\pm i(k)} = \frac{1}{c}\int\!\!\iiint |\psi_{\pm i(k)}|^2 \left(\frac{\partial L_{\pm i(k)}}{\partial |\psi_{\pm i(k)}|} - \partial^\nu \frac{\partial L_{\pm i(k)}}{\partial\partial_\nu |\psi_{\pm i(k)}|}\right)\delta|\psi_k| d\Omega = \delta\left(\Delta \mathcal{S}_{\pm i(k)Q}\right) \qquad (2.5.3)$$

that is not null for the dependence of the hydrodynamic Lagrangean $L_{\pm i(k)}$ by the quantum potential.

Thence, for the quantum hydrodynamic evolution the variational principle reads

$$\delta \mathcal{S}_{\pm i(k)} - \delta\left(\Delta \mathcal{S}_{\pm i(k)Q}\right) = 0 \qquad (2.5.4)$$

that in the classical limit (i.e., for $\hbar \to 0$ ,$V_{qu} \to 0$ and

$$\frac{\partial\left(lim_{\hbar\to 0} L_{\pm i(k)}\right)}{\partial |\psi_{\pm i(k)}|} = 0 \qquad (2.5.5)$$

$$\frac{\partial\left(lim_{\hbar\to 0} L_{\pm i(k)}\right)}{\partial^\nu |\psi_{\pm i(k)}|} = 0 \ ) \qquad (2.5.6)$$

leads to the classical extremal principle

$$lim_{\hbar\to 0}\, \delta \mathcal{S}_{\pm i(k)} = lim_{\hbar\to 0}\, \delta\left(\Delta \mathcal{S}_{\pm i(k)Q}\right) = 0 \ . \qquad (2.5.7)$$

Moreover, generally speaking, by using (2.2.17), for the general superposition of state (2.2.1) the variation of the action reads

$$\delta \mathcal{S}_{\pm i} = -\frac{1}{c}\int\!\!\iiint |\psi_{\pm i}|^2 \sum_{\tilde k} \begin{pmatrix} \left(\tilde p_{\pm i(\tilde k)} \dfrac{\partial \dot q_{\pm i(\tilde k)}}{\partial q^\nu}\right)\delta q^\nu \\ - \left(\dfrac{\partial \tilde L_{\pm i(\tilde k)}}{\partial |\psi_{\pm i(\tilde k)}|} - \partial^\nu \dfrac{\partial \tilde L_{\pm i(\tilde k)}}{\partial\partial^\nu |\psi_{\pm i(\tilde k)}|}\right)\delta|\psi_{\tilde k}|\end{pmatrix} d\Omega \qquad (2.5.8)$$

$$= \delta\left(\Delta \mathcal{S}_{\pm iQmix} + \Delta \mathcal{S}_{\pm iQ}\right) = \delta\left(\Delta \mathcal{S}_{\pm i}\right)$$

where

$$\delta\left(\Delta\mathcal{S}_{\pm iQ_{mix}}\right) = -\frac{1}{c}\int\iiint |\psi_{\pm i}|^2 \sum_{\tilde{k}} \tilde{p}_{\pm i(k)} \frac{\partial \dot{q}_{\pm i(k)}}{\partial q^{\mu}} \delta|\psi_{\pm i(k)}| d\Omega \qquad (2.5.9)$$

is due to the mixing of the quantum superposition of states and where

$$\Delta\mathcal{S}_{\pm iQ} = \frac{1}{c}\int\iiint |\psi_{\pm i}|^2 \sum_{\tilde{k}} \left( \frac{\partial \tilde{L}_{\pm i(k)}}{\partial |\psi_{\pm i(\tilde{k})}|} - \partial^{\mu} \frac{\partial \tilde{L}_{\pm i(k)}}{\partial \partial^{\mu} |\psi_{\pm i(\tilde{k})}|} \right) \delta|\psi_{k}| d\Omega \qquad (2.5.10)$$

Thence, by from (2.5.8), the minimum action principle can be generalized to

$$\delta\mathcal{S}_{\pm i} - \delta\left(\Delta\mathcal{S}_{\pm i}\right) = 0 \qquad (2.5.11)$$

that, for the decoherence (decay to an eigenstate) at the macroscopic scale leads to

$$\begin{aligned} lim_{macro}\, \delta\mathcal{S}_{\pm i} = lim_{macro}\, \delta\left(\Delta\mathcal{S}_{\pm i}\right) &= lim_{\hbar\to 0}\, lim_{dec}\, \delta\left(\Delta\mathcal{S}_{\pm iQ} + \Delta\mathcal{S}_{\pm iQ_{mix}}\right) \\ &= lim_{\hbar\to 0}\, \delta\left(\Delta\mathcal{S}_{\pm iQ_{(\tilde{k})}} + lim_{dec}\, \Delta\mathcal{S}_{\pm iQ_{mix}}\right) \\ &= lim_{\hbar\to 0}\, \delta\left(\Delta\mathcal{S}_{\pm iQ_{(\tilde{k})}}\right) = 0 \end{aligned} \qquad (2.5.12)$$

that converges to the extremal classical principle.

**2.6. The hydrodynamic DFWE in the curved space-time**

In order to derive the gravitational equation for the Dirac field, we assume that the hydrodynamic representation of the Dirac bi-spinor field, depicted by four mass densities moving with momenta $p_{\pm\mu} = \begin{pmatrix} p_{\pm 1\mu} \\ p_{\pm 2\mu} \end{pmatrix} = -\partial_{\mu} S_{\pm} = -\partial_{\mu} \begin{pmatrix} S_{\pm 1} \\ S_{\pm 2} \end{pmatrix}$ under the action of the quantum potential (2.0.11), has its own gravitational effect by producing the curvature of the space-time that makes stationary the overall action comprehending the gravitational field [19]). This assumption is supported the by the fact the hydrodynamic approach leads to the derivation of the Einstein equation with the correct electromagnetic energy-impulse tensor when applied to the boson field of the photon [19] (see appendix A).

Also the experimental demonstration of the physical existence of the quantum potential energy by the Bohm-Aharonov effect supports this approach.

One possible generalization of the Dirac equation to the curved space-time can be obtained by assuming the physics covariance of the equation. In reference [27] it has been shown that the physics covariance condition plays the same role as the inertial to gravitational mass equivalence postulate in the classical general relativity.

The covariant Dirac equation in curved space-time, respect affine and spin connections, reads [28]

$$\left( i\hbar\gamma^{\mu} \left( \partial_{\mu} - \frac{i}{4}\sigma^{ab}\omega_{ab\mu} + \frac{ie}{\hbar}A_{\mu} \right) - mc \right) \Psi = 0 \qquad (2.6.1)$$

where

$$\dagger^{ab} = \frac{i}{2}[\mathrm{x}^a, \mathrm{x}^b] \quad (2.6.2)$$

and

$$\check{S}_{ab\sim} = f_b^{\mathrm{r}} e_{a\mathrm{s}} \Gamma^{\mathrm{s}}_{\sim\mathrm{r}} - f_b^{\mathrm{r}} \partial_{\sim} e_{a\mathrm{r}}\,, \quad (2.6.3)$$

where $\Gamma^{\mathrm{s}}_{\sim\mathrm{r}}$ is the Christoffel symbol that reads

$$\Gamma^{\mathrm{s}}_{\sim\mathrm{r}} = \frac{1}{2} g^{\mathrm{sx}} \left( \partial_{\mathrm{r}} g_{\mathrm{x}\sim} + \partial_{\sim} g_{\mathrm{xr}} - \partial_{\mathrm{x}} g_{\sim\mathrm{r}} \right) \quad (2.6.4)$$

and $\square$, and $f_b^{\mathrm{r}}$ , are the *vielbein* and the inverse *vielbein*, respectively.

By assuming the covariant derivative for affine and spinor connections

$$D_{\sim} = \partial_{\sim} - \frac{i}{4} \dagger^{ab} \check{S}_{ab\sim} \quad (2.6.5)$$

equation (2.6.1) more simply reads

$$\left( i\hbar \mathrm{x}^{\sim} \left( D_{\sim} + \frac{ie}{\hbar} A_{\sim} \right) - mc \right) \Psi = 0 \quad (2.6.6)\ .$$

leading, by following the procedure in section 2.0, to

$$\frac{i\hbar}{mc} \dagger^{\sim} \left( D_{\sim} + \frac{ie}{\hbar} A_{\sim} \right) \text{Œ}_{+} = \text{Œ}_{-} \quad (2.6.7)$$

$$\frac{i\hbar}{mc} \tilde{\dagger}^{\sim} \left( D_{\sim} + \frac{ie}{\hbar} A_{\sim} \right) \text{Œ}_{-} = \text{Œ}_{+} \quad (2.6.8)$$

to the covariant equation

$$\left( g^{\sim\text{€}} + \mathrm{r}_{\pm}^{\sim\text{€}} \right) \left( \partial_{\sim} - \frac{i}{4} \dagger^{ab} \check{S}_{ab\sim} + \frac{ie}{\hbar} A_{\sim} \right) \left( \partial_{\text{€}} - \frac{i}{4} \dagger^{ab} \check{S}_{ab\text{€}} + \frac{ie}{\hbar} A_{\text{€}} \right) \text{Œ}_{\pm} = -\frac{m^2 c^2}{\hbar^2} \text{Œ}_{\pm}\,, \quad (2.6.9)$$

and of equation (2.0.8)

$$\left( \begin{array}{l} \left( \partial_{\sim} - \frac{i}{4} \dagger^{ab} \check{S}_{ab\sim} + \frac{ie}{\hbar} A_{\sim} \right) \left( g^{\sim\text{€}} \left( \partial_{\text{€}} - \frac{i}{4} \dagger^{ab} \check{S}_{ab\text{€}} + \frac{ie}{\hbar} A_{\text{€}} \right) \right) \\ + g^{\sim\mathrm{s}} g^{\text{€}\mathrm{r}} \mathrm{r}_{\pm\mathrm{sr}} \left( \frac{ie}{2\hbar} F_{\sim\text{€}} - \frac{i}{4} \dagger^{ab} \left( \check{S}_{ab\sim} \partial_{\text{€}} + \partial_{\sim} \check{S}_{ab\text{€}} \right) \right) \end{array} \right) \text{Œ}_{\pm} = -\left( \frac{m^2 c^2}{\hbar^2} + \right) \text{Œ}_{\pm} \quad (2.6.10)$$

where, due to the antisymmetry of $\mathrm{r}_{\pm}^{\sim\text{€}}$ , the properties

$$\mathrm{r}_{\pm}^{\sim\text{€}} \dagger^{ab} \check{S}_{ab\sim} \dagger^{ab} \check{S}_{ab\text{€}} = 0\,. \quad (2.6.11)$$

$$\mathrm{r}_{\pm}^{\sim\text{€}} \dagger^{ab} \check{S}_{ab\sim} A_{\text{€}} + \mathrm{r}_{\pm}^{\sim\text{€}} A_{\sim} \dagger^{ab} \check{S}_{ab\text{€}} = 0 \quad (2.6.12)$$

have been used. Moreover, by using (2.0.9) and equating the real part of (2.6.10), the HHJE in curved space-time reads (see Appendix B)

$$\left(\partial_{\mu} S_{\alpha}-\left(Re\left\{\frac{\hbar}{4}\gamma^{ab}\omega_{ab\mu}\right\}-eA_{\mu}\right)\right)g^{\mu\nu}\left(\partial_{\nu} S_{\alpha}-\left(Re\left\{\frac{\hbar}{4}\gamma^{ab}\omega_{ab\nu}\right\}-eA_{\nu}\right)\right) \qquad (2.6.13)$$

$$=\left(m^{2}c^{2}-mV_{qu\alpha}\right)$$

where (see Appendix B)

$$V_{qu}=-\frac{\hbar^{2}}{m}\left[\begin{array}{l}\left(\begin{array}{c}\dfrac{\partial_{\mu}\left(g^{\mu\nu}\partial_{\nu}|\Psi_{\alpha 1}|\right)}{|\Psi_{\alpha 1}|}\\ \dfrac{\partial_{\mu}\left(g^{\mu\nu}\partial_{\nu}|\Psi_{\alpha 2}|\right)}{|\Psi_{\alpha 2}|}\end{array}\right)+\dfrac{\hbar^{2}}{2}Im\left\{\gamma^{ab}\omega_{ab\mu}\right\}g^{\mu\nu}\left(\begin{array}{c}\dfrac{\partial_{\nu}|\Psi_{\alpha 1}|}{|\Psi_{\alpha 1}|}\\ \dfrac{\partial_{\nu}|\Psi_{\alpha 2}|}{|\Psi_{\alpha 2}|}\end{array}\right)\\ +\partial_{\mu}g^{\mu\nu}\left(\dfrac{Im\left\{\gamma^{ab}\omega_{ab\nu}\right\}}{4}\right)+g^{\mu\nu}\,Im\left\{\dfrac{1}{4}\gamma^{ab}\omega_{ab\mu}\right\}Im\left\{\dfrac{1}{4}\gamma^{ab}\omega_{ab\nu}\right\}\end{array}\right]$$

$$+Re\left(g^{\mu s}g^{\nu r}\left(\begin{array}{l}\dfrac{ie}{2\hbar}\left(\begin{array}{c}r_{\alpha s r 1 j}\dfrac{|\Psi_{\alpha j}|}{|\Psi_{\alpha 1}|}\\ r_{\alpha s r 2 j}\dfrac{|\Psi_{\alpha j}|}{|\Psi_{\alpha 2}|}\end{array}\right)F_{\mu\nu}\\ -\dfrac{i}{4}r_{\alpha s r}\gamma^{ab}\left(\omega_{ab\mu}\left(\begin{array}{c}\dfrac{\partial_{\nu}|\Psi_{\alpha 1}|}{|\Psi_{\alpha 1}|}\\ \dfrac{\partial_{\nu}|\Psi_{\alpha 2}|}{|\Psi_{\alpha 2}|}\end{array}\right)+\left(\begin{array}{c}\dfrac{\partial_{\mu}\omega_{ab\nu}|\Psi_{\alpha 1}|}{|\Psi_{\alpha 1}|}\\ \dfrac{\partial_{\mu}\omega_{ab\nu}|\Psi_{\alpha 2}|}{|\Psi_{\alpha 2}|}\end{array}\right)\right)\end{array}\right)\right) \qquad (2.6.14)$$

As far as it concerns the current (2.0.12) the covariant form reads

$$J^{(\alpha)\mu}=-\left(g^{\mu\nu}+r_{\alpha}^{\mu\nu}\right)\left(\frac{|\Psi_{\alpha}|^{2}}{m}\left(\partial_{\nu}S_{\alpha}+\frac{\hbar}{4}\gamma^{ab}\omega_{ab\nu}+eA_{\nu}\right)\right) \qquad (2.6.15)$$

so that the total fermion current $J^{\mu}$ reads

$$J^{\mu}=J^{(+)\mu}+J^{(-)\mu}$$

$$=\frac{\hbar}{2im}\left[\begin{array}{l}\Psi_{+}^{\dagger}\tilde{\gamma}^{\mu}\gamma^{\nu}\left(\partial_{\nu}-\dfrac{i}{4}\gamma^{ab}\omega_{ab\nu}+\dfrac{e}{i\hbar}A_{\nu}\right)\Psi_{+}\\ -\Psi_{+}\tilde{\gamma}^{\mu}\gamma^{\nu}\left(\partial_{\nu}+\dfrac{i}{4}\gamma^{ab}\omega_{ab\nu}-\dfrac{e}{i\hbar}A_{\nu}\right)\Psi_{+}^{\dagger}\\ +(\Psi_{-}^{\dagger}\gamma^{\mu}\tilde{\gamma}^{\nu}\left(\partial_{\nu}-\dfrac{i}{4}\gamma^{ab}\omega_{ab\nu}+\dfrac{e}{i\hbar}A_{\nu}\right)\Psi_{-}\\ -\Psi_{-}\gamma^{\mu}\tilde{\gamma}^{\nu}\left(\partial_{\nu}+\dfrac{i}{4}\gamma^{ab}\omega_{ab\nu}-\dfrac{e}{i\hbar}A_{\nu}\right)\Psi_{-}^{\dagger})\end{array}\right] \qquad (2.6.16)$$

that by using (2.6.7-8), leads to

$$
\begin{aligned}
J^{˜} &= -\frac{\hbar}{2im}\begin{bmatrix} Œ_{+}{}^{\dagger} \text{†}^{˜} Œ_{-} - Œ_{+} \text{†}^{˜} Œ_{-}{}^{\dagger} \\ + Œ_{-}{}^{\dagger} \text{†}^{˜} Œ_{+} + Œ_{-} \text{†}^{˜} Œ_{+}{}^{\dagger} \end{bmatrix} \\
&= -\frac{\hbar}{2im}\left[ \overline{\Psi} \text{X}^{˜} \Psi + \overline{\Psi *} \text{X}^{˜} \Psi * \right] = J^{(+)˜} + J^{(-)˜} = -\frac{\hbar}{im} \overline{\Psi} \text{X}^{˜} \Psi
\end{aligned}
\qquad (2.6.17)
$$

in agreement with the standard results for the DFWE that show that the connection acting on the current $J^{˜}$ is the Levi-Civita connection where the condition for its conservation reads

$$
D_{˜} J^{˜} = 0 \qquad (2.6.18)
$$

where $\square$ is the co-variant derivative.
For a direct check, equation (2.6.18) can be derived by equating the imaginary part of equation (2.6.11) (the demonstration is not given here).

## 2. 7. The hydrodynamic Hamilton-Jacobi equation for eigenstates in curved space-time

By using (2.6.13) for the k-th eigenstates it follows that

$$
p_{\pm i(k)˜} = \left( m\text{X}\ \dot{q}_{\pm i(k)}{}^{˜} \sqrt{1 - \frac{V_{qu\pm(k)}}{mc^2}} - Re\left\{ \frac{\hbar}{4} \text{†}^{ab} \text{Š}_{ab˜} \right\} + eA_{˜} \right), \qquad (2.7.1)
$$

and that

$$
\begin{aligned}
L_{\pm i(k)} &= -\frac{mc^2}{\text{X}} \sqrt{1 - \frac{V_{qu\pm(k)}}{mc^2}} - \left( Re\left\{ \frac{\hbar}{4} \text{†}^{ab} \text{Š}_{ab˜} \right\} - eA_{˜} \right) \dot{q}_{\pm i(k)}{}^{˜} \\
&= L_{\pm i(k)class} + L_{\pm i(k)Q}
\end{aligned}
\quad . \qquad (2.7.2)
$$

where now

$$
L_{\pm i(k)class} = -\frac{mc^2}{\text{X}} - \left( Re\left\{ \frac{\hbar}{4} \text{†}^{ab} \text{Š}_{ab˜} \right\} - eA_{˜} \right) \dot{q}_{\pm i(k)}{}^{˜} , \qquad (2.7.3)
$$

$$
\begin{aligned}
L_{\pm i(k)Q} &= a_{\left(V_{qu\pm i(k)}\right)} \frac{mc^2}{\text{X}} \\
&= -\left( 1 - \sqrt{1 - \frac{V_{qu\pm i(k)}}{mc^2}} \right) \left( L_{\pm i(k)class} + \left( Re\left\{ \frac{\hbar}{4} \text{†}^{ab} \text{Š}_{ab˜} \right\} - eA_{˜} \right) \dot{q}_{\pm i(k)}{}^{˜} \right)
\end{aligned}
\qquad (2.7.4)
$$

$$p_{\sim_{\pm i(k)}class} = \left( m\gamma_{\pm i}\ \dot{q}_{\sim_{\pm i(k)}} - \left( Re\left\{ \frac{\hbar}{4} \dagger^{ab} \check{S}_{ab\sim} \right\} - eA_{\sim} \right) \right)$$

$$= \frac{p_{\sim_{\pm i(k)}} + \left( Re\left\{ \frac{\hbar}{4} \dagger^{ab} \check{S}_{ab\sim} \right\} - eA_{\sim} \right)\left( 1 - \sqrt{1 - \frac{V_{qu\pm i(k)}}{mc^2}} \right)}{\sqrt{1 - \frac{V_{qu\pm i(k)}}{mc^2}}}$$

$$= -\frac{\partial_{\sim} S_{\pm i(k)} - \left( Re\left\{ \frac{\hbar}{4} \dagger^{ab} \check{S}_{ab\sim} \right\} - eA_{\sim} \right) a_{\left(V_{qu\pm i(k)}\right)}}{1 - a_{\left(V_{qu\pm i(k)}\right)}} \qquad (2.7.5)$$

$$p_{\sim_{\pm i(k)}Q} = -a_{\left(V_{qu\pm i(k)}\right)} m\gamma_{\pm i}\ \dot{q}_{\sim_{\pm i(k)}}$$

$$= -a_{\left(V_{qu\pm i(k)}\right)} \left( p_{\sim_{\pm i(k)}class} + \left( Re\left\{ \frac{\hbar}{4} \dagger^{ab} \check{S}_{ab\sim} \right\} - eA_{\sim} \right) \right) \qquad (2.7.6)$$

As far as it concerns the EITD $T_{\pm i(k)\sim}^{\text{€}} = |Œ_{\pm i(k)}|^2\, \mathsf{T}_{\pm i(k)\sim}^{\text{€}}$, we obtain that

$$\mathsf{T}_{\pm i(k)\sim}^{\text{€}} = \mathsf{T}_{\pm i(k)Class\sim}^{\text{€}} + L_{\pm i(k)Class}\mathsf{u}_{\sim}^{\text{€}} + \mathsf{T}_{\pm i(k)Q\sim}^{\text{€}} =$$

$$m\mathsf{X}\begin{pmatrix} \dot{q}_{\pm i(k)\sim}\left( \sqrt{1 - \frac{V_{qu\pm(\pm k)}}{mc^2}}\,\dot{q}_{\pm i(k)}^{\text{€}} - \left( Re\left\{ \frac{\hbar}{4} \dagger^{ab} \check{S}_{ab}^{\text{€}} \right\} - eA^{\text{€}} \right) \right) \\ -\dot{q}_{\pm i(k)\mathsf{r}}\left( \sqrt{1 - \frac{V_{qu\pm(\pm k)}}{mc^2}}\,\dot{q}_{\pm i(k)}^{\mathsf{r}} - \left( Re\left\{ \frac{\hbar}{4} \dagger^{ab} \check{S}_{ab}^{\mathsf{r}} \right\} - eA^{\mathsf{r}} \right) \right) \mathsf{u}_{\sim}^{\text{€}} \end{pmatrix}. \qquad (2.7.7)$$

where

$$\mathsf{T}_{\pm i(k)Class\sim}^{\text{€}} = \frac{mc^2}{\gamma} u_{\pm i(k)\sim} \left( u_{\pm i(k)}^{\text{€}} - \frac{1}{mc}\left( Re\left\{ \frac{\hbar}{4} \dagger^{ab} \check{S}_{ab}^{\text{€}} \right\} - eA^{\text{€}} \right) \right) \qquad (2.7.8)$$

where

$$\mathsf{T}_{\pm i(k)Q\sim}^{\text{€}} = -a_{\left(V_{qu\pm i(k)}\right)} \frac{mc^2}{\mathsf{X}_{\pm i(k)}} \left( u_{\pm i(k)\sim} u_{\pm i(k)}^{\text{€}} - u_{\pm i(k)\mathsf{r}} u_{\pm i(k)}^{\mathsf{r}} \mathsf{u}_{\sim}^{\text{€}} \right). \qquad (2.7.9)$$

Finally, by using the identities (2.7.1) the identity (2.7.7) can be recast as follows

$$\text{T}_{\pm i(k)\sim\text{€}} = g_{\text{€}|}\,\text{T}_{\pm i(k)\sim}{}^{|} = -\left(-g_{\text{€}|}\,\dot{q}_{\sim\pm i(k)}\,p^{|}{}_{\pm i(k)} + g_{\text{r}|}\,\dot{q}^{|}_{\pm i(k)}\,p^{\text{r}}{}_{\pm i(k)}\,g_{\sim\text{€}}\right)$$

$$= c^2\left[\frac{\partial S_{\pm i(k)}}{\partial t} - c\left(Re\left\{\frac{\hbar}{4}\text{†}^{ab}\text{Š}_{ab0}\right\} - eA_0\right)\right]^{-1}$$

$$\begin{pmatrix} g_{\text{€}|}\left(p_{\sim\pm i(k)} + \left(Re\left\{\frac{\hbar}{4}\text{†}^{ab}\text{Š}_{ab\sim}\right\} - eA_{\sim}\right)\right)p^{|}{}_{\pm i(k)} \\ -g_{\text{r}|}\left(p^{|}{}_{\pm i(k)} + \left(Re\left\{\frac{\hbar}{4}\text{†}^{ab}\text{Š}_{ab}{}^{|}\right\} - eA^{|}\right)\right)p^{\text{r}}{}_{\pm i(k)}\,g_{\sim\text{€}} \end{pmatrix}, \qquad (2.7.10)$$

that, as a function of the DFW field, reads

$$\text{T}_{\pm i(k)\sim}{}^{\text{€}} = -\frac{\hbar c^2}{2i}\left(\frac{\partial}{\partial t} ln[\frac{\text{Œ}_{\pm i(k)}}{\text{Œ}_{\pm i(k)}{}^{*}}] - \frac{2i}{\hbar}\left(Re\left\{\frac{\hbar}{4}\text{†}^{ab}\text{Š}_{ab0}\right\} - eA_0\right)\right)^{-1}$$

$$\begin{pmatrix} \left(\dfrac{\partial\, ln[\frac{\text{Œ}_{\pm i(k)}}{\text{Œ}_{\pm i(k)}{}^{*}}]}{\partial q^{\sim}} + \dfrac{2i}{\hbar}\left(Re\left\{\dfrac{\hbar}{4}\text{†}^{ab}\text{Š}_{ab\sim}\right\} - eA_{\sim}\right)\right)\dfrac{\partial\, ln[\frac{\text{Œ}_{\pm i(k)}}{\text{Œ}_{\pm i(k)}{}^{*}}]}{\partial q_{\text{€}}} \\ -g_{\text{r}|}\left(\dfrac{\partial\, ln[\frac{\text{Œ}_{\pm i(k)}}{\text{Œ}_{\pm i(k)}{}^{*}}]}{\partial q_{|}} + \dfrac{2i}{\hbar}\left(Re\left\{\dfrac{\hbar}{4}\text{†}^{ab}\text{Š}_{ab}{}^{|}\right\} - eA^{|}\right)\right)\dfrac{\partial\, ln[\frac{\text{Œ}_{\pm i(k)}}{\text{Œ}_{\pm i(k)}{}^{*}}]}{\partial q_{\text{r}}}\text{u}_{\sim}{}^{\text{€}} \end{pmatrix} \qquad (2.7.12)$$

where $\text{Œ}_{\pm i(k)}$ represents the eigenstates of the DFWE in curved space-time that for *positive* $\text{Œ}^{(+)}_{\pm i(k)}$ and *negative* $\text{Œ}^{(-)}_{\pm i(k)}$ *solutions* reads, respectively,

$$\text{Œ}^{(+)}_{\pm i(k)} = \text{Œ}_{\pm i(k'>0)} = \text{r}_{\pm i(k)}\, exp(\frac{iS_{\pm i(k)}}{\hbar}) \qquad (2.7.13)$$

$$\text{Œ}^{(-)}_{\pm i(k)} = \text{Œ}_{\pm i(k'<0)} = \text{s}^{*}{}_{\pm i(k)}\, exp(\frac{iS_{\pm i(-k)}}{\hbar})] \,. \qquad (2.7.14)$$

### 2.8. The electromagnetic field

Since we are considering fermions that bring the electric charge, the overall Einstein equation also contains the coupling with electromagnetic field

$$F^{\sim\text{€}}{}_{,\text{€}} = -4\text{ƒ}\; J^{\sim}{}_{tot} \qquad (2.8.1)$$

where

$$\boxed{\qquad\qquad\qquad}\,, \qquad (2.8.2)$$

where

$$A_\mu = \left(\frac{\phi}{c},\ -A_i\right) \qquad (2.8.3)$$

is the potential 4-vector and where $J^\mu_{tot}$ is the current of the charged particles (if only fermions are present ) . Moreover, since the equations for the EM field are the same both in classical and quantum case, so that the hydrodynamic description coincides with the classical one (i.e., the quantum potential is null), it follows that EITD for the EM field $T_{\nu\mu f}$ equals the classical form and reads [19, 29] (see Appendix A).

$$T_{\nu\mu\, em} = \frac{1}{4\pi}\left( -F_{\mu\alpha} F_{\nu}^{\ \alpha} + \frac{1}{4} F_{\alpha\beta} F^{\alpha\beta} g_{\mu\nu} \right) \qquad (2.8.4)$$

## 3. The minimum action in curved space-time and the gravity equation for the DFWE

In this section we derive the gravity equation, by applying the minimum action principle to the quantum hydrodynamic evolution associated to the fields $\psi_{\pm i}$.
Given that the quantum hydrodynamic equations in the Minkowskian space-time, satisfies the minimum action principle (2.5.11), when we consider the covariant formulation in the curved space-time, such variation takes a contribution from the variability of the metric tensor. When we consider the gravity and we assume that the geometry of the space-time is that one which makes null the overall variation of the action (2.5.11), we have a condition that leads to the definition of the GE.
By considering the variation of the action due both to the curvilinear coordinates and to the functional dependence by $\psi_{\pm i}$ [19], it follows that

$$\delta \mathcal{S}_{\pm i} = \frac{1}{c}\int\!\!\iiint |\psi_{\pm i}|^2 \sum_k \left( \begin{array}{l} \left( \dfrac{1}{\sqrt{-g}} \left( \begin{array}{l} \dfrac{\partial \sqrt{-g}\,\tilde{L}_{\pm i(k)}}{\partial g^{\mu\nu}} - \dfrac{\partial}{\partial q^{\alpha}} \dfrac{\partial \sqrt{-g}\,\tilde{L}_{\pm i(k)}}{\partial \dfrac{\partial g^{\mu\nu}}{\partial q^{\alpha}}} \\ - \dfrac{\partial}{\partial |\psi|} \dfrac{\partial \sqrt{-g}\,\tilde{L}_{\pm i(k)}}{\partial \dfrac{\partial g^{\mu\nu}}{\partial |\psi_k|}} \end{array} \right) \right) \delta g^{\mu\nu} \\ + \tilde{p}_{\pm i(k)\nu}\, \dot{q}_{\pm i(k);\mu}^{\ \nu}\, \delta q^{\mu} + \left( \begin{array}{l} \dfrac{\partial \tilde{L}_{\pm i(k)}}{\partial |\psi_{\pm ik}|} \\ - \partial^{\mu} \dfrac{\partial \tilde{L}_{\pm i(k)}}{\partial \partial^{\mu} |\psi_{\pm ik}|} \end{array} \right) \delta |\psi_k| \end{array} \right) \sqrt{-g}\, d\Omega \qquad (3.1)$$

and, from (2.5.11),

$$\text{u}\mathcal{S}_D = \text{u}\mathcal{S}_{\pm i} - \text{u}\left(\Delta\mathcal{S}_{\pm i}\right) = \frac{1}{c}\int\iiint /\text{Œ}_{\pm i}/^2 \sum_k \begin{pmatrix} \dfrac{\partial\sqrt{-g}\,\tilde{L}_{\pm i(k)}}{\partial g^{\sim\text{€}}} - \dfrac{\partial}{\partial q^{\}}}\dfrac{\partial\sqrt{-g}\,\tilde{L}_{\pm i(k)}}{\partial\dfrac{\partial g^{\sim\text{€}}}{\partial q^{\}}}} \\ -\dfrac{\partial}{\partial/\text{Œ}_k/}\dfrac{\partial\sqrt{-g}\,\tilde{L}_{\pm i(k)}}{\partial\dfrac{\partial g^{\sim\text{€}}}{\partial/\text{Œ}_{\pm ik}/}} \end{pmatrix}\text{u}\,g^{\sim\text{€}}\,d\Omega . \quad (3.2)$$

If we postulate that the variation of action of the gravitational field [30]

$$\text{u}\mathcal{S}_g = \frac{c^3}{16f\,G}\int\iiint\left(R_{\sim\text{€}} - \frac{1}{2}Rg_{\sim\text{€}}\right)\text{u}\,g^{\sim\text{€}}\sqrt{-g}\,d\Omega \quad (3.3)$$

offsets that one produced both by the DFWE field, $\text{u}\mathcal{S}_D$, and by the EM field [30]

$$\text{u}\mathcal{S}_{em} = \frac{1}{c}\int\iiint T_{\text{€}\sim\ em}\,\text{u}\,g^{\sim\text{€}}\sqrt{-g}\,d\Omega \quad (3.4)$$

so that

$$\begin{aligned} &\text{u}\mathcal{S}_g + \sum_{\pm,i}\left(\text{u}\mathcal{S}_{\pm i} - \text{u}\left(\Delta\mathcal{S}_{\pm i}\right)\right) + \text{u}\mathcal{S}_{em} \\ &= \int\iiint\left(R_{\sim\text{€}} - \frac{1}{2}Rg_{\sim\text{€}} - \frac{8f\,G}{c^4}\left(\sum_{\pm,i}/\text{Œ}_{\pm i}/^2\,\text{‡}_{\pm i\sim\text{€}} + T_{\text{€}\sim\ em}\right)\right)\text{u}\,g^{\sim\text{€}}\sqrt{-g}\,d\Omega = 0 \end{aligned} , \quad (3.5)$$

we obtain the gravitational equation for the DFWE field

$$R_{\sim\text{€}} - \frac{1}{2}Rg_{\sim\text{€}} = \frac{8f\,G}{c^4}\left(\sum_{\pm,i}/\text{Œ}_{\pm i}/^2\,\text{‡}_{\pm i\sim\text{€}} + T_{\text{€}\sim\ em}\right) \quad (3.6)$$

where by posing

$$\begin{aligned} \text{‡}_{\pm i\sim\text{€}} &= \frac{1}{2}\frac{\displaystyle\sum_{k=0}\begin{pmatrix} \text{r}_{\pm ik}\,exp(\frac{iS_{\pm i(k)}}{\hbar})\left(\text{‡}_{\pm i(k)\sim\text{€}\,class} + \text{‡}_{\pm i(k)\sim\text{€}\,Q} - \text{‡}_{\pm i(k)\sim\text{€}\,mix}\right) \\ +\text{s}^*_{\pm ik}\,exp(\frac{iS_{\pm i(-k)}}{\hbar})\left(\text{‡}_{\pm i(-k)\sim\text{€}\,class} + \text{‡}_{\pm i(-k)\sim\text{€}\,Q} - \text{‡}_{\pm i(-k)\sim\text{€}\,mix}\right) \end{pmatrix}}{\displaystyle\sum_{k=0}\left(\text{r}_{\pm ik}\,exp(\frac{iS_{\pm i(k)}}{\hbar}) + \text{s}^*_{\pm ik}\,exp(\frac{iS_{\pm i(-k)}}{\hbar})\right)} \\ &+\frac{1}{2}\frac{\displaystyle\sum_{k=0}\begin{pmatrix} \text{r}^*_{\pm ik}\,exp(-\frac{iS_{\pm i(k)}}{\hbar})\left(\text{‡}_{\pm i(k)\sim\text{€}\,class} + \text{‡}_{\pm i(k)\sim\text{€}\,Q} + \left(\text{‡}_{\pm i(k)\sim\text{€}\,mix}\right)^*\right) \\ +\text{s}_{\pm ik}\,exp(-\frac{iS_{\pm i(-k)}}{\hbar})\left(\text{‡}_{\pm i(-k)\sim\text{€}\,class} + \text{‡}_{\pm i(-k)\sim\text{€}\,Q} + \left(\text{‡}_{\pm i(-k)\sim\text{€}\,mix}\right)^*\right) \end{pmatrix}}{\displaystyle\sum_{k=0}\text{r}^*_{\pm ik}\,exp(-\frac{iS_{\pm i(k)}}{\hbar}) + \text{s}_{\pm ik}\,exp(-\frac{iS_{\pm i(-k)}}{\hbar})\,)]} \\ &+\text{‡}_{\pm i\sim\text{€}\,curv} \\ &= \text{‡}_{\pm i\sim\text{€}\,class} + \text{‡}_{\pm i\sim\text{€}\,Q} + \text{‡}_{\pm i\sim\text{€}\,mix} + \text{‡}_{\pm i\sim\text{€}\,curv} \end{aligned} , \quad (3.7)$$

it follows that

$$\text{‡}_{\pm i(k)\sim\text{€}\,class} = 2\frac{1}{\sqrt{-g}}\begin{pmatrix} \dfrac{\partial}{\partial g^{\sim\text{€}}} - \dfrac{\partial}{\partial q^{\}}}\dfrac{\partial}{\partial \dfrac{\partial g^{\sim\text{€}}}{\partial q^{\}}}} \\ -\dfrac{\partial}{\partial /\text{Œ}_{\pm i(k)}/}\dfrac{\partial}{\partial \dfrac{\partial g^{\sim\text{€}}}{\partial /\text{Œ}_{\pm i(k)}/}} \end{pmatrix}\sqrt{-g}\, p_{\sim_{\pm i(k)class}}\, \dot{q}^{\sim}{}_{\pm i} \quad (3.8)$$

$$\text{‡}_{\pm i(k)\sim\text{€}\,Q} = 2\frac{1}{\sqrt{-g}}\begin{pmatrix} \dfrac{\partial}{\partial g^{\sim\text{€}}} - \dfrac{\partial}{\partial q^{\}}}\dfrac{\partial}{\partial \dfrac{\partial g^{\sim\text{€}}}{\partial q^{\}}}} \\ -\dfrac{\partial}{\partial /\text{Œ}_{\pm i(k)}/}\dfrac{\partial}{\partial \dfrac{\partial g^{\sim\text{€}}}{\partial /\text{Œ}_{\pm i(k)}/}} \end{pmatrix}\sqrt{-g}\, p_{\sim_{\pm i(k)Q}}\, \dot{q}^{\sim}{}_{\pm i} \quad (3.9)$$

$$\text{‡}_{\pm i(k)\sim\text{€}\,mix} = 2\frac{1}{\sqrt{-g}}\begin{pmatrix} \dfrac{\partial}{\partial g^{\sim\text{€}}} - \dfrac{\partial}{\partial q^{\}}}\dfrac{\partial}{\partial \dfrac{\partial g^{\sim\text{€}}}{\partial q^{\}}}} \\ -\dfrac{\partial}{\partial /\text{Œ}_{\pm i(k)}/}\dfrac{\partial}{\partial \dfrac{\partial g^{\sim\text{€}}}{\partial /\text{Œ}_{\pm i(k)}/}} \end{pmatrix}\sqrt{-g}\, p_{\sim_{\pm i(k)mix}}\, \dot{q}^{\sim}{}_{\pm i} \quad (3.10)$$

where

$$p_{\sim_{\pm i(k)mix}} = -\frac{\hbar}{i}\partial_{\sim}\, ln\, \text{r}_{\pm ik} \quad , \qquad (3.11)$$

$$p_{\sim_{\pm i(-k)mix}} = -\frac{\hbar}{i}\partial_{\sim}\, ln\, \text{s}^{*}{}_{\pm ik} \qquad (3.12)$$

and that

$$\text{‡}_{\pm i\sim\text{€}\,curv} = \sum_{k}\left(\text{‡}_{\pm i(k)\sim\text{€}\,curv} + \text{‡}_{\pm i(-k)\sim\text{€}\,curv}\right) \qquad (3.13)$$

where

$$
\begin{aligned}
\ddagger_{\pm i(k)\sim\text{€}_{curv}} = -\frac{1}{2}
&\left(
\begin{array}{l}
p_{\sim_{\pm i(k)tot}}\,\dot{q}^{\sim}{}_{\pm i}
\left(
\begin{array}{l}
\dfrac{\partial}{\partial g^{\sim\text{€}}}-\dfrac{\partial}{\partial q^{\}}}\dfrac{\partial}{\partial\dfrac{\partial g^{\sim\text{€}}}{\partial q^{\}}}}\\
-\dfrac{\partial}{\partial /\text{Œ}_{\pm i(k)}/}\dfrac{\partial}{\partial\dfrac{\partial g^{\sim\text{€}}}{\partial /\text{Œ}_{\pm i(k)}/}}
\end{array}
\right)\Xi_{\pm i(k)}\\
+\left(
\begin{array}{l}
\dfrac{\partial\Xi_{\pm i(\pm k)}}{\partial q^{\}}}\dfrac{\partial p_{\sim_{\pm i(k)tot}}\,\dot{q}^{\sim}{}_{\pm i}}{\dfrac{\partial g^{\sim\text{€}}}{\partial q^{\}}}}+\dfrac{\partial p_{\sim_{\pm i(k)tot}}\,\dot{q}^{\sim}{}_{\pm i}}{\partial q^{\}}}\dfrac{\partial\Xi_{\pm ik)}}{\dfrac{\partial g^{\sim\text{€}}}{\partial q^{\}}}}\\
+\dfrac{\partial\Xi_{\pm i(\pm k)}}{\partial /\text{Œ}_{\pm i(k)}/}\dfrac{\partial p_{\sim_{\pm i(k)tot}}\,\dot{q}^{\sim}{}_{\pm i}}{\dfrac{\partial g^{\sim\text{€}}}{\partial /\text{Œ}_{\pm i(k)}/}}+\dfrac{\partial p_{\sim_{\pm i(k)tot}}\,\dot{q}^{\sim}{}_{\pm i}}{\partial /\text{Œ}_{\pm i(k)}/}\dfrac{\partial\Xi_{\pm i(k)}}{\dfrac{\partial g^{\sim\text{€}}}{\partial /\text{Œ}_{\pm i(k)}/}}
\end{array}
\right)
\end{array}
\right) \qquad (3.14)\\
-\frac{1}{2}
&\left(
\begin{array}{l}
p_{\sim_{\pm i(k)tot}}\,\dot{q}^{\sim}{}_{\pm i}
\left(
\begin{array}{l}
\dfrac{\partial}{\partial g^{\sim\text{€}}}-\dfrac{\partial}{\partial q^{\}}}\dfrac{\partial}{\partial\dfrac{\partial g^{\sim\text{€}}}{\partial q^{\}}}}\\
-\dfrac{\partial}{\partial /\text{Œ}_{\pm i(k)}/}\dfrac{\partial}{\partial\dfrac{\partial g^{\sim\text{€}}}{\partial /\text{Œ}_{\pm i(k)}/}}
\end{array}
\right)\Xi^{*}{}_{\pm i(k)}\\
+\left(
\begin{array}{l}
\dfrac{\partial\Xi^{*}{}_{\pm i(k)}}{\partial q^{\}}}\dfrac{\partial p_{\sim_{\pm i(k)tot}}\,\dot{q}^{\sim}{}_{\pm i}}{\dfrac{\partial g^{\sim\text{€}}}{\partial q^{\}}}}+\dfrac{\partial p_{\sim_{\pm i(k)tot}}\,\dot{q}^{\sim}{}_{\pm i}}{\partial q^{\}}}\dfrac{\partial\Xi^{*}{}_{\pm i(k)}}{\dfrac{\partial g^{\sim\text{€}}}{\partial q^{\}}}}\\
+\dfrac{\partial\Xi^{*}{}_{\pm i(k)}}{\partial /\text{Œ}_{k}/}\dfrac{\partial p_{\sim_{\pm i(k)tot}}\,\dot{q}^{\sim}{}_{\pm i}}{\dfrac{\partial g^{\sim\text{€}}}{\partial /\text{Œ}_{\pm i(k)}/}}+\dfrac{\partial p_{\sim_{\pm i(k)tot}}\,\dot{q}^{\sim}{}_{\pm i}}{\partial /\text{Œ}_{\pm i(k)}/}\dfrac{\partial\Xi^{*}{}_{\pm i(k)}}{\dfrac{\partial g^{\sim\text{€}}}{\partial /\text{Œ}_{\pm i(k)}/}}
\end{array}
\right)
\end{array}
\right)
\end{aligned}
$$

where

$$p_{\sim_{\pm i(k)tot}} = p_{\sim_{\pm i(k)class}} + p_{\sim_{\pm i(k)Q}} + p_{\sim_{\pm i(k)mix}} \qquad (3.15)$$

$$\Xi_{\pm i(k)} = \frac{\mathrm{X}_{\pm i(k)}\,exp[\dfrac{iS_{\pm i(k)}}{\hbar}]}{\sum\limits_{k_j=0}\left(\mathrm{r}_{\pm ik_j}\,exp(\dfrac{iS_{\pm i(k_j)}}{\hbar})+\mathrm{s}^{*}{}_{\pm ik_j}\,exp(\dfrac{iS_{\pm i(-k_j)}}{\hbar})\right)} \qquad (3.16)$$

where

$$\mathrm{X}_{\pm i(k)} = \mathrm{r}_{\pm ik} \qquad (3.17)$$

$$\mathrm{X}_{\pm i(-k)} = \mathrm{s}^{*}{}_{\pm i(k)}\,. \qquad (3.18)$$

Moreover, given that [19]

$$\frac{1}{2c}\int\iiint\begin{pmatrix}\sum_{\pm,i}/\text{Œ}_{\pm i}/^{2}\ \text{‡}_{\pm i\sim€}\\ +T_{€\sim\ em}\end{pmatrix}\text{u}\, g^{\sim€}\sqrt{-g}\,d\Omega$$

$$=-\frac{1}{c}\int\iiint\begin{pmatrix}\sum_{\pm,i}/\text{Œ}_{\pm i}/^{2}\ \text{‡}_{\pm i\sim€}\\ +T_{€\sim\ em}\end{pmatrix}_{,€}<^{\sim}\sqrt{-g}\,d\Omega \qquad (3.19)$$

where

$$\text{u}\, g^{\sim€} = <^{\sim,€} + <^{€;\sim}, \qquad (3.20)$$

for the infinitesimal transformation of coordinates

$$q'^{\sim} = q^{\sim} + <^{\sim}, \qquad (3.21)$$

in the Minkowskian case (3.5) leads to

$$\sum_{\pm,i}\left(\text{u}\,\mathcal{S}_{\pm i}-\text{u}\left(\Delta\mathcal{S}_{\pm i}\right)\right)+\text{u}\,\mathcal{S}_{em}$$

$$=-\frac{1}{c}\int\iiint\left(\sum_{\pm,i}/\text{Œ}_{\pm i}/^{2}\ \text{‡}_{\pm i\sim€}+T_{€\sim\ em}\right)_{,€}<^{\sim}\sqrt{-g}\,d\Omega=0 \qquad (3.22)$$

and thence, for the arbitrariness of $<^{\sim}$, to

$$\left(\sum_{\pm,i}/\text{Œ}_{\pm i}/^{2}\ \text{‡}_{\pm i\sim€}+T_{€\sim\ em}\right)_{,€}=0, \qquad (3.23)$$

that in the decoherent limit leads to

$$lim_{dec}\left(\sum_{\pm,i}/\text{Œ}_{\pm i}/^{2}\ \text{‡}_{\pm i\sim€}+T_{€\sim\ em}\right)_{,€}=\left(\sum_{\pm,i}/\text{Œ}_{\pm i(k)}/^{2}\ \text{‡}_{\pm i(k)\sim€}+T_{€\sim\ em}\right)_{,€}=0 \qquad (3.24)$$

Moreover, since in the decoherent macroscopic Minkowskian limit it holds:

i. $\text{‡}_{\sim€_{mix}}=0$

ii. $\text{‡}_{\sim€_{curv}}=0$

iii. $\frac{V_{qu\pm(k)}}{mc^{2}}\rightarrow 0$ and, hence, $a_{\left(V_{qu\pm i(k)}\right)}\rightarrow 0$ as well as $\text{‡}_{\pm i\sim€_{Q}}\rightarrow 0$

(3.24) leads to

$$\left(\sum_{\pm,i}/\text{Œ}_{\pm i(k)}/^{2}\ \text{‡}_{\pm i(k)\sim€}+T_{€\sim\ em}\right)_{,€}=\left(\sum_{\pm,i}/\text{Œ}_{\pm i(k)}/^{2}\ \text{‡}_{\pm i(k)\sim€_{class}}+T_{€\sim\ em}\right)_{,€}=0. \qquad (3.25)$$

Furthermore, given that in the Minkowskian decoherent macroscopic limit it holds (see (C.11) in Appendix C)

$$lim_{dec}\left(\sum_{\pm i}/\text{Œ}_{\pm i}/^{2}\ \text{T}_{\pm i\sim}{}^{€}+T_{\sim}{}^{€}{}_{em}\right)_{,€}=\left(\sum_{\pm i}/\text{Œ}_{\pm i(k)}/^{2}\ \text{T}_{\pm i(k)\sim}{}^{€}+T_{\sim}{}^{€}{}_{em}\right)_{,€},$$

$$=\left(\sum_{\pm i}/\text{Œ}_{\pm i(k)}/^{2}\left(\text{T}_{\pm i(k)_{class}\sim}{}^{€}+L_{\pm i(k)_{class}}\text{u}_{\sim}{}^{€}\right)+T_{\sim}{}^{€}{}_{em}\right)_{,€}=0 \qquad (3.26)$$

by comparing (3.2.5) with (3.2.6), it follows that

$$\sum_{\pm,i} /\text{Œ}_{\pm i(k)}/^{2}\, \text{‡}_{\pm i(k)\sim}{}^{\text{€}}{}_{class} + C_{\left(\dot{q}_{\pm i(k)}\right)} \text{u}_{\sim}{}^{\text{€}} = \sum_{\pm i} /\text{Œ}_{\pm i(k)}/^{2} \left( \text{T}_{\pm i(k)class\sim}{}^{\text{€}} + L_{\pm i(k)class} \text{u}_{\sim}{}^{\text{€}} \right) \quad (3.27)$$

and, by defining $\Lambda$ such as

$$\Lambda \text{u}_{\sim}{}^{\text{€}} = C_{(k)} \text{u}_{\sim}{}^{\text{€}} - \sum_{\pm i} /\text{Œ}_{\pm i(k)}/^{2}\, L_{\pm i(k)class} \text{u}_{\sim}{}^{\text{€}} , \quad (3.28)$$

finally, that

$$\sum_{\pm i} /\text{Œ}_{\pm i(k)}/^{2}\, \text{‡}_{\pm i(k)\sim}{}^{\text{€}}{}_{class} = \sum_{\pm i} /\text{Œ}_{\pm i(k)}/^{2}\, \text{T}_{\pm i(k)class\sim}{}^{\text{€}} - \Lambda \text{u}_{\sim}{}^{\text{€}} \quad (3.29)$$

where the classical CC $\Lambda$ is independent by $k$.

**3.1. The GE for the DFWE eigenstates**

By using (2.7.5-6, 3.8-9) for the $k$ eigenstate, it follows that

$$\text{‡}_{\pm i(k)\sim\text{€}Q} = -a_{(V_{qu\pm i})} \left( \text{‡}_{\pm i(k)\sim\text{€}class} + \Theta_{\sim\text{€}} + \begin{pmatrix} p_{\text{r}\,\pm i(k)class} - eA_{\text{r}} \\ + Re\left\{ \dfrac{\hbar}{4} \text{†}^{ab} \text{Š}_{ab\text{r}} \right\} \end{pmatrix} \dot{q}^{\text{r}}{}_{\pm i(k)} \Delta_{\pm i(k)\sim\text{€}} \right)$$

$$= -a_{(V_{qu\pm i})} \begin{pmatrix} T_{\pm i(k)class\sim\text{€}} - \dfrac{\Lambda}{/\text{Œ}_{\pm i(k)}/^{2}} g_{\sim\text{€}} + \Theta_{\sim\text{€}} \\ + \left( L_{\pm i(k)class} + \begin{pmatrix} Re\left\{ \dfrac{\hbar}{4} \text{†}^{ab} \text{Š}_{ab\text{r}} \right\} \\ -eA_{\text{r}} \end{pmatrix} \dot{q}^{\text{r}}{}_{\pm i(k)} \right) \Delta_{\pm i(k)\sim\text{€}} \end{pmatrix} \quad (3.1.1)$$

where

$$\Theta_{\sim\text{€}} = 2 \frac{1}{\sqrt{-g}} \begin{pmatrix} \dfrac{\partial}{\partial g^{\sim\text{€}}} - \dfrac{\partial}{\partial q^{\text{\}}}} \dfrac{\partial}{\partial \dfrac{\partial g^{\sim\text{€}}}{\partial q^{\text{\}}}}} \\ - \dfrac{\partial}{\partial /\text{Œ}_{\pm i(k)}/} \dfrac{\partial}{\partial \dfrac{\partial g^{\sim\text{€}}}{\partial /\text{Œ}_{\pm i(k)}/}} \end{pmatrix} \sqrt{-g} \begin{pmatrix} Re\left\{ \dfrac{\hbar}{4} \text{†}^{ab} \text{Š}_{ab\text{r}} \right\} \\ -eA_{\text{r}} \end{pmatrix} \quad (3.1.2)$$

$$\Delta_{\pm i(k)\sim€} = \frac{2}{a_{(V_{qu\pm i(k)})}} \left( \begin{array}{l} \left( \dfrac{\partial}{\partial g^{\sim€}} - \dfrac{\partial}{\partial q^{\}}} \dfrac{\partial}{\dfrac{\partial g^{\sim€}}{\partial q^{\}}}} - \dfrac{\partial}{\partial /Œ_{\pm ik}/} \dfrac{\partial}{\dfrac{\partial g^{\sim€}}{\partial /Œ_{k}/}} \right) a_{(V_{qu\pm i(k)})} \\ - \left( \begin{array}{l} \dfrac{\partial a_{(V_{qu\pm i(k)})}}{\partial q^{\}}} \dfrac{\partial ln\left( L_{\pm i(k)class} + \begin{pmatrix} Re\left\{ \dfrac{\hbar}{4} † ^{ab}Š_{abr} \right\} \\ -eA_{r} \end{pmatrix} \dot{q}^{r}{}_{\pm i(k)} \right)}{\dfrac{\partial g^{\sim€}}{\partial q^{\}}}} \\ + \dfrac{\partial ln\left( L_{\pm i(k)class} + \begin{pmatrix} Re\left\{ \dfrac{\hbar}{4} † ^{ab}Š_{abr} \right\} \\ -eA_{r} \end{pmatrix} \dot{q}^{r}{}_{\pm i(k)} \right)}{\partial q^{\}}} \dfrac{\partial a_{(V_{qu\pm i(k)})}}{\dfrac{\partial g^{\sim€}}{\partial q^{\}}}} \\ + \dfrac{\partial a_{(V_{qu\pm i(k)})}}{\partial /Œ_{\pm ik}/} \dfrac{\partial ln\left( L_{\pm i(k)class} + \begin{pmatrix} Re\left\{ \dfrac{\hbar}{4} † ^{ab}Š_{abr} \right\} \\ -eA_{r} \end{pmatrix} \dot{q}^{r}{}_{\pm i(k)} \right)}{\dfrac{\partial g^{\sim€}}{\partial /Œ_{\pm ik}/}} \\ + \dfrac{\partial ln\left( L_{\pm i(k)class} + \begin{pmatrix} Re\left\{ \dfrac{\hbar}{4} † ^{ab}Š_{abr} \right\} \\ -eA_{r} \end{pmatrix} \dot{q}^{r}{}_{\pm i(k)} \right)}{\partial /Œ_{\pm ik}/} \dfrac{\partial a_{(V_{qu\pm i(k)})}}{\dfrac{\partial g^{\sim€}}{\partial /Œ_{\pm ik}/}} \end{array} \right) \end{array} \right),$$

(3.1.3)

and, being $‡_{\pm i\sim€_{mix}} = ‡_{\pm i(k)\sim€_{mix}} = 0$, $‡_{\sim€_{curv}} = 0$ and $\Xi_{\pm i(k)} = 1$ for the eigenstates, the GE reads

$$R_{\sim€} - \frac{1}{2} R g_{\sim€} = \frac{8ƒ G}{c^{4}} \left( \begin{array}{l} \sum\limits_{\pm i} \left( \begin{array}{l} T_{\pm i(k)\sim€_{class}} \left( 1 - a_{(V_{qu\pm i(k)})} \right) \\ + a_{(V_{qu\pm i(k)})} \left( \begin{array}{l} \Lambda g_{\sim€} + /Œ_{\pm i(k)}/^{2}\, \Theta_{\sim€} \\ - /Œ_{\pm i(k)}/^{2} \left( \begin{array}{l} L_{\pm i(k)class} \\ + \begin{pmatrix} Re\left\{ \dfrac{\hbar}{4} † ^{ab}Š_{abr} \right\} \\ -eA_{r} \end{pmatrix} \dot{q}^{r}{}_{\pm i(k)} \end{array} \right) \Delta_{\pm i(k)\sim€} \end{array} \right) \end{array} \right) \\ + T_{€\sim\, em} - \Lambda g_{\sim€} \end{array} \right)$$

(3.1.4)

Finally, by separating both $\Delta_{\sim€}$ and in the isotropic and stress part as following

$$\Delta_{\pm i(k)\sim€} = \frac{\Delta_{\pm i(k)\}\}}}{4}\, g_{\sim€} + \Delta_{s\pm i(k)\sim€}\,, \qquad (3.1.5)$$

$$\Theta_{\sim€} = \frac{\Theta_{\}\}}}{4} + \Theta_{s\sim€} \qquad (3.1.6)$$

the (3.1.4) reads

$$R_{\sim€} - \frac{1}{2} R g_{\sim€} = \frac{8f\,G}{c^{4}} \begin{pmatrix} \sum\limits_{\pm i} \begin{pmatrix} T_{\pm i(k)\sim€\, class}\left(1 - a_{(V_{qu\pm i(k)})}\right) \\ +a_{(V_{qu})} \begin{pmatrix} \Lambda g_{\sim€} + |Œ_{\pm i(k)}|^{2}\left(\dfrac{\Theta_{\}\}}}{4} g_{\sim€} + \Theta_{s\sim€}\right) \\ -|Œ_{\pm i(k)}|^{2} \begin{pmatrix} L_{\pm i(k)class} + \\ \begin{pmatrix} Re\left\{\dfrac{\hbar}{4} \dagger\, {}^{ab}\text{Š}_{abr}\right\} \\ -eA_{r} \end{pmatrix} \dot{q}^{r}_{\pm i(k)} \end{pmatrix} \dfrac{\Delta_{\pm i(k)\}\}}}{4} g_{\sim€} \\ -|Œ_{\pm i(k)}|^{2} \begin{pmatrix} L_{\pm i(k)class} + \\ \begin{pmatrix} Re\left\{\dfrac{\hbar}{4} \dagger\, {}^{ab}\text{Š}_{abr}\right\} \\ -eA_{r} \end{pmatrix} \dot{q}^{r}_{\pm i(k)} \end{pmatrix} \Delta_{s\pm i(k)\sim€} \end{pmatrix} \end{pmatrix} \\ +T_{€\sim\, em} - \Lambda g_{\sim€} \end{pmatrix} \qquad (3.1.7)$$

Equation (3.1.4, 3.1.7, 3.2.1) can be expressed as a function of the DFWE field $\Psi$ by using the relations (2.0.11, 2.1.1, 2.2.3, 2.3.1, 2.7.11) and the following ones

$$\begin{aligned} L_{\pm i(k)} &= -c^{2} \begin{pmatrix} \partial_{t} S_{\pm i(k)} - eW \\ + Re\left\{\dfrac{\hbar c}{4} \dagger\, {}^{ab}\text{Š}_{ab0}\right\} \end{pmatrix}^{-1} g_{\sim r}\, p^{r}_{\pm i(k)} \begin{pmatrix} p^{\sim}_{\pm i(k)} - eA_{\sim} \\ + Re\left\{\dfrac{\hbar}{4} \dagger\, {}^{ab}\text{Š}_{ab\sim}\right\} \end{pmatrix} \\ &= -c^{2} \begin{pmatrix} \partial_{t} S_{\pm i(k)} - eW \\ + Re\left\{\dfrac{\hbar c}{4} \dagger\, {}^{ab}\text{Š}_{ab0}\right\} \end{pmatrix}^{-1} g_{\sim r}\, \partial^{r} S_{\pm i(k)} \begin{pmatrix} \partial^{\sim} S_{\pm i(k)} - eA_{\sim} \\ + Re\left\{\dfrac{\hbar}{4} \dagger\, {}^{ab}\text{Š}_{ab\sim}\right\} \end{pmatrix} \qquad (3.1.8) \\ &= -c^{2}\, \frac{i\hbar}{2} \begin{pmatrix} \partial_{t}\, ln[\dfrac{Œ_{\pm ik}}{Œ_{\pm ik}{}^{*}}] - eW \\ + Re\left\{\dfrac{\hbar c}{4} \dagger\, {}^{ab}\text{Š}_{ab0}\right\} \end{pmatrix}^{-1} g_{\sim r}\, \partial^{r}\, ln[\frac{Œ_{\pm ik}}{Œ_{\pm ik}{}^{*}}] \begin{pmatrix} \partial^{\sim}\, ln[\dfrac{Œ_{\pm ik}}{Œ_{\pm ik}{}^{*}}] - eA_{\sim} \\ + Re\left\{\dfrac{\hbar}{4} \dagger\, {}^{ab}\text{Š}_{ab\sim}\right\} \end{pmatrix} \end{aligned}$$

where from (2.26-27) it has been used the identity

$$\dot{q}_{\pm i(k)\sim} = c^2 \frac{p_{\pm i(k)\sim} - eA_{\sim} + Re\left\{\frac{\hbar c}{4}\dagger^{\,ab}\text{Š}_{ab\sim}\right\}}{\left(E_{\pm i(k)} - e\text{W} + Re\left\{\frac{\hbar c}{4}\dagger^{\,ab}\text{Š}_{ab0}\right\}\right)}$$

$$= -c^2 \frac{\partial_{\sim} S_{\pm i(k)} + eA_{\sim} - Re\left\{\frac{\hbar c}{4}\dagger^{\,ab}\text{Š}_{ab\sim}\right\}}{\left(\partial_t S_{\pm i(k)} - e\text{W} + Re\left\{\frac{\hbar c}{4}\dagger^{\,ab}\text{Š}_{ab0}\right\}\right)} \qquad (3.1.9)$$

$$= -c^2 \frac{\partial_{\sim}\, ln[\frac{\text{Œ}_{\pm ik}}{\text{Œ}_{\pm ik}*}] + \frac{2i}{\hbar}\left(eA_{\sim} - Re\left\{\frac{\hbar c}{4}\dagger^{\,ab}\text{Š}_{ab\sim}\right\}\right)}{\left(\partial_t\, ln[\frac{\text{Œ}_{\pm ik}}{\text{Œ}_{\pm ik}*}] - \frac{2i}{\hbar}\left(e\text{W} - Re\left\{\frac{\hbar c}{4}\dagger^{\,ab}\text{Š}_{ab0}\right\}\right)\right)}$$

### 3.2. The GE of the DFWE field

Finally, in the matrix form, the GE can be written as a function of the general DFWE field (2.2.1) and reads

$$R_{\sim\text{€}} - \frac{1}{2}Rg_{\sim\text{€}} = \frac{8f\,G}{c^4}\left(Tr\left(\ddagger_{\sim\text{€}\,class} > ]\, g_{\sim\text{€}} - ]_{\,Q}\, g_{\sim\text{€}} + \text{U}\ddagger_{\sim\text{€}\,stress}\right) + T_{\text{€}\sim\; em}\right) \quad (3.2.1)$$

where

$$] = \Lambda\text{u}_{hk} \qquad (3.2.2)$$

,

$$\ddagger_{\sim\text{€}\,class} = \sum_{\pm i}\left(\begin{array}{l} \text{g}_{\pm i(k)}\left(1 - a_{(V_{qu\pm i(k)})}\right)/\text{Œ}_{\pm i(k)}/^2\; \text{T}_{\pm i(k)\sim\text{€}\,class} \\ +\text{g}_{\pm i(-k)}\left(1 - a_{(V_{qu\pm i(-k)})}\right)/\text{Œ}_{\pm i(-k)}/^2\; \text{T}_{\pm i(-k)\sim\text{€}\,class} \end{array}\right)\text{u}_{hk}\,, \qquad (3.2.3)$$

where

$$\text{g}_{\pm i(k)} = \frac{1}{2}\left(\Xi_{\pm i(k)} + \Xi*_{\pm i(k)}\right), \qquad (3.2.4)$$

$$]_{\,Q} = ]_{\,Q(k)} + ]_{\,Q(-k)} \qquad (3.2.5)$$

where

$$\text{]}_{Q(k)} = -\sum_{\pm i} \frac{1}{2} \left( \begin{array}{l} \left( \begin{array}{l} \Xi_{\pm i(k)} \left( \begin{array}{l} a_{\pm i(k)} \left( \begin{array}{l} \Lambda + /\text{Œ}_{\pm i(k)}/^{2} \dfrac{\Theta_{\}\}}}{4} \\ - /\text{Œ}_{\pm i(k)}/^{2} \left( \begin{array}{l} p_{r\ \pm i(k) class} - eA_{r} \\ + Re\left\{ \dfrac{\hbar}{4} \text{†}^{ab} \text{Š}_{abr} \right\} \end{array} \right) \dot{q}^{r}{}_{\pm i} \dfrac{\Delta_{\pm i(k)\}\}}}{4} \end{array} \right) \\ + /\text{Œ}_{\pm i(k)}/^{2}\ \text{‡}_{\pm i(k)||mix} \end{array} \right) \\ + \Xi^{*}_{\pm i(k)} \left( \begin{array}{l} a_{\pm i(k)} \left( \begin{array}{l} \Lambda + /\text{Œ}_{\pm i(k)}/^{2} \dfrac{\Theta_{\}\}}}{4} \\ - /\text{Œ}_{\pm i(k)}/^{2} \left( \begin{array}{l} p_{r\ \pm i(k) class} - eA_{r} \\ + Re\left\{ \dfrac{\hbar}{4} \text{†}^{ab} \text{Š}_{abr} \right\} \end{array} \right) \dot{q}^{r}{}_{\pm i} \dfrac{\Delta_{\pm i(k)\}\}}}{4} \end{array} \right) \\ - /\text{Œ}_{\pm i(k)}/^{2}\ \text{‡}_{\pm i(k)||mix} \end{array} \right) \end{array} \right) \\ + /\text{Œ}_{\pm i(k)}/^{2} \dfrac{\text{‡}_{\pm i\}\}curv}}{8} \end{array} \right) \text{u}_{hk},$$

(3.2.6)

and

$$\text{U‡}_{\sim\text{€}\, stress} = \text{U‡}_{(+k)\sim\text{€}\, stress} < \text{U‡}_{(-k)\sim\text{€}\, stress} \qquad (3.2.7)$$

where

$$\text{U‡}_{(k)\sim\text{€}\, stress} = \sum_{\pm i} \frac{/\text{Œ}_{\pm i(k)}/^{2}}{2} \left( \begin{array}{l} \left( \begin{array}{l} \Xi_{\pm i(k)} \left( \begin{array}{l} a_{\pm i(k)} \left( \begin{array}{l} p_{r\ \pm i(k) class} - eA_{r} \\ + Re\left\{ \dfrac{\hbar}{4} \text{†}^{ab} \text{Š}_{abr} \right\} \end{array} \right) \dot{q}^{r}{}_{\pm i} \dfrac{\Delta_{s\pm i(k)\sim\text{€}}}{4} \\ - a_{\pm i(k)} \Theta_{s\sim\text{€}} - \text{‡}_{s\pm i(k)\sim\text{€}\, mix} \end{array} \right) \\ + \Xi^{*}_{\pm i(k)} \left( \begin{array}{l} a_{\pm i(k)} \left( \begin{array}{l} p_{r\ \pm i(k) class} - eA_{r} \\ + Re\left\{ \dfrac{\hbar}{4} \text{†}^{ab} \text{Š}_{abr} \right\} \end{array} \right) \dot{q}^{r}{}_{\pm i} \dfrac{\Delta_{s\pm i(k)\sim\text{€}}}{4} \\ - a_{\pm i(k)} \Theta_{s\sim\text{€}} + \text{‡}_{s\pm i(k)\sim\text{€}\, mix} \end{array} \right) \end{array} \right) \\ + \dfrac{\text{‡}_{s\pm i\sim\text{€}\, curv}}{2} \end{array} \right) \text{u}_{hk} \qquad (3.2.8)$$

where both $\text{‡}_{\pm i(k)\sim\text{€}\, mix}$ and $\text{‡}_{s\pm i\sim\text{€}\, curv}$ have been splitted into the isotropic and stress parts as follows

$$\text{‡}_{\pm i(k)\sim\text{€}\, mix} = \frac{\text{‡}_{\pm i(k)\}\}\, mix}}{4} g_{\sim\text{€}} + \text{‡}_{s\pm i(k)\sim\text{€}\, mix} \qquad (3.2.9)$$

$$\text{‡}_{\pm i\sim\text{€}_{curv}} = \frac{\text{‡}_{\pm i\}\}\, curv}}{4} g_{\sim\text{€}} + \text{‡}_{s\pm i\sim\text{€}_{curv}} . \qquad (3.2.10)$$

It is worth noting that equation (3.1.7) represents the decoherent limit of (3.2.1).
Equations (3.2.1-7) can be expressed as a function of the DFWE field $\Psi$ by using the relations (2.1.12, 2,2.3, 2.4.8, 2.6.14) and, from (2.1.7-8),

$$\dot{q}_{\pm i\sim} = -c^2 \frac{\partial_{\sim}\, ln[\frac{\text{Œ}_{\pm i}}{\text{Œ}_{\pm i}{}^*}] + \frac{2i}{\hbar}\left( eA_{\sim} - Re\left\{ \frac{\hbar c}{4} \dagger^{ab} \text{Š}_{ab\sim} \right\} \right)}{\left( \partial_t\, ln[\frac{\text{Œ}_{\pm i}}{\text{Œ}_{\pm i}{}^*}] - \frac{2i}{\hbar}\left( e\text{W} - Re\left\{ \frac{\hbar c}{4} \dagger^{ab} \text{Š}_{ab0} \right\} \right) \right)} . \qquad (3.2.11)$$

Actually, the hydrodynamic approach has been used as a "Trojan horse" to find the GE (3.2.1) of the DFWE where the non-physical states are excluded by writing it as a function of the original bi-spinor field $\Psi$ (corresponding just to the hydrodynamic irrotational states) .
The main difference with the classical Einstein equation is given by the terms $a_{(V_{qu})}$, $\text{U‡}_{\sim\text{€}_{stress}}$, $\text{‡}_{\pm i\sim\text{€}_{mix}}$ and $\text{‡}_{\pm i\sim\text{€}_{curv}}$ whose quantum-mechanical origin can be highlighted by passing to the macroscopic classical scale being $lim_{\hbar\to 0} a_{(V_{qu})} = 0$ , $lim_{macro} \text{U‡}_{\sim\text{€}_{stress}} = 0$, $lim_{macro} \text{‡}_{\pm i\sim\text{€}_{curv}} = 0$ and $lim_{macro} \text{‡}_{\pm i\sim\text{€}_{mix}} = 0$.

## 4. The GE-DFWE-EM system of quantum evolution

Once the system of evolutionary equations for the charged classical fermion field is defined in curved space-time

$$R_{\sim\text{€}} - \frac{1}{2} R g_{\sim\text{€}} = \frac{8f\, G}{c^4} \left( Tr\left( \text{‡}_{\sim\text{€}_{class}} > ]\, g_{\sim\text{€}} - ]_Q\, g_{\sim\text{€}} + \text{U‡}_{\sim\text{€}_{stress}} \right) + T_{\text{€}\sim\ em} \right) \qquad (4.1)$$

$$\left( i\hbar \text{X}^{\sim} \left( \partial_{\sim} - \frac{i}{4} \dagger^{ab} \text{Š}_{ab\sim} + \frac{ie}{\hbar} A_{\sim} \right) - mc \right) \Psi = 0 \qquad (4.2)$$

$$F^{\sim\text{€}}{}_{,\text{€}} = -4f\ J^{\sim} \qquad (4.3)$$

where the cosmological gravity tensor (CGT)

$$]\, g_{\sim\text{€}} < ]_Q\, g_{\sim\text{€}} - \text{U‡}_{\sim\text{€}_{stress}} \qquad (4.4)$$

is given by the classical cosmological constant $]\, g_{\sim\text{€}}$ plus the quantum CGT $]_Q\, g_{\sim\text{€}} > \text{U‡}_{\sim\text{€}_{stress}}$,
it is possible to proceed with the quantization of the fermion field.
From the general point of view, we observe that the quantization of the Dirac field in curved space-time, with the metric defined by the GE (4.1), owns the following properties:

1. In the limit of flat space-time, the standard quantum electrodynamics (QED) is recovered (see section 5.).
2. Since all ordinary elementary particles (low energy vacuum states) own by far mass densities very much smaller than the Planck mass in a sphere of Planck length, the first order of approximation of the gravitational evolutionary system of equations can lead to precise results by using the perturbative approach.
3. For fermions in very high curved space-time or of Planckian mass, the superposition of the Dirac fields eigenstates is quite different by the Fourier superposition of the Minkowskian case and the full treatment (e.g., high energy QFT) strongly diverges by the traditional results.

### 4.1 The perturbative approach to the GE-DFWE-EM evolution

For particles very far from the Planckian mass density $\frac{m_p}{l_p^{\,3}} = \frac{c^5}{\hbar G^2}$, it is possible to solve the system of the GE equation (4.1-4.3) by a perturbative iteration

$$R_{\sim€}\left(\text{V}_{\sim€}\right) \cong R^{(0)}{}_{\sim€}\left(\text{V}_{\sim€}\right) + R^{(1)}{}_{\sim€}\left(\text{V}_{\sim€}\right) + R^{(2)}{}_{\sim€}\left(\text{V}_{\sim€}\right) + \dots + R^{(n)}{}_{\sim€}\left(\text{V}_{\sim€}\right), \qquad (4.1.1)$$

$$\Psi = \Psi^{(0)} + \Psi^{(1)} + \Psi^{(2)} + \dots + \Psi^{(n)} \qquad (4.1.2)$$

for a small change of the space-time geometry $\text{V}^{a}{}_{\sim}$ given by the vielbein $e^{a}{}_{\sim}$ transformation

$$e^{a}{}_{\sim} = \text{u}^{a}{}_{\sim} + \frac{\text{V}^{a}{}_{\sim}}{2}, \qquad (4.1.3)$$

where ($\text{V}_{\sim€}\text{V}^{\sim€} << 1$), that leads (see appendix D) to the metric tensor variation

$$g_{\sim€} \cong \text{y}_{\sim€} + \frac{\text{y}_{\sim a}\text{V}^{a}{}_{€} + \text{y}_{€a}\text{V}^{a}{}_{\sim}}{2} = \begin{bmatrix} 1 & 0 & 0 & 0 \\ 0 & -1 & 0 & 0 \\ 0 & 0 & -1 & 0 \\ 0 & 0 & 0 & -1 \end{bmatrix} + \text{V}_{sim\,\sim€} \qquad (4.1.4)$$

where

$$\text{V}_{sim\,\sim€} = \frac{\text{y}_{\sim a}\text{V}^{a}{}_{€} + \text{y}_{€a}\text{V}^{a}{}_{\sim}}{2} \qquad (4.1.5)$$

and where

$$y_{a€} v^{a}{}_{\sim} = \begin{bmatrix} v^{0}{}_{0} & v^{0}{}_{1} & v^{0}{}_{2} & v^{0}{}_{3} \\ -v^{1}{}_{0} & -v^{1}{}_{1} & -v^{1}{}_{2} & -v^{1}{}_{3} \\ -v^{2}{}_{0} & -v^{2}{}_{1} & -v^{2}{}_{2} & -v^{2}{}_{3} \\ -v^{3}{}_{0} & -v^{3}{}_{1} & -v^{3}{}_{2} & -v^{3}{}_{3} \end{bmatrix}, \tag{4.1.6}$$

that for the successive steps of approximation $h^{(n)a}{}_{\sim}$

$$v^{a}{}_{\sim} = h^{(1)a}{}_{\sim} + h^{(2)a}{}_{\sim} + \ldots + h^{(n)a}{}_{\sim} \tag{4.1.7}$$

leads to

$$\begin{aligned} v_{sim\sim€} &= \frac{y_{\sim a} h^{(1)a}{}_{€} + y_{\sim a} h^{(1)a}{}_{€}}{2} + \frac{y_{\sim a} h^{(2)a}{}_{€} + y_{\sim a} h^{(2)a}{}_{€}}{2} \\ &= h^{(1)}_{sim\sim€} + h^{(2)}_{sim\sim€} + \ldots.. + h^{(n)}_{sim\sim€} \end{aligned} \tag{4.1.8}$$

where $y_{€\sim}$ satisfies the flat static solution $R^{(0)}_{\sim€\,(y_{€\sim})} = 0$, of the zero order GE

$$R^{(0)}_{\sim€} - \frac{1}{2} R^{(0)} = 0, \tag{4.1.9}$$

$\Psi^{(0)}$ is the solution of the zero order DFWE

$$\left( i\hbar x^{\sim} \left( \partial_{\sim} + \frac{ie}{\hbar} A_{\sim} \right) - mc \right) \Psi^{(0)} = 0 \quad , \tag{4.1.10}$$

$\Psi^{(1)}$ the solution of the first order DFWE (see appendix D)

$$\left( i\hbar x^{\sim} \left( \partial_{\sim} + \frac{ie}{\hbar} A_{\sim} \right) - mc \right) \Psi^{(1)} \cong \frac{\hbar}{8} x^{\sim} \dagger^{ab} \left( \partial_{\sim} v_{ab} \right) \Psi^{(0)} \quad , \tag{4.1.11}$$

and $h^{(1)}_{sim\sim€}$ is the solution of the first order GE

$$R^{(1)}_{\sim€\left(h^{(1)}_{sim\sim€}\right)} - \frac{1}{2} R^{(1)}_{\sim€\left(h^{(1)}_{sim\sim€}\right)} g_{\sim€} = \frac{8f G}{c^{4}} \left( Tr\left( \ddagger^{(0)}_{\sim€_{class}} - ]\, g_{\sim€} - ]^{(0)}_{Q} g_{\sim€} + U\ddagger^{(0)}_{\sim€_{stress}} \right) + T_{€\sim\;em} \right) \tag{4.1.12}$$

where the Christoffel symbol reads [19]

$$\begin{aligned} \Gamma^{s}_{\sim r} &= \frac{1}{2} y^{sx} \left( \partial_{r} v_{simx\sim} + \partial_{\sim} v_{simxr} - \partial_{x} v_{sim\sim r} \right) \\ &\cong \frac{1}{2} y^{sx} \left( \left( \partial_{r} h^{(1)}_{simx\sim} + \partial_{\sim} h^{(1)}_{simxr} - \partial_{x} h^{(1)}_{sim\sim r} \right) + \ldots.. \right) \end{aligned} \tag{4.1.13}$$

allowing to obtain the Ricci tensor from the expression [19]

$$R_{\sim€\,(h_{\sim€})} = \left( \partial_{l} \Gamma^{l}_{\sim€} - \partial_{€} \Gamma^{l}_{\sim l} + \Gamma^{l}_{\sim€} \Gamma^{m}_{lm} - \Gamma^{m}_{\sim l} \Gamma^{l}_{€m} \right). \tag{4.1.14}$$

Moreover, by using the zero-order relations $V_{qu0(k)} = 0$, $r_{(V_{qu0(k)})} = 0$, $\ddagger^{(0)}_{(k)\sim€_{mix}} = 0$,the components $\ddagger^{(0)}_{\sim€_{class}}$, $U\ddagger^{(0)}_{\sim€_{stress}}$ and $]^{(0)}_{Q} g_{\sim€}$ read, respectively,

$$‡^{(0)}_{\sim€_{class}} = \sum_{\pm i}\left( g^{(0)}_{\pm i(k)} /Œ^{(0)}_{\pm i(k)}/^2 \, T^{(0)}_{\pm i(k)\sim€_{class}} + g^{(0)}_{\pm i(-k)} /Œ^{(0)}_{\pm i(-k)}/^2 \, T^{(0)}_{\pm i(-k)\sim€_{class}} \right) µ_{hk} \qquad (4.1.15)$$

where $T^{(0)}_{\pm i(k)\sim€_{class}} = T_{\pm i(k)\sim€_{class}(Œ^{(0)}_{\pm i})}$ , where

$$T_{\pm i(k)\sim€_{class}} = \frac{mc^2}{\gamma} u_{\sim\,\pm i(k)} \left( u_{€\,\pm i(k)} - \frac{1}{mc}\left( Re\left\{ \frac{\hbar}{4} † ^{ab}Š_{ab€} \right\} - eA_{€} \right)\right) \qquad (4.1.16)$$

$$u_{\pm i(k)\sim} = \frac{ p_{\pm i(k)\sim} + Re\left\{ \frac{\hbar}{4} † ^{ab}Š_{ab\sim} \right\} - eA_{\sim} }{ mc \sqrt{1 - \frac{V_{qu\pm(k)}}{mc^2}} }$$

$$= \frac{ \frac{i\hbar}{2} g_{\sim r} \partial^{r} \, ln[\frac{Œ_{\pm i(k)}}{Œ_{\pm i(k)}{}^{*}}] + Re\left\{ \frac{\hbar}{4} † ^{ab}Š_{ab\sim} \right\} - eA_{\sim} }{ mc \sqrt{1 - \frac{V_{qu\pm(k)}}{mc^2}} } \qquad (4.1.17)$$

where $g^{(0)}_{\pm i(k)} = g_{\pm i(k)(Œ^{(0)}_{\pm i})}$, and, being $‡_{\sim€_{curv}(Œ_{0(k)})} = 0$ ,

$$U‡^{(0)}_{\sim€_{stress}} = 0 \qquad (4.1.18)$$

$$]^{(0)}_{Q} = 0 \qquad (4.1.19)$$

the first-order GE reads

$$R^{(1)}_{\sim€\left(h^{(1)}_{sim\sim€}\right)} - \frac{1}{2} R^{(1)}_{\sim€\left(h^{(1)}_{sim\sim€}\right)} g_{\sim€} = \frac{8ƒG}{c^4}\left( Tr\left( ‡^{(0)}_{\sim€_{class}} - ] \, g_{\sim€} \right) + T_{€\sim\ em} \right) \qquad (4.1.20)$$

leading to the decoherent GE limit

$$lim_{dec}\left( R_{\sim€\left(h^{(1)}_{sim\sim€}\right)} - \frac{1}{2} R_{\left(h^{(1)}_{sim\sim€}\right)} g_{\sim€} \right) = R^{dec}_{\sim€\left(h^{(1)}_{sim\sim€}\right)} - \frac{1}{2} R^{dec}{}_{\left(h^{(1)}_{sim\sim€}\right)} g^{dec}_{\sim€}$$

$$= \frac{8ƒG}{c^4}\left( \sum_{\pm i} /Œ^{(0)}_{\pm i(k)}/^2 \, T^{(0)}_{\pm i(k)\sim€_{class}} + T_{€\sim\ em} - \Lambda g_{\sim€} \right) \qquad . \qquad (4.1.21)$$

where it has been used the identity $g_{\pm i(k)} = 1$ for the eigenstates.

By performing the limit for $\hbar \rightarrow 0$ (when the DFWE disembogues into the KGE (2.4.5), equation (4.1.20) leads to the classical limit that coincides with the Einstein equation. (see ref. [19]).

The first contribution of the quantum part of the CGT (i.e., $\left( U‡_{\sim€_{stress}} - ]_{Q} \, g_{\sim€} \right)$) in (4.1), in the macroscopic (decoherent) limit, comes from the second order of approximation (i.e., $v_{sim\sim€} = h^{(1)}_{sim\sim€} + h^{(2)}_{sim\sim€}$ )

$$R^{(2)}_{\sim€\left(v_{sim\sim€}\right)} - \frac{1}{2} R^{(2)}_{\sim€\left(v_{sim\sim€}\right)} g_{\sim€} = \frac{8ƒG}{c^4}\left( Tr\left( ‡^{(1)}_{\sim€_{class}} - ] \, g_{\sim€} - ]^{(1)}_{Q} g_{\sim€} + U‡^{(1)}_{\sim€_{stress}} \right) + T_{€\sim\ em} \right)$$

(4.1.22)

where the fermion EITD $\ddagger^{(1)}_{\sim\text{€}_{class}}$ is calculated by using $\text{Œ}_{\pm ik} \cong \text{Œ}^{(0)}_{\pm ik} + \text{Œ}^{(1)}_{\pm ik}$ obtained by (4.1.11), where the CPTD $]^{(1)}_{Q}$ reads

$$]^{(1)}_{Q} = ]_{Q(\text{Œ}^{(0)}_{\pm ik}+\text{Œ}^{(1)}_{\pm ik})} \equiv \Lambda^{(1)}_{\pm i(k)Q}\, u_{jk} \qquad (4.1.23)$$

and where

$$U\ddagger^{(1)}_{\sim\text{€}_{stress}} = U\ddagger_{\sim\text{€}_{stress}(\text{Œ}^{(0)}_{\pm ik}+\text{Œ}^{(1)}_{\pm ik})} \equiv \Delta\ddagger^{(1)}_{\sim\text{€}_{stress}(k)}\, u_{jk} \qquad (4.1.24)$$

Moreover, being $lim_{dec}\ddagger_{\sim\text{€}_{curv}} = 0$, the macroscopic limit of the GE (to be used on cosmological scale) reads

$$R^{(2)dec}_{\sim\text{€}\left(v_{sim\sim\text{€}}\right)} - \frac{1}{2} R^{(2)dec}_{\left(v_{sim\sim\text{€}}\right)} g_{\sim\text{€}} = \frac{8f\, G}{c^4}\left( \sum_{\pm i} /\text{Œ}^{(1)}_{\pm i(\tilde{k})}/^2 \begin{pmatrix} \left(1-a^{(1)}_{\pm i(\tilde{k})}\right) T^{(1)}_{\pm i(k)\sim\text{€}_{class}} \\ +\Lambda^{(1)}_{\pm i(\tilde{k})Q}\, g_{\sim\text{€}} + \Delta\ddagger^{(1)}_{\sim\text{€}_{stress}(k)} \end{pmatrix} - \Lambda_{\sim\text{€}} + T_{\sim\text{€}\;em} \right)$$

(4.1.25)

where $k$ is the eigenstates to which the wave-function collapses due to the decoherence process, where

$$T^{(1)}_{\pm i(k)\sim\text{€}_{class}} = T_{\pm i(k)\sim\text{€}_{class}(\text{Œ}^{(0)}_{\pm ik}+\text{Œ}^{(1)}_{\pm ik})}, \qquad (4.1.26)$$

and where

$$\Lambda^{(1)}_{(k)Q} = \Lambda^{(1)}_{(k)Q} + \Lambda^{(1)}_{(-k)Q} \qquad (4.1.27)$$

where, by using (2.7.3), it follows that

$$\Lambda^{(1)}_{(k)Q} = \Lambda_{(k)Q\left(\text{Œ}_{\pm ik}=\text{Œ}^{(0)}_{\pm ik}+\text{Œ}^{(1)}_{\pm ik}\right)}$$

$$= lim_{\text{Œ}=\text{Œ}^{(0)}+\text{Œ}^{(1)}} \sum_{\pm i} /\text{Œ}_{\pm i(k)}/^2\, a_{\pm i(k)} \frac{\Delta_{\pm i(\pm k)\}\}}}{4} \begin{pmatrix} L_{\pm i(k)_{class}} + \\ \begin{pmatrix} Re\left\{\frac{\hbar}{4}\dagger^{ab}\check{S}_{abr}\right\} \\ -eA_{r} \end{pmatrix} \dot{q}^{r}_{\pm i(k)} \end{pmatrix}, \qquad (4.1.28)$$

$$= lim_{\text{Œ}=\text{Œ}^{(0)}+\text{Œ}^{(1)}} \sum_{\pm i} /\text{Œ}_{\pm i(k)}/^2\, a_{\pm i(k)} \frac{\Delta_{\pm i(\pm k)\}\}}}{4} \sqrt{g_{rs}\,\dot{q}^{s}_{\pm i(k)}\dot{q}^{r}_{\pm i(k)}}$$

and

$$\Delta\ddagger^{(1)}_{\sim\text{€}_{stress}} = \Delta\ddagger^{(1)}_{\sim\text{€}_{stress}(k)} + \Delta\ddagger^{(1)}_{\sim\text{€}_{stress}(-k)} \qquad (4.1.29)$$

where

$$\Delta \ddagger^{(1)}_{\sim\text{€}\,stress(k)} = \sum_{\pm i} /\text{Œ}_{\pm i(k)}/^{2}\, a_{\pm i(k)} \Delta_{s\pm i(k)\sim\text{€}} \left( L_{\pm i(k)class} + \begin{pmatrix} Re\left\{\frac{\hbar}{4}\dagger^{ab}\check{S}_{abr}\right\} \\ -eA_{r} \end{pmatrix} \dot{q}^{r}_{\pm i(k)} \right), \quad (4.1.30)$$

$$= \sum_{\pm i} /\text{Œ}_{\pm i(k)}/^{2}\, a_{\pm i(k)} \Delta_{s\pm i(k)\sim\text{€}} \sqrt{g_{rs}\dot{q}^{s}_{\pm i(k)}\dot{q}^{r}_{\pm i(k)}}$$

The second order GE (4.1.25) contains the cosmological pressure tensor density (CPTD) $\Lambda^{(1)}_{(k)Q}$ and the stress gravity term $\Delta\ddagger^{(1)}_{\sim\text{€}\,stress}$ that go to zero if $\hbar$ is set to zero. Thence, $\Lambda^{(1)}_{(k)Q}$ and $\Delta\ddagger^{(1)}_{\sim\text{€}\,stress}$ actually are quantum-mechanical decoherent contribution to the Newtonian gravity.

Since both terms $r^{(1)}_{(V_{qu1(k)})}$ and $\Delta^{(1)}_{s\sim\text{€}(k)}$ become relevant in very high-curvature space-time, the quantum-mechanical correction to the gravity should give noticeable effects in the neighbourhood of very massive object such as black holes and in the rotation of twin neutron stars or black holes.

## 5. Cosmological tensor density of the quantum Dirac field

In the classical treatment, except for the photon, the Einstein equation is not coupled to the particles fields, but just to the energy impulse tensor of classical bodies.
The GE (4.1) is analytically coupled to the classical fermion field and it takes into account the energy of the (non-local) quantum potential for determining the geometry of the space-time.
The most important effect of the quantum potential is the generation of the CGT.

If we put to zero the quantum potential in the decoherent GE (3.1.7) (as it happens in the classical limit (i.e., $\hbar = 0$ and $V_{qu} = 0$ )), the CGT (4.4) reduces to the classical cosmological constant.
Thence, the gravitational effect of the quantum potential leads to a not null CGT that allows the quantum vacuum to be in the stable physical phase [9] even if the classical cosmological constant is null.

Given that the contribution of the quantum potential modifies both the classical metric tensor of general relativity and the quantization rules (see Appendix E), it also affects the field quantization.
In this section, we will derive the CPTD of the quantum fermion field in in the quasi Minkowskian space-time in order to evaluate its mean value on cosmological scale.

Since the EM field owns a null CGT (the classical theory coincides with the quantum hydrodynamic description (i.e., $V_{qu} = 0$ for the photon) and the gravitational coupling is the same of the classical general relativity), for sake of simplicity we derive the CPTD by considering the case of uncharged fermions in the quasi-Minkowskian limit whose Fourier decomposition [31] reads

$$\Psi = \sum_{r=1}^{2} \iint \frac{d^{3}p}{(2f)^{3}} \frac{1}{2E_{p}} \left( b_{r(p)} u^{(r)}_{(p)} exp[-i\frac{p_{r}q^{r}}{\hbar}] + d^{\dagger}_{r(p)} v^{(r)}_{(p)} exp[i\frac{p_{r}q^{r}}{\hbar}] \right) \quad (5.1)$$

$$\overline{\Psi} = \sum_{r=1}^{2} \iint \frac{d^{3}p}{(2f)^{3}} \frac{1}{2E_{p}} \left( b^{\dagger}_{r(p)} \overline{u}^{(r)}_{(p)} exp[i\frac{p_{r}q^{r}}{\hbar}] + d_{r(p)} \overline{v}^{(r)}_{(p)} exp[-i\frac{p_{r}q^{r}}{\hbar}] \right) \quad (5.2),$$

where $r$=1,2 and where

$$u^{(r)}_{(p)} = \begin{pmatrix} u^{(r)}_{1(p)} \\ u^{(r)}_{2(p)} \end{pmatrix} = \frac{1}{\sqrt{E_p + m}} \begin{pmatrix} \left(E_p + m\right) \mathrm{t}^{(r)} \\ (\dagger .p)\, \mathrm{t}^{(r)} \end{pmatrix} \tag{5.3}$$

$$v^{(r)}_{(p)} = \begin{pmatrix} v^{(r)}_{1(p)} \\ v^{(r)}_{2(p)} \end{pmatrix} = \frac{1}{\sqrt{E_p + m}} \begin{pmatrix} (\dagger .p)\, \mathrm{t}^{(r)} \\ \left(E_p + m\right) \mathrm{t}^{(r)} \end{pmatrix} \tag{5.4}$$

where

$$\mathrm{t}^{(1)} = \begin{pmatrix} 1 \\ 0 \end{pmatrix} \tag{5.5}$$

$$\mathrm{t}^{(2)} = \begin{pmatrix} 0 \\ 1 \end{pmatrix}. \tag{5.6}$$

Moreover, by using the discrete form of field Fourier decomposition (i.e., $\iint \frac{d^3k}{(2f\ )^3} \rightarrow \frac{1}{\sqrt{V}} \sum_{k=0}$ )

$$\Psi = \begin{pmatrix} \text{Œ}_1 \\ \text{Œ}_2 \end{pmatrix} = \frac{1}{\sqrt{V}} \sum_{p=0}^{\infty} \frac{1}{2\hbar \text{Š}_p} \sum_{r=1}^{2} \begin{pmatrix} b_{r(p)} u^{(r)}_{(p)}\, exp[-i \frac{p_{\mathrm{r}} q^{\mathrm{r}}}{\hbar}] \\ +d^{\dagger}_{r(p)} v^{(r)}_{(p)}\, exp[\,i \frac{p_{\mathrm{r}} q^{\mathrm{r}}}{\hbar}] \end{pmatrix}$$

$$= \frac{1}{\sqrt{V}} \sum_{p=0}^{\infty} \frac{1}{2\hbar \text{Š}_p} \sum_{r=1}^{2} \begin{pmatrix} b_{r(p)} \begin{pmatrix} u^{(r)}_{1(p)} \\ u^{(r)}_{2(p)} \end{pmatrix} exp[-i \frac{p_{\mathrm{r}} q^{\mathrm{r}}}{\hbar}] \\ +d^{\dagger}_{r(p)} \begin{pmatrix} v^{(r)}_{1(p)} \\ v^{(r)}_{2(p)} \end{pmatrix} exp[\,i \frac{p_{\mathrm{r}} q^{\mathrm{r}}}{\hbar}] \end{pmatrix} \tag{5.7}$$

it follows that

$$\text{Œ}_{\pm} = \frac{\text{Œ}_1 \pm \text{Œ}_2}{\sqrt{2}} = \frac{1}{\sqrt{V}} \sum_{p=0}^{\infty} \frac{1}{2\hbar \text{Š}_p} \sum_{r=1}^{2} \begin{pmatrix} b_{r(p)} \frac{u^{(r)}_{1(p)} \pm u^{(r)}_{2(p)}}{\sqrt{2}} exp[-i \frac{p_{\mathrm{r}} q^{\mathrm{r}}}{\hbar}] \\ +d^{\dagger}_{r(p)} \frac{v^{(r)}_{1(p)} \pm v^{(r)}_{2(p)}}{\sqrt{2}} exp[\,i \frac{p_{\mathrm{r}} q^{\mathrm{r}}}{\hbar}] \end{pmatrix} \tag{5.8}$$

that compared to the Minkowskian equation (2.2.1)

$$\text{Œ}_{\pm} = \sum_{k=0} \mathrm{r}_{\pm k}\, exp[-i \frac{p_{\mathrm{r}} q^{\mathrm{r}}}{\hbar}] + \mathrm{s}^{\dagger}_{\pm k}\, exp[\,i \frac{p_{\mathrm{r}} q^{\mathrm{r}}}{\hbar}]\,, \tag{5.9}$$

leads to

$$\mathrm{r}_{\pm p} = \frac{1}{2\hbar \text{Š}_p} \sum_{r} b_{r(p)} \frac{u^{(r)}_{1(p)} \pm u^{(r)}_{2(p)}}{\sqrt{2}} \tag{5.10}$$

$$\mathrm{s}^{\dagger}_{\pm p} = \frac{1}{2\hbar \check{S}_p}\sum_r d^{\dagger}_{r(p)} \frac{v_{1(p)}^{(r)} \pm v_{2(p)}^{(r)}}{\sqrt{2}} \tag{5.11}$$

Once $\mathrm{r}_{\pm p}$ and $\mathrm{s}_{\pm p}$ are defined as a function of the creation and annihilation operators of the quantum Dirac field we can obtain the expression of the expectation value of the CPTD that reads

$$< 0_p / \Lambda_{(k)Q} / 0_p > = \sum_{\pm i} \int^{\check{S}_{k_{max}}} \frac{\Lambda_{\pm i(k)Q}}{(2f)^2 c^2} \sqrt{\frac{\left(\hbar \check{S}_p - eW\right)^2}{c^2} - m^2c^2\left(1 - \frac{V_{qu\pm i(p)}}{mc^2}\right)}\, d\check{S}_{\pm i(p)} \tag{5.12}$$

where, from (3.1.7), the decoherent macroscopic CPTD $\Lambda_{\pm i(k)Q}$ reads

$$\Lambda_{\pm i(k)Q} = -/Œ_{\pm i(k)}/^2 a_{\pm i(k)} \left( \frac{\Theta_{\}\}}}{4} - \frac{\Delta_{\pm i(k)\}\}}}{4} \left( \begin{matrix} L_{\pm i(k)class} - eA_{\mathrm{r}}\dot{q}^{\mathrm{r}}{}_{\pm i(k)} \\ + Re\left\{\frac{\hbar}{4} \dagger^{ab}\check{S}_{ab\mathrm{r}}\right\}\dot{q}^{\mathrm{r}}{}_{\pm i(k)} \end{matrix} \right) \right) \tag{5.13}$$

and where (for uncharged fermion $e = 0$)

$$\check{S}_{\pm i(p)} = \frac{eW}{\hbar} + \sqrt{\frac{c^2 p_{\pm i}^2}{\hbar^2} + \frac{m^2c^4}{\hbar^2}\left(1 - \frac{V_{qu\pm i(p)}}{mc^2}\right)}, \tag{5.1.4}$$

obtained from the HHJE (2.6.13) and the identities: (2.1.1) , $p_{\pm i_{\sim}} = (p_{\pm i_0}, -p_{\pm i_j})$ , $p_{\pm i}^2 = p_{\pm i_j} p_{\pm i_j}$ and $\frac{\hbar}{c}\check{S}_{\pm i} = p_{\pm i_0}$ .

In order to have the explicit form of the quantum operator CPTD as a function of the creation and annihilation operators of the Dirac field, the ordering problem in the expressions $/Œ_{\pm i(\pm k)}/^2$ and $V_{qu\pm i(\pm p)}$ must be defined. The criterion of ordering the creation and annihilation operators is defined by the necessity to avoid divergences [31]. Thence, the normal ordering is assumed here for the CPTD that reads

$$< 0_p / \Lambda_{(k)Q} / 0_p > =$$

$$= \sum_{\pm i} \int^{\check{S}_{k_{max}}} \frac{:\Lambda_{\pm i(k)Q}:}{(2f)^2 c^2} \sqrt{\frac{\left(\hbar :\check{S}_{\pm i(p)}: - eW\right)^2}{c^2} - m^2c^2\left(1 - \frac{:V_{qu\pm i(p)}:}{mc^2}\right)}\, d :\check{S}_{\pm i(p)}: \tag{5.15}$$

where

$$:\check{S}_{\pm i(p)}:= \frac{eW}{\hbar} + \sqrt{\frac{c^2 p_{\pm i}^2}{\hbar^2} + \frac{m^2c^4}{\hbar^2}\left(1 - \frac{:V_{qu\pm i(p)}:}{mc^2}\right)} \tag{5.16}$$

Moreover, since at zero order of the Minkowskian limit, the mode *k* of the quantum potential (2.6.14) for both positive and negative states (for uncharged fermions (i.e., $e = 0$)) reads

$$:V_{qu\pm i(+k)}:= -\frac{\hbar^2}{m}\frac{1}{:/Œ_{\pm i\,(+k)}/:\sqrt{-g}}\partial_{\sim}\sqrt{-g}\,g^{\sim\text{€}}\partial_{\text{€}}\,/\left(\mathrm{r}_{\pm k}\, exp[\,ik_{\mathrm{r}}q^{\mathrm{r}}\,]\right)/$$

$$= -\frac{\hbar^2}{m}\frac{1}{:/Œ_{\pm i\,(+k)}/:\sqrt{-g}}\partial_{\sim}\sqrt{-g}\,g^{\sim\text{€}}\partial_{\text{€}}\,:\left(\sqrt{\mathrm{r}_{\pm k}\mathrm{r}^{\dagger}{}_{\pm k}}\right):= 0 \tag{5.17}$$

and

$$:V_{qu\pm i(-k)}:=-\frac{\hbar^2}{m}\frac{1}{:/\text{Œ}_{\pm i\ (-k)}/:\sqrt{-g}}\partial_{\tilde{}}\sqrt{-g}\,g^{\tilde{}\text{€}}\partial_{\text{€}}\,/\left(\text{s}^{\dagger}{}_{\pm k}\,exp[-ik_{\text{r}}q^{\text{r}}\ ]\right)/$$
$$=-\frac{\hbar^2}{m}\frac{1}{:/\text{Œ}_{\pm i\ (-k)}/:\sqrt{-g}}\partial_{\tilde{}}\sqrt{-g}\,g^{\tilde{}\text{€}}\partial_{\text{€}}\ :\left(\sqrt{\text{s}^{\dagger}{}_{\pm k}\text{s}_{\pm k}}\right):=0 \quad , \qquad (5.18)$$

it follows that

$$\left(:\check{S}_{\pm i(p)}:-\frac{e\text{W}}{\hbar}\right)^2=\left(\check{S}_{\pm i(p)}-\frac{e\text{W}}{\hbar}\right)^2=\frac{c^2p_{\pm i}^2}{\hbar^2}+\frac{m^2c^4}{\hbar^2}, \qquad (5.19)$$

(showing that, at the zero order, the standard QFT identity is recovered) and that

$$:a_{\pm i(k)(:\hat{V}_{qu}:)}:=0 \qquad (5.20)$$
$$:\Delta_{||(k)}:=0 \qquad (5.21)$$
$$:\Lambda_{(k)Q}:=0 \qquad (5.22)$$
$$<0_p/:\Lambda_{(k)Q}:/0_p>=0 \qquad (5.23)$$

where (5.23) shows that the Minkowskian ST (very far from matter) does not contribute to the cosmological constant.

### 5.1. Weak gravity perturbation Hamiltonian of the quantum Dirac field

Given the weak-gravity coupling (see appendix D)

$$J=\frac{\hbar}{8}\text{x}^{\tilde{}}\dagger^{ab}\left(\partial_{\tilde{}}\text{V}_{ab}\right), \qquad (5.1.1)$$

where $\text{V}_{ab}$ can be calculated by solving the gravitational equation (4.1) by using the zero order of the Dirac field $\Psi_0$, the interaction term of the field Lagrangean $\mathcal{L}_I$ reads

$$\mathcal{L}_I=\overline{\Psi}J\Psi=\frac{\hbar}{8}\left(\overline{\Psi}\text{x}^{\tilde{}}\left(\partial_{\tilde{}}\text{V}_{ab}\right)\Psi\right) \qquad (5.1.2)$$

leading to a correction to the Minkowskian zero order

$$:H_0:=\frac{1}{2}\sum_{r=1}^{2}\iint\frac{d^3p}{(2f\ )^3}\left(b_{r(p)}^{\dagger}b_{r(p)}+d_{r(p)}^{\dagger}d_{r(p)}\right), \qquad (5.1.3)$$

given by the perturbative interaction Hamiltonian [31]

$$H_1=-\int d^3q\quad\overline{\Psi}J\Psi$$
$$=-\int\frac{d^3k}{(2f\ )^3 2\check{S}_k}\begin{pmatrix}\left(b_{r(p)}^{\dagger}\overline{u}_{(p)}^{(r)}+d_{r(p)}^{\dagger}\text{v}_{(p)}^{(r)}\right)\overline{J}(t,k)exp[i\frac{p_{\text{r}}q^{\text{r}}}{\hbar}]\\+\left(b_{r(p)}u_{(p)}^{(r)}+d_{r(p)}\overline{\text{v}}_{(p)}^{(r)}\right)J(t,k)exp[-i\frac{p_{\text{r}}q^{\text{r}}}{\hbar}]\end{pmatrix} \qquad (5.1.4)$$

where

$$J(t,k) = \int d^3q \; J(t,q) e^{-ikq} = \frac{\hbar}{8} \text{x} \,\tilde{}\, \dagger^{ab} \int d^3q \; e^{-ikq} \partial_{\sim} \text{V}_{ab} \tag{5.1.5}$$

$$\overline{J}(t,k) = \int d^3q \; J(t,q) e^{+ikq} \tag{5.1.6}$$

In appendix F it is shown that the Newtonian gravity leads to a null gravitational perturbation to the quantum Dirac field. This fact agrees with the experimental outputs of the Minkowskian QED that leads to the prediction of the anomalous magnetic moment with six significant digits [31]. The gravitational corrections coming from higher order terms can be a test for the theory by predicting the anomalous magnetic moment with higher precision.
Moreover, (5.1.4) may give an important contribution in very heavy fermion particles.

**6. Discussion**

The hydrodynamic representation of the DFWE field makes it equivalent to mass distributions submitted to the non-local quantum potential (2.6.14) obeying to the GE (4.1).
In agreement with the basic principle of the general relativity, the assumption of the theory is that the energy of the quantum potential (expression of the quantum-mechanical properties of the vacuum) contributes to the curvature of the space-time.
On the basis of this postulate, the quantum-mechanical non-local effects (e.g., the uncertainty principle) come into the gravity leading to the theoretical appearance of the CGT in the GE and to the removal of the point singularity of black holes in general relativity [32].
In the classical treatment, the Einstein equation for massive particles is not coupled to any matter field, but just to the energy impulse tensor of the classical bodies and does not contain any information about how it couples with their quantum-mechanical fields (i.e., their phases).
On the other hand, the GE-DFWE-EM (4.1-3) system owns a complete gravitational coupling and can be further quantized leading to a quantum gravity based on the description of the physical vacuum that includes its quantum-mechanical properties (e.g the quantum non-local potential that generates repulsion against the mass density concentration in smaller and smaller volume).

**6.1. The GE and the quantum gravity**

Even if the quantization of the DFWE field in the curved ST is not treated in this work, the inspection of some features of the quantum gravity to the light of the GE (4.1) deserves a mention.
Since the action of the GE (4.1) is basically given by the standard Einstein-Hilbert action plus terms stemming by the energy of the non-local quantum potential of massive DFWE, the outputs of the quantum "pure gravity" practically remains valid for the GE presented in this work.
As shown in [9], one interesting aspect of the quantum pure gravity is that the vacuum does not make a transition to the *collapsed branched polymer phase*, if and only if there exists a vanishing small cosmological constant (i.e., $] \neq 0$).
Therefore the presence of the term $]_Q$ in the GE (4.1), in principle, allows to pose $] \; \text{N} \, 0$ (as strongly supported by Einstein [6]) with the GE that reads.

$$R_{\sim\text{€}} - \frac{1}{2} R g_{\sim\text{€}} + \frac{8f\,G}{c^4} Tr\left( ]_Q \right) g_{\sim\text{€}} = \frac{8f\,G}{c^4} \left( Tr\left( \ddagger_{\sim\text{€}\,class} + \text{U}\ddagger_{\sim\text{€}\,stress} \right) + T_{\text{€}\sim\,em} \right) \tag{6.1.1}$$

Moreover, since the matter makes the space-time curved and, hence,

$$a_{\pm i(\pm k)} \neq 0 \tag{6.1.2}$$

$$\Delta_{||(k)} \neq 0 \tag{6.1.3}$$

$$\text{U}\ddagger_{\sim\text{€}\,stress} \neq 0 \tag{6.1.4}$$

and

$$ ]_{Q} \neq 0 , \tag{6.1.5} $$

it follows that the matter itself stabilizes the vacuum in the physical strong gravity phase [9].

On the other hand, a perfect Minkowskian vacuum (i.e., without matter) will make transition to the non-physical *collapsed branched polymer phase* with no sensible continuum limit [9], leading to no-space and no-time as we know.

**6.2. Analogy with the Brans-Dicke gravity**

The output the work highlights an interesting analogy with the Brans-Dicke [13-15] gravity that solves the problem of the cosmological constant [16] as well as those of the inflation [17] and dark energy [18].

If we look in detail to the EITD $/\text{Œ}_{\pm i}/^{2}\,\ddagger_{\pm i\sim\text{€}\,class}$ , from the macroscopic decoherent limit (see (3.29)

$$ lim_{dec} \sum_{\pm i} /\text{Œ}_{\pm i}/^{2}\,\ddagger_{\pm i\sim\text{€}\,class} = \sum_{\pm i} T_{\pm i(k)\sim\text{€}\,class} , \tag{6.2.1} $$

we can see that, by using (2.3.9), it leads to the expression

$$ \begin{aligned} \frac{8f\,G}{c^{4}} T_{\pm i\sim\text{€}\,class} &= \frac{8f\,G}{c^{4}} \sum_{\pm i} /\text{Œ}_{\pm i}/^{2}\, c^{2} \left( \frac{\partial S_{\pm i}}{\partial t} \right)^{-1} \left( g^{\sim\text{s}} \frac{\partial S_{\pm i}}{\partial q^{\text{s}}} \frac{\partial S_{\pm i}}{\partial q^{\text{€}}} \right) \\ &= -\frac{8f\,G}{c^{4}} \sum_{\pm i} /\text{Œ}_{\pm i}/^{2}\, \frac{\hbar^{2}c^{2}}{4E} \left( g^{\sim\text{s}} \left( \frac{\partial_{\text{s}}\text{Œ}_{\pm i}}{\text{Œ}_{\pm i}} - \frac{\partial_{\text{s}}\text{Œ}_{\pm i}{}^{*}}{\text{Œ}_{\pm i}{}^{*}} \right) \left( \frac{\partial_{\text{€}}\text{Œ}_{\pm i}}{\text{Œ}_{\pm i}} - \frac{\partial_{\text{€}}\text{Œ}_{\pm i}{}^{*}}{\text{Œ}_{\pm i}{}^{*}} \right) \right) \\ &= -\frac{8f}{c^{4}} \sum_{\pm i} \frac{G}{\text{Œ}_{\pm i}^{2}} /\text{Œ}_{\pm i}/^{2}\, \frac{\hbar^{2}c^{2}}{4E} \begin{pmatrix} g^{\sim\text{s}} \partial_{\text{s}}\text{Œ}_{\pm i} \partial_{\text{€}}\text{Œ}_{\pm i} + \ldots. \\ \ldots. + \left( \frac{\text{Œ}_{\pm i}}{\text{Œ}_{\pm i}{}^{*}} \right)^{2} g^{\sim\text{s}} \partial_{\text{s}}\text{Œ}_{\pm i}{}^{*} \partial_{\text{€}}\text{Œ}_{\pm i}{}^{*} \end{pmatrix} \end{aligned} \tag{6.2.2} $$

that is the summation over the spinor terms, each one owing the form of the scalar Brans-Dicke gravity [13] in absence of external potentials (for sake of simplicity we have considered the case of uncharged fermions).

If is worth noting that the DWFE gravity, endorses the concept of the effective gravitational constant of the Brans-Dicke model that in the spinor-tensor form (6.1.2) leads to effective gravitational constants for each spinor $G_{eff\pm i} = \dfrac{G}{\text{Œ}_{\pm i}^{2}}$ .

**6.3. Quantum-mechanical gravity and the foundations of the quantum theory**

The output (2.5.12) shows that in order to have the macroscopic behaviour in addition to the smallness of the Planck constants, the decoherence process induced by uncorrelated fluctuations is necessary: The decoherence process of the quantum states is necessary to obtain the classical extremal principle and the axiomatic description of the macroscopic classical physics.

In the quantum hydrodynamic model (Madelung – De Broglie-Bohm pilot wave approach) the passage to the classic equation of motion is obtained just by the condition of assuming the Planck's constant vanishing compared to the action scales of the problem [22]. Actually, in the quantum hydrodynamic model, the decoherence remains undefined since it shows itself in the open-system stochastic generalization of quantum systems [22-26]. The decoherence process does not appear in the quantum hydrodynamic theory since fluctuations are absent. Actually, in the classical limit for $\hbar \rightarrow 0$ there is the change of the "mathematical nature" of the quantum hydrodynamic motion equations, produced by the disappearance of the quantum potential that generates the eigenstates and the coherent evolution of their superposition [22-24 ].

In presence of a noisy environment (when the quantum potential energy is much smaller than the energy amplitude of fluctuations) the coherent superposition of states cannot be maintained in time and only the eigenstates configurations are stable [24, 33]
The approach based on the minimum action condition (2.5.11) shows that the Madelung–De Broglie-Bohm pilot wave approach needs to be completed by the existence of the decoherence process

### 6.4. Experimental tests

The result (5.23) basically shows that in the Minkowskian space-time (i.e., the vacuum very far from particles) the CPTD expectation value is vanishing regardless its zero-point energy density.
The contribution to the CPTD appears at second order in the perturbative development of the GE. The macroscopic CPTD expectation value is vanishing in the region of space-time with (Newtonian) weak gravity and it increases at higher gravity as quantum-mechanical corrections. In fact, by observing that, due to the high density of mass distribution in black holes (e.g., $V_{qu\pm i}$ acquires a value comparable to the mass energy $mc^2$ (e.g., $\frac{V_{qu\pm i(\pm p)}}{mc^2} \to 1$ and $a_{\pm i(\pm k)} \to 1$ [32]) it turns out that the largest contribution to the measured cosmological constant of the universe comes from the high-gravity regions of space (e.g., black holes, neutron stars and super massive BHs (SMBHs) at the center of the galaxies).
On the other hand, from the ordinary baryonic mass, a very small bit of contribution to the CPTD may come from the space very close to the center of ordinary elementary particles, where the gravity reaches the largest value.
Thence, the quantum-mechanical properties of the vacuum lead to a great lowering of the expectation value of the CPTD on cosmological scale with an order of magnitude that can be in agreement with the astronomical observations [20].
From the experimental point of view, the *quantum-mechanical corrections* to the Newtonian gravity, $a_{(\tilde{k})} \frac{mc^2}{\chi} g_{\sim r} u^{r}_{class} u^{€}_{class}$ , $\Lambda_{(k)Q} g_{\sim €}$ and $\Delta‡_{\sim € \, stress}$ in (4.1.25), must be primarily detectable in the rotation of twin black holes or neutron stars (by means of the interferometric detection of the angular and frequency-dependent response functions [34] of gravitational waves) in the motion of stars around the SMBHs at the center of the galaxies (i.e., in the bulge) and in the inter-galactic interaction.
Moreover, since the Newtonian gravity leads to a null first order gravitational perturbation of the quantum Dirac field (see appendix F) the prediction of the anomalous magnetic moment with six significant digits [31] in the Minkowskian QED is confirmed and explained by the theory. The gravitational corrections coming from higher order terms (that can be relevant in high gravity and for very heavy fermions) can be a test for the theory that should also allow to predict the anomalous magnetic moment with higher precision.

### 7. Conclusion

The quantum hydrodynamic representation of the Dirac-Fock-Weil equation, describing the bi-spinor fields $\Psi = \begin{pmatrix} Œ_1 \\ Œ_2 \end{pmatrix}$ as the motion of four mass densities $/Œ_{\pm 1}/^2, /Œ_{\pm 2}/^2$ (where $Œ_{\pm} = \frac{Œ_1 \pm Œ_2}{\sqrt{2}} = \begin{pmatrix} Œ_{\pm 1} \\ Œ_{\pm 2} \end{pmatrix} = \begin{pmatrix} /Œ_{\pm 1}/ \, exp[\, i \frac{S_{\pm 1}}{\hbar} \,] \\ /Œ_{\pm 2}/ \, exp[\, i \frac{S_{\pm 2}}{\hbar} \,] \end{pmatrix}$.) owing the hydrodynamic moments $p_{\pm 1 \sim} = -\partial_{\sim} S_{\pm 1}$ $p_{\pm 2 \sim} = -\partial_{\sim} S_{\pm 2}$ and subject to the quantum potential, has been used to derive the correspondent gravity equation by using the minimum action principle. The gravity equation associated to the DFWE field takes into account the gravitational effects of the energy of the non-local theory-defined quantum potential.

The self-generation of the CPTD $]_Q$ leads to the attractive hypothesis that the matter itself generates the physical stable vacuum phase in which it is embedded.

The GE of the fermion field shows that the measured cosmological constant is the effect of the cosmological pressure tensor density that emerges as 2nd order quantum-mechanical correction to the Newton gravity.

The macroscopic CPTD $\Lambda_{(k)Q}$ it is not null if, and only if, massive particles are present but it tends to zero in the flat vacuum so that its spatial mean, on the cosmological scale, have a much smaller order of magnitude than that one deriving *tout court* by accounting the zero-point vacuum energy density, and possibly agrees with the measured cosmological constant .

The GE of the classical DFWE field shows that the CPTD $\Lambda_{(k)Q}$ , as well as the quantum CGT, are 2nd order macroscopic "quantum-mechanical" gravitational effects that become more and more relevant higher and higher the space-time curvature. This output shows that the gravity near super massive black holes at the center of the galaxies (e.g., in the bulge), between galaxies on cosmological scale and between twin black holes can be deeply influenced by the quantum-mechanical effects of the gravity.

The spinor-tensor gravitational coupling of the fermion field owns (for each spinor) terms typical of the Brans-Dicke modified gravity in absence of external potentials.

The gravitational corrections to the quantum fermion field can be tested by predicting the anomalous magnetic moment with higher precision and in determining the states of very heavy fermions in high-energy QED.

## References

1. Ortiz, M.E. Black Holes, the Wheeler-DeWitt equation and the semiclassical Approximation. arXiv **1994**, arXiv:9410030v1.
2. Barcelò, C. Revisiting the semiclassical gravity scenario for gravitational collapse. arXiv **2009**, arXiv:0909.4157v1.
3. Hawking, S.W. Particle creation by black holes. Commun. Math. Phys. **1975**, 43, 199–220. [CrossRef].
4. Parikh, M.K.;Wilczek, F. Hawking radiation as tunneling. Phys. Rev. Lett. **2000**, 85, 5042–5045. [CrossRef] [PubMed]
5. Banerjee, R.; Majhi, B.R. Hawking black body spectrum from tunneling mechanism. Phys. Lett. B **2009**, 675, 243–245. [CrossRef]
6. Einstein, A. Zum kosmologischen Problem der allgemeinen Relativit¨atstheorie. Sitzungsber. Preuss. Akad. Wiss. **1931**, 142, 235–237.
7. Hamber, H.W. Gravitational scaling dimensions. Phys. Rev. D **2000**, 61, 124008. [CrossRef] Symmetry **2019**, 11, 322 26 of 26
8. Hamber, H.W.; Williams, R.M. Higher derivative quantum gravity on a simplicial lattice. Nucl. Phys. B **1984**, 248, 392.
9. Hamber, H.W. Vacuum Condensate Picture of Quantum Gravity. Symmetry **2019**, 11, 87.
10. Rugh, S.E.; Zinkernagel, H. The quantum vacuum and the Cosmological Constant Problem. arXiv **2000**, arXiv:0012253v1.
11. Carroll, S.M.; Press, W.H.; Turner, E.L. The Cosmological Constant. Ann. Rev. Astron. Astrophys. **1992**, 30, 499–542.
12. Landau, L.D.; Lifšits, E.M. Course of Theoretical Physics Italian Edition. Mir Mosca Editori Riuniti **1976**, 351–66.

13. Pakravan, J.; Takook, M.V. Thermodynamics of nonlinearly charged black holes in the Brans-Dicke modified gravity. J. Theor. Appl. Phys. **2018**, 12, 147–157.
14. Ma, C.-P.; Bertschinger, E. Cosmological Perturbation Theory in the Synchronous and Conformal Newtonian Gauges. Astrophys. J. **1995**, 455, 7.
15. Bernardeau, F.; Colombi, S.; Gaztanaga, E.; Scoccimarro, R. Large-Scale Structure of the Universe and Cosmological Perturbation Theory. Phys. Rept. **2002**, 367, 1.
16. Klebanov, I.R.; Susskind, L.; Banks, T. Wormholes and the cosmological constant. Nucl. Phys. B **1989**, 317, 665–692. -
17. Herrera, R.; Contreras, C.; del Campo, S. The Starobinsky inflationary model in a Jordan–Brans–Dicke-type theory. Class. Quantum. Grav. **1995**, 12, 1937.
18. Hrycyna, O.; Szydlowski, M. Brans–Dicke theory and the emergence of K CDM model. Phys. Rev. D **2013**, 88, 064018.
19. Chiarelli, P., The Gravity of the Classical Klein-Gordon Field, Symmetry, 11, 322 (**2019**); doi 103390/Sym11030322.
20. Chiarelli, P., The Cosmological Constant: the $2^{nd}$ order Quantum-Mechanical Correction to the Newton Gravity, OALIB, **6**: e2466 (**2019**) http:dx.doi.org/10.4236/oalib.1102466.
21. Chiarelli, P., The Lagrangean hydrodynamic representation of Dirac equation, To Phys. J., 01-02, **2018** pp.146-61.
22. Birula, I.B.; Cieplak, M.; Kaminski, J. Theory of Quanta; Oxford University Press: New York, NY, USA, **1992**; pp. 87–115.
23. Chiarelli, P., Quantum decoherence induced by fluctuations, OALIB, **3**, e2466 (**2016**) http:dx.doi.org/10.4236/oalib.1102466.
24. Chiarelli, P. Can fluctuating quantum states acquire the classical behaviour on large scale? J. Adv. Phys. **2013**, 2, 139–163.
25. Chiarelli, P. Stochastically-induced quantum-to-classical transition: The Lindeman relation, maximum density at He lambda point and water-ice transition, in Advances and trends in Physical Science Research Ed 1, (**2018**), ISBN 978-81934224-0-3; Book Publisher International, London, W1b 3HH, UK.
26. Chiarelli, P., The Classical Mechanics from Stochastic Quantum Dynamics, in Advances and trends in Physical Science Research Ed 1, (**2018**), ISBN 978-81934224-0-3; Book Publisher International, London, W1b 3HH, UK.
27. Chiarelli, P., The Gravity of the Classical Klein-Gordon Field, arXiv:1711.06093v9 [physics.gen-ph] (**2019**).
28. Pollock, M., D., On the Dirac Equation in curved Space-Time, Acta Physica Polonica B, **41**-8 (**2010**).
29. Landau, L.D. and Lifsits E.M., Course of Theoretical Physics, vol.2, Italian edition, Mir Mosca, Editori Riuniti, Rome **1976**, p. 152.
30. Landau, L.D. and Lifšits, E.M. Course of Theoretical Physics. Italian Edition, Mir Mosca, Editori Riuniti, Rome **1976**, pp. 386-446.
31. Le Bellac, M. Quantum and Statistical Field Theory. Oxford Science Publications, Oxford, **1986**, p 410
32. Chiarelli, P. The quantum lowest limit to the black hole mass. Phys. Sci. Int. J. **2016**, 9, 1–25. [CrossRef]
33. Cerruti, N.R., Lakshminarayan, A., Lefebvre, T.H. and Tomsovic, S. (**2000**) Exploring Phase Space Localization of Chaotic Eigenstates via Parametric Variation. *Physical Review E*, **63**, 016208. http://dx.doi.org/10.1103/PhysRevE.63.016208
34. Corda C. (**2009**) Interferometric Detection of Gravitational Waves: The DefinitiveTest for General Relativity. International Journal of Modern Physics D, 18, 2275-2282. https://doi.org/10.1142/S0218271809015904

# Appendix A

The EITD of the boson field, given the KGE

$$Œ_{;\sim}^{;\sim} = \left( g^{\sim€} \partial_{€} Œ \right)_{;\sim} = \frac{1}{\sqrt{-g}} \partial_{\sim} \sqrt{-g} \left( g^{\sim€} \partial_{€} Œ \right) = 0 \qquad \text{(A.1)}$$

reads [19]

$$T_{\sim€} = |Œ|^2 \, \mathsf{T}_{\sim€} = |Œ|^2 \, c^2 \left( \frac{\partial S}{\partial t} \right)^{-1} \left( \frac{\partial S}{\partial q^{\sim}} \frac{\partial S}{\partial q^{€}} - g_{rs} \frac{\partial S}{\partial q_{s}} \frac{\partial S}{\partial q^{r}} g_{\sim€} \right). \qquad \text{(A.2)}$$

Given the wave function of a photon

$$Œ \propto \frac{A}{c} \sqrt{\frac{Š}{\hbar}} \qquad \left[ l^{-3/2} \right] \qquad \text{(A.3)}$$

where $|Œ|^2$ represents the number of photons per volume and where $A$ is the vector potential plane wave of the photon in the Minkowskian case

$$A = A_0 \, exp[ -ik_{\sim} q^{\sim} ] = A_0 \, exp[ i \frac{S}{\hbar} ] \, , \qquad \text{(A.4)}$$

from (A.2(it follows that

$$T_{\sim€} \propto |A|^2 \, k_{\sim} k_{€} \qquad \text{(A.5)}$$

that compared with the output of the EM theory for the photon [30]

$$T_{€\sim \, em} = \frac{|E|^2}{4f} \frac{c^2}{Š^2} k_{\sim} k_{€} \propto \frac{|A|^2}{4f} k_{\sim} k_{€} \qquad \text{(A.5)}$$

leads to

$$T_{\sim€} \propto 4f \, T_{€\sim \, em} \, . \qquad \text{(A.6)}$$

Moreover, since for the photon the hydrodynamic Lagrangean reads

$$L = L_{class} = -p_{(k)\sim} \dot{q}^{\sim}_{(k)} = -\hbar g_{\sim r} k^{r} \dot{q}^{\sim} = -\frac{\hbar Š}{|k|^2} g_{\sim r} k^{r} k^{\sim} = 0 \, , \qquad \text{(A.7)}$$

it follows that

$$‡_{\sim€ \, class} = T_{(\tilde{k})\sim€ \, class} \qquad \text{(A.4)}$$

$$]_{\sim€ \, Q} = -\frac{|Œ_{k}|^2}{2} \left( \frac{‡_{(k)|| \, mix}}{4} - \frac{‡_{(k)|| \, mix}}{4} \right) g_{\sim€} \, u_{hk} = 0 \qquad \text{(A.8)}$$

$$\text{U‡}_{\text{~€}\,stress} = -\frac{/\text{Œ}_k/^2}{2}\left(\text{‡}_{s(k)\text{~€}\,mix} - \text{‡}_{s(k)\text{~€}\,mix}\right)\text{u}_{hk} = 0 \qquad \text{(A.9)}$$

From which we can see that the quantum CGT, as well as the CPTD, for the photon is null.

# Appendix B

By using the identity (2.0.9), the equation (2.6.10) reads

$$\left(\begin{array}{l} \partial_{\text{~}} g^{\text{~€}}\partial_{\text{€}} + \dfrac{ie}{\hbar}\left(\begin{array}{l}\left(-\dfrac{i}{4}\text{†}^{ab}\text{Š}_{ab\text{~}} + \dfrac{ie}{\hbar}A_{\text{~}}\right) g^{\text{~€}}\partial_{\text{€}} \\ +\partial_{\text{~}} g^{\text{~€}}\left(-\dfrac{i}{4}\text{†}^{ab}\text{Š}_{ab\text{€}} + \dfrac{ie}{\hbar}A_{\text{€}}\right)\end{array}\right) \\ +\left(-\dfrac{i}{4}\text{†}^{ab}\text{Š}_{ab\text{~}} + \dfrac{ie}{\hbar}A_{\text{~}}\right) g^{\text{~€}}\left(-\dfrac{i}{4}\text{†}^{ab}\text{Š}_{ab\text{€}} + \dfrac{ie}{\hbar}A_{\text{€}}\right)\end{array}\right)\left(\begin{array}{l} /\text{Œ}_{\pm 1}/\,exp\left[i\dfrac{S_{\pm 1}}{\hbar}\right] \\ /\text{Œ}_{\pm 2}/\,exp\left[i\dfrac{S_{\pm 2}}{\hbar}\right]\end{array}\right)$$

$$= -\left(\frac{m^2c^2}{\hbar^2} + g^{\text{~s}} g^{\text{€r}}\,\text{r}_{\pm\text{sr}}\left(\frac{ie}{2\hbar}F_{\text{~€}} - \frac{i}{4}\text{†}^{ab}\left(\text{Š}_{ab\text{~}}\partial_{\text{€}} + \partial_{\text{~}}\text{Š}_{ab\text{€}}\right)\right)\right)\left(\begin{array}{l} /\text{Œ}_{\pm 1}/\,exp\left[i\dfrac{S_{\pm 1}}{\hbar}\right] \\ /\text{Œ}_{\pm 2}/\,exp\left[i\dfrac{S_{\pm 2}}{\hbar}\right]\end{array}\right) \qquad \text{(B.1)}$$

that, by developing the derivatives, leads to

$$
\begin{pmatrix}
\begin{pmatrix}
exp\left[i\frac{S_{\pm1}}{\hbar}\right]\partial_{\sim}\left(g^{\sim\text{€}}\partial_{\text{€}}\,/\text{Œ}_{\pm1}/+\frac{i}{\hbar}g^{\sim\text{€}}\partial_{\text{€}}S_{\pm1}\right)\\
exp\left[i\frac{S_{\pm2}}{\hbar}\right]\partial_{\sim}\left(g^{\sim\text{€}}\partial_{\text{€}}\,/\text{Œ}_{\pm2}/+\frac{i}{\hbar}g^{\sim\text{€}}\partial_{\text{€}}S_{\pm2}\right)
\end{pmatrix}\\
+\begin{pmatrix}
\frac{i}{\hbar}exp\left[i\frac{S_{\pm1}}{\hbar}\right]\partial_{\sim}S_{\pm1}\left(g^{\sim\text{€}}\partial_{\text{€}}\,/\text{Œ}_{\pm1}/+\frac{i}{\hbar}/\text{Œ}_{\pm1}/\,g^{\sim\text{€}}\partial_{\text{€}}S_{\pm1}\right)\\
\frac{i}{\hbar}exp\left[i\frac{S_{\pm2}}{\hbar}\right]\partial_{\sim}S_{\pm2}\left(g^{\sim\text{€}}\partial_{\text{€}}\,/\text{Œ}_{\pm2}/+/\text{Œ}_{\pm2}/\,\frac{i}{\hbar}g^{\sim\text{€}}\partial_{\text{€}}S_{\pm2}\right)
\end{pmatrix}\\
+\begin{pmatrix}
\left(-\frac{i}{4}\dagger^{ab}\check{S}_{ab\sim}+\frac{ie}{\hbar}A_{\sim}\right)g^{\sim\text{€}}
\begin{pmatrix}
exp\left[i\frac{S_{\pm1}}{\hbar}\right]\left(\partial_{\text{€}}\,/\text{Œ}_{\pm1}/+\frac{i}{\hbar}/\text{Œ}_{\pm1}/\,\partial_{\text{€}}S_{\pm1}\right)\\
exp\left[i\frac{S_{\pm2}}{\hbar}\right]\left(\partial_{\text{€}}\,/\text{Œ}_{\pm2}/+\frac{i}{\hbar}/\text{Œ}_{\pm2}/\,\partial_{\text{€}}S_{\pm2}\right)
\end{pmatrix}\\
+g^{\sim\text{€}}\left(-\frac{i}{4}\dagger^{ab}\check{S}_{ab\text{€}}+\frac{ie}{\hbar}A_{\text{€}}\right)
\begin{pmatrix}
exp\left[i\frac{S_{\pm1}}{\hbar}\right]\left(\partial_{\sim}\,/\text{Œ}_{\pm1}/+\frac{i}{\hbar}/\text{Œ}_{\pm1}/\,\partial_{\sim}S_{\pm1}\right)\\
exp\left[i\frac{S_{\pm2}}{\hbar}\right]\left(\partial_{\sim}\,/\text{Œ}_{\pm2}/+\frac{i}{\hbar}/\text{Œ}_{\pm2}/\,\partial_{\sim}S_{\pm2}\right)
\end{pmatrix}\\
+\begin{pmatrix}
/\text{Œ}_{\pm1}/\,exp\left[i\frac{S_{\pm1}}{\hbar}\right]\\
/\text{Œ}_{\pm2}/\,exp\left[i\frac{S_{\pm2}}{\hbar}\right]
\end{pmatrix}\partial_{\sim}g^{\sim\text{€}}\left(-\frac{i}{4}\dagger^{ab}\check{S}_{ab\text{€}}+\frac{ie}{\hbar}A_{\text{€}}\right)
\end{pmatrix}\\
-g^{\sim\text{€}}\left(-\frac{1}{4}\dagger^{ab}\check{S}_{ab\sim}+\frac{e}{\hbar}A_{\sim}\right)\left(-\frac{1}{4}\dagger^{ab}\check{S}_{ab\text{€}}+\frac{e}{\hbar}A_{\text{€}}\right)
\begin{pmatrix}
/\text{Œ}_{\pm1}/\,exp\left[i\frac{S_{\pm1}}{\hbar}\right]\\
/\text{Œ}_{\pm2}/\,exp\left[i\frac{S_{\pm2}}{\hbar}\right]
\end{pmatrix}
\end{pmatrix}
$$

$$
=-\left(\frac{m^2c^2}{\hbar^2}+g^{\sim\text{s}}g^{\text{€r}}\text{r}_{\pm\text{sr}}\left(\frac{ie}{2\hbar}F_{\sim\text{€}}-\frac{i}{4}\dagger^{ab}\left(\check{S}_{ab\sim}\partial_{\text{€}}+\partial_{\sim}\check{S}_{ab\text{€}}\right)\right)\right)
\begin{pmatrix}
/\text{Œ}_{\pm1}/\,exp\left[i\frac{S_{\pm1}}{\hbar}\right]\\
/\text{Œ}_{\pm2}/\,exp\left[i\frac{S_{\pm2}}{\hbar}\right]
\end{pmatrix}
\qquad (B.2)
$$

and, by dividing each component by $exp\left[i\frac{S_{\pm i}}{\hbar}\right]$, where i =1,2 , to

$$
\begin{pmatrix}
\begin{pmatrix}
\partial_{\text{~}}\left(g^{\text{~€}}\partial_{\text{€}}\,\text{/Œ}_{\text{±}1}\text{/}+\frac{i}{\hbar}g^{\text{~€}}\partial_{\text{€}}S_{\text{±}1}\right)\\
\partial_{\text{~}}\left(g^{\text{~€}}\partial_{\text{€}}\,\text{/Œ}_{\text{±}2}\text{/}+\frac{i}{\hbar}g^{\text{~€}}\partial_{\text{€}}S_{\text{±}2}\right)
\end{pmatrix}
+
\begin{pmatrix}
\frac{i}{\hbar}\partial_{\text{~}}S_{\text{±}1}\left(g^{\text{~€}}\partial_{\text{€}}\,\text{/Œ}_{\text{±}1}\text{/}+\frac{i}{\hbar}\text{/Œ}_{\text{±}1}\text{/}\,g^{\text{~€}}\partial_{\text{€}}S_{\text{±}1}\right)\\
\frac{i}{\hbar}\partial_{\text{~}}S_{\text{±}2}\left(g^{\text{~€}}\partial_{\text{€}}\,\text{/Œ}_{\text{±}2}\text{/}+\frac{i}{\hbar}\text{/Œ}_{\text{±}2}\text{/}\,g^{\text{~€}}\partial_{\text{€}}S_{\text{±}2}\right)
\end{pmatrix}
\\
+\begin{pmatrix}
\left(-\frac{i}{4}\text{†}^{ab}\text{Š}_{ab\text{~}}+\frac{ie}{\hbar}A_{\text{~}}\right)g^{\text{~€}}
\begin{pmatrix}
\left(\partial_{\text{€}}\,\text{/Œ}_{\text{±}1}\text{/}+\frac{i}{\hbar}\text{/Œ}_{\text{±}1}\text{/}\,\partial_{\text{€}}S_{\text{±}1}\right)\\
\left(\partial_{\text{€}}\,\text{/Œ}_{\text{±}2}\text{/}+\frac{i}{\hbar}\text{/Œ}_{\text{±}2}\text{/}\,\partial_{\text{€}}S_{\text{±}2}\right)
\end{pmatrix}\\
+g^{\text{~€}}\left(-\frac{i}{4}\text{†}^{ab}\text{Š}_{ab\text{€}}+\frac{ie}{\hbar}A_{\text{€}}\right)
\begin{pmatrix}
\left(\partial_{\text{~}}\,\text{/Œ}_{\text{±}1}\text{/}+\frac{i}{\hbar}\text{/Œ}_{\text{±}1}\text{/}\,\partial_{\text{~}}S_{\text{±}1}\right)\\
\left(\partial_{\text{~}}\,\text{/Œ}_{\text{±}2}\text{/}+\frac{i}{\hbar}\text{/Œ}_{\text{±}2}\text{/}\,\partial_{\text{~}}S_{\text{±}2}\right)
\end{pmatrix}\\
+\begin{pmatrix}\text{/Œ}_{\text{±}1}\text{/}\\ \text{/Œ}_{\text{±}2}\text{/}\end{pmatrix}\partial_{\text{~}}g^{\text{~€}}\left(-\frac{i}{4}\text{†}^{ab}\text{Š}_{ab\text{€}}+\frac{ie}{\hbar}A_{\text{€}}\right)
\end{pmatrix}\\
-g^{\text{~€}}\left(-\frac{1}{4}\text{†}^{ab}\text{Š}_{ab\text{~}}+\frac{e}{\hbar}A_{\text{~}}\right)\left(-\frac{1}{4}\text{†}^{ab}\text{Š}_{ab\text{€}}+\frac{e}{\hbar}A_{\text{€}}\right)\begin{pmatrix}\text{/Œ}_{\text{±}1}\text{/}\\ \text{/Œ}_{\text{±}2}\text{/}\end{pmatrix}
\end{pmatrix}.
\tag{B.3}
$$

$$
=-\left(\frac{m^2c^2}{\hbar^2}+g^{\text{~s}}g^{\text{€r}}\,\text{r}_{\text{±sr}}\left(\frac{ie}{2\hbar}F_{\text{~€}}-\frac{i}{4}\text{†}^{ab}\left(\text{Š}_{ab\text{~}}\partial_{\text{€}}+\partial_{\text{~}}\text{Š}_{ab\text{€}}\right)\right)\right)\begin{pmatrix}\text{/Œ}_{\text{±}1}\text{/}\\ \text{/Œ}_{\text{±}2}\text{/}\end{pmatrix}
$$

Moreover, by grouping the terms and equating the real part of the equation, it follows that

$$
\begin{pmatrix}
\begin{pmatrix}\partial_{\text{~}}\left(g^{\text{~€}}\partial_{\text{€}}\,/\text{Œ}_{\pm1}/\right)\\ \partial_{\text{~}}\left(g^{\text{~€}}\partial_{\text{€}}\,/\text{Œ}_{\pm2}/\right)\end{pmatrix}
+\begin{pmatrix}\frac{1}{\hbar^2}/\text{Œ}_{\pm1}/\partial_{\text{~}}S_{\pm1}\left(g^{\text{~€}}\partial_{\text{€}}S_{\pm1}\right)\\ \frac{1}{\hbar^2}/\text{Œ}_{\pm2}/\partial_{\text{~}}S_{\pm2}\left(g^{\text{~€}}\partial_{\text{€}}S_{\pm2}\right)\end{pmatrix}\\
+\begin{pmatrix}
\frac{1}{\hbar}\left(Re\left\{\frac{1}{4}\text{†}^{ab}\text{Š}_{ab\text{~}}\right\}-\frac{e}{\hbar}A_{\text{~}}\right)g^{\text{~€}}\begin{pmatrix}\left(/\text{Œ}_{\pm1}/\partial_{\text{€}}S_{\pm1}\right)\\ \left(/\text{Œ}_{\pm2}/\partial_{\text{€}}S_{\pm2}\right)\end{pmatrix}\\
+\frac{1}{\hbar}g^{\text{~€}}\left(Re\left\{\frac{1}{4}\text{†}^{ab}\text{Š}_{ab\text{€}}\right\}-\frac{e}{\hbar}A_{\text{€}}\right)\begin{pmatrix}\left(/\text{Œ}_{\pm1}/\partial_{\text{~}}S_{\pm1}\right)\\ \left(/\text{Œ}_{\pm2}/\partial_{\text{~}}S_{\pm2}\right)\end{pmatrix}\\
+\begin{pmatrix}/\text{Œ}_{\pm1}/\\ /\text{Œ}_{\pm2}/\end{pmatrix}\partial_{\text{~}}g^{\text{~€}}\left(-\frac{Im\left\{\text{†}^{ab}\text{Š}_{ab\text{€}}\right\}}{4}\right)\\
-Im\left\{\frac{1}{4}\text{†}^{ab}\text{Š}_{ab\text{~}}\right\}g^{\text{~€}}\begin{pmatrix}\left(\partial_{\text{€}}\,/\text{Œ}_{\pm1}/\right)\\ \left(\partial_{\text{€}}\,/\text{Œ}_{\pm2}/\right)\end{pmatrix}\\
-g^{\text{~€}}\,Im\left\{\frac{1}{4}\text{†}^{ab}\text{Š}_{ab\text{€}}\right\}\begin{pmatrix}\left(\partial_{\text{~}}\,/\text{Œ}_{\pm1}/\right)\\ \left(\partial_{\text{~}}\,/\text{Œ}_{\pm2}/\right)\end{pmatrix}
\end{pmatrix}\\
-g^{\text{~€}}\left(Re\left\{\frac{1}{4}\text{†}^{ab}\text{Š}_{ab\text{~}}\right\}-\frac{e}{\hbar}A_{\text{~€}}\right)\left(Re\left\{\frac{1}{4}\text{†}^{ab}\text{Š}_{ab\text{€}}\right\}-\frac{e}{\hbar}A\right)\begin{pmatrix}/\text{Œ}_{\pm1}/\\ /\text{Œ}_{\pm2}/\end{pmatrix}\\
-g^{\text{~€}}\,Im\left\{\frac{1}{4}\text{†}^{ab}\text{Š}_{ab\text{~}}\right\}Im\left\{\frac{1}{4}\text{†}^{ab}\text{Š}_{ab\text{€}}\right\}\begin{pmatrix}/\text{Œ}_{\pm1}/\\ /\text{Œ}_{\pm2}/\end{pmatrix}
\end{pmatrix}
$$

$$
=-\left(\frac{m^2c^2}{\hbar^2}+Re\left\{g^{\text{~s}}g^{\text{€r}}\,\text{r}_{\pm\text{sr}}\left(\frac{ie}{2\hbar}F_{\text{~€}}-\frac{i}{4}\text{†}^{ab}\left(\text{Š}_{ab\text{~}}\partial_{\text{€}}+\partial_{\text{~}}\text{Š}_{ab\text{€}}\right)\right)\right\}\right)\begin{pmatrix}/\text{Œ}_{\pm1}/\\ /\text{Œ}_{\pm2}/\end{pmatrix} \tag{B.4}
$$

and, by dividing each component by $/\text{Œ}_{\pm i}/$, where i =1,2 , that

$$
\begin{aligned}
&\left(
\begin{aligned}
&\left(\begin{pmatrix} \dfrac{\partial_{\sim}\left(g^{\sim\text{€}}\partial_{\text{€}}\,|\text{Œ}_{\pm1}|\right)}{|\text{Œ}_{\pm1}|} \\ \dfrac{\partial_{\sim}\left(g^{\sim\text{€}}\partial_{\text{€}}\,|\text{Œ}_{\pm2}|\right)}{|\text{Œ}_{\pm2}|} \end{pmatrix}
+\begin{pmatrix} \dfrac{1}{\hbar^{2}}\partial_{\sim}S_{\pm1}\left(g^{\sim\text{€}}\partial_{\text{€}}S_{\pm1}\right) \\ \dfrac{1}{\hbar^{2}}\partial_{\sim}S_{\pm2}\left(g^{\sim\text{€}}\partial_{\text{€}}S_{\pm2}\right) \end{pmatrix}\right) \\
&+\left(\begin{aligned}
&\frac{1}{\hbar}\left(Re\left\{\frac{1}{4}\dagger^{ab}\check{S}_{ab\sim}\right\}-\frac{e}{\hbar}A_{\sim}\right)g^{\sim\text{€}}\begin{pmatrix}\left(\partial_{\text{€}}S_{\pm1}\right)\\ \left(\partial_{\text{€}}S_{\pm2}\right)\end{pmatrix} \\
&+\frac{1}{\hbar}g^{\sim\text{€}}\left(Re\left\{\frac{1}{4}\dagger^{ab}\check{S}_{ab\text{€}}\right\}-\frac{e}{\hbar}A_{\text{€}}\right)\begin{pmatrix}\left(\partial_{\sim}S_{\pm1}\right)\\ \left(\partial_{\sim}S_{\pm2}\right)\end{pmatrix} \\
&+\partial_{\sim}g^{\sim\text{€}}\left(-\frac{Im\left\{\dagger^{ab}\check{S}_{ab\text{€}}\right\}}{4}\right) \\
&-\frac{1}{2}Im\left\{\dagger^{ab}\check{S}_{ab\sim}\right\}g^{\sim\text{€}}\begin{pmatrix}\dfrac{\partial_{\text{€}}\,|\text{Œ}_{\pm1}|}{|\text{Œ}_{\pm1}|}\\ \dfrac{\partial_{\text{€}}\,|\text{Œ}_{\pm2}|}{|\text{Œ}_{\pm2}|}\end{pmatrix}
\end{aligned}\right) \\
&-g^{\sim\text{€}}\left(Re\left\{\frac{1}{4}\dagger^{ab}\check{S}_{ab\sim}\right\}-\frac{e}{\hbar}A_{\sim}\right)\left(Re\left\{\frac{1}{4}\dagger^{ab}\check{S}_{ab\text{€}}\right\}-\frac{e}{\hbar}A_{\text{€}}\right) \\
&-g^{\sim\text{€}}\,Im\left\{\frac{1}{4}\dagger^{ab}\check{S}_{ab\sim}\right\}Im\left\{\frac{1}{4}\dagger^{ab}\check{S}_{ab\text{€}}\right\}
\end{aligned}
\right) \\
&=-\left(\frac{m^{2}c^{2}}{\hbar^{2}}+Re\left\{g^{\sim\text{s}}g^{\text{€r}}\left(
\begin{aligned}
&\frac{ie}{2\hbar}\begin{pmatrix}\text{r}_{\text{sr}1j}\dfrac{|\text{Œ}_{\pm j}|}{|\text{Œ}_{\pm1}|}\\ \text{r}_{\text{sr}2j}\dfrac{|\text{Œ}_{\pm j}|}{|\text{Œ}_{\pm2}|}\end{pmatrix}F_{\sim\text{€}} \\
&-\frac{i}{4}\text{r}_{\text{sr}}\dagger^{ab}\left(\check{S}_{ab\sim}\begin{pmatrix}\dfrac{\partial_{\text{€}}\,|\text{Œ}_{\pm1}|}{|\text{Œ}_{\pm1}|}\\ \dfrac{\partial_{\text{€}}\,|\text{Œ}_{\pm2}|}{|\text{Œ}_{\pm2}|}\end{pmatrix}+\begin{pmatrix}\dfrac{\partial_{\sim}\check{S}_{ab\text{€}}\,|\text{Œ}_{\pm1}|}{|\text{Œ}_{\pm1}|}\\ \dfrac{\partial_{\sim}\check{S}_{ab\text{€}}\,|\text{Œ}_{\pm2}|}{|\text{Œ}_{\pm2}|}\end{pmatrix}\right)
\end{aligned}\right)\right\}\right)
\end{aligned}
$$

$$
\left(\left(\begin{pmatrix} \partial_{\sim} S_{\pm 1}\left(g^{\sim\text{€}}\partial_{\text{€}} S_{\pm 1}\right) \\ \partial_{\sim} S_{\pm 2}\left(g^{\sim\text{€}}\partial_{\text{€}} S_{\pm 2}\right) \end{pmatrix}
+2\hbar\left(Re\left\{\frac{1}{4}\dagger^{ab}\check{S}_{ab\sim}\right\}-\frac{e}{\hbar}A_{\sim}\right)g^{\sim\text{€}}\begin{pmatrix}\partial_{\text{€}} S_{\pm 1} \\ \partial_{\text{€}} S_{\pm 2}\end{pmatrix}
-\hbar^{2}g^{\sim\text{€}}\left(Re\left\{\frac{1}{4}\dagger^{ab}\check{S}_{ab\sim}\right\}-\frac{e}{\hbar}A_{\sim}\right)\left(Re\left\{\frac{1}{4}\dagger^{ab}\check{S}_{ab\text{€}}\right\}-\frac{e}{\hbar}A_{\text{€}}\right)\right)
-\hbar^{2}\partial_{\sim}g^{\sim\text{€}}\left(\frac{Im\left\{\dagger^{ab}\check{S}_{ab\text{€}}\right\}}{4}\right)
-\hbar^{2}g^{\sim\text{€}}\,Im\left\{\frac{1}{4}\dagger^{ab}\check{S}_{ab\sim}\right\}Im\left\{\frac{1}{4}\dagger^{ab}\check{S}_{ab\text{€}}\right\}\right)
$$

$$
=-\left(m^{2}c^{2}+Re\left\{g^{\sim\mathsf{s}}g^{\text{€}\mathsf{r}}\left(\frac{ie}{2\hbar}\begin{pmatrix}\mathsf{r}_{\pm\mathsf{sr}1j}\dfrac{/\text{Œ}_{\pm j}/}{/\text{Œ}_{\pm 1}/} \\ \mathsf{rr}_{\pm\mathsf{sr}2j}\dfrac{/\text{Œ}_{\pm j}/}{/\text{Œ}_{\pm 2}/}\end{pmatrix}F_{\sim\text{€}}
-\frac{i}{4}\mathsf{r}_{\pm\mathsf{sr}}\dagger^{ab}\left(\check{S}_{ab\sim}\begin{pmatrix}\dfrac{\partial_{\text{€}}\,/\text{Œ}_{\pm 1}/}{/\text{Œ}_{\pm 1}/} \\ \dfrac{\partial_{\text{€}}\,/\text{Œ}_{\pm 2}/}{/\text{Œ}_{\pm 2}/}\end{pmatrix}+\begin{pmatrix}\dfrac{\partial_{\sim}\check{S}_{ab\text{€}}\,/\text{Œ}_{\pm 1}/}{/\text{Œ}_{\pm 1}/} \\ \dfrac{\partial_{\sim}\check{S}_{ab\text{€}}\,/\text{Œ}_{\pm 2}/}{/\text{Œ}_{\pm 2}/}\end{pmatrix}\right)\right)\right\}
-\hbar^{2}\begin{pmatrix}\dfrac{\partial_{\sim}\left(g^{\sim\text{€}}\partial_{\text{€}}\,/\text{Œ}_{\pm 1}/\right)}{/\text{Œ}_{\pm 1}/} \\ \dfrac{\partial_{\sim}\left(g^{\sim\text{€}}\partial_{\text{€}}\,/\text{Œ}_{\pm 2}/\right)}{/\text{Œ}_{\pm 2}/}\end{pmatrix}
+\frac{\hbar^{2}}{2}Im\left\{\dagger^{ab}\check{S}_{ab\sim}\right\}g^{\sim\text{€}}\begin{pmatrix}\dfrac{\partial_{\text{€}}\,/\text{Œ}_{\pm 1}/}{/\text{Œ}_{\pm 1}/} \\ \dfrac{\partial_{\text{€}}\,/\text{Œ}_{\pm 2}/}{/\text{Œ}_{\pm 2}/}\end{pmatrix}\right)
$$

$$\begin{pmatrix} \partial_{\sim}\left(S_{\pm1}+\left(Re\left\{\frac{\hbar}{4}\dagger^{ab}\text{Š}_{ab\sim}\right\}-eA_{\sim}\right)\right)\left(g^{\sim\text{€}}\partial_{\text{€}}\left(S_{\pm1}+\left(Re\left\{\frac{\hbar}{4}\dagger^{ab}\text{Š}_{ab\sim}\right\}-eA_{\sim}\right)\right)\right) \\ \partial_{\sim}\left(S_{\pm2}+\left(Re\left\{\frac{\hbar}{4}\dagger^{ab}\text{Š}_{ab\sim}\right\}-eA_{\sim}\right)\right)\left(g^{\sim\text{€}}\partial_{\text{€}}\left(S_{\pm2}+\left(Re\left\{\frac{\hbar}{4}\dagger^{ab}\text{Š}_{ab\sim}\right\}-eA_{\sim}\right)\right)\right) \end{pmatrix}$$

$$= -\begin{pmatrix} m^2c^2 + Re\left\{ g^{\sim\text{s}} g^{\text{€r}} \begin{pmatrix} \dfrac{ie}{2\hbar}\begin{pmatrix} \text{r}_{\pm\text{sr}1j}\dfrac{/\text{Œ}_{\pm j}/}{/\text{Œ}_{\pm1}/} \\ \text{r}_{\pm\text{sr}2j}\dfrac{/\text{Œ}_{\pm j}/}{/\text{Œ}_{\pm2}/} \end{pmatrix} F_{\sim\text{€}} \\ -\dfrac{i}{4}\text{r}_{\pm\text{sr}}\dagger^{ab}\left(\text{Š}_{ab\sim}\begin{pmatrix} \dfrac{\partial_{\text{€}}\,/\text{Œ}_{\pm1}/}{/\text{Œ}_{\pm1}/} \\ \dfrac{\partial_{\text{€}}\,/\text{Œ}_{\pm2}/}{/\text{Œ}_{\pm2}/} \end{pmatrix} + \begin{pmatrix} \dfrac{\partial_{\sim}\text{Š}_{ab\text{€}}\,/\text{Œ}_{\pm1}/}{/\text{Œ}_{\pm1}/} \\ \dfrac{\partial_{\sim}\text{Š}_{ab\text{€}}\,/\text{Œ}_{\pm2}/}{/\text{Œ}_{\pm2}/} \end{pmatrix}\right) \end{pmatrix} \right\} \\ -\hbar^2\begin{pmatrix} \dfrac{\partial_{\sim}\left(g^{\sim\text{€}}\partial_{\text{€}}\,/\text{Œ}_{\pm1}/\right)}{/\text{Œ}_{\pm1}/} \\ \dfrac{\partial_{\sim}\left(g^{\sim\text{€}}\partial_{\text{€}}\,/\text{Œ}_{\pm2}/\right)}{/\text{Œ}_{\pm2}/} \end{pmatrix} + \dfrac{\hbar^2}{2}Im\left\{\dagger^{ab}\text{Š}_{ab\sim}\right\}g^{\sim\text{€}}\begin{pmatrix} \dfrac{\partial_{\text{€}}\,/\text{Œ}_{\pm1}/}{/\text{Œ}_{\pm1}/} \\ \dfrac{\partial_{\text{€}}\,/\text{Œ}_{\pm2}/}{/\text{Œ}_{\pm2}/} \end{pmatrix} \\ +\hbar^2\partial_{\sim}g^{\sim\text{€}}\left(\dfrac{Im\left\{\dagger^{ab}\text{Š}_{ab\text{€}}\right\}}{4}\right) \\ +\hbar^2 g^{\sim\text{€}}\,Im\left\{\dfrac{1}{4}\dagger^{ab}\text{Š}_{ab\sim}\right\}Im\left\{\dfrac{1}{4}\dagger^{ab}\text{Š}_{ab\text{€}}\right\} \end{pmatrix}, \quad \text{(B.5)}$$

leading to the final expression

$$g^{\sim\text{€}}\left(\partial_{\sim}\begin{pmatrix} S_{\pm1} \\ S_{\pm2} \end{pmatrix}+\left(Re\left\{\frac{\hbar}{4}\dagger^{ab}\text{Š}_{ab\sim}\right\}-eA_{\sim}\right)\right)\left(\partial_{\text{€}}\begin{pmatrix} S_{\pm1} \\ S_{\pm2} \end{pmatrix}+\left(Re\left\{\frac{\hbar}{4}\dagger^{ab}\text{Š}_{ab\text{€}}\right\}-eA_{\text{€}}\right)\right) \quad \text{(B.6)}$$

$$= -\left(m^2c^2 + mV_{qu}\right)$$

that, more synthetically can read

$$g^{\sim\text{€}}\left(\partial_{\sim}S_{\pm}+\left(Re\left\{\frac{\hbar}{4}\dagger^{ab}\text{Š}_{ab\sim}\right\}-eA_{\sim}\right)\right)\left(\partial_{\text{€}}S_{\pm}+\left(Re\left\{\frac{\hbar}{4}\dagger^{ab}\text{Š}_{ab\text{€}}\right\}-eA_{\text{€}}\right)\right) \quad \text{(B.7)}$$

$$= -\left(m^2c^2 + mV_{qu}\right)$$

where

$$V_{qu} = -\frac{\hbar^2}{m}\left(\begin{matrix}\left(\begin{matrix}\dfrac{\partial_{\text{~}}\left(g^{\text{~€}}\partial_{\text{€}}\,/\text{Œ}_{\pm 1}/\right)}{/\text{Œ}_{\pm 1}/}\\ \dfrac{\partial_{\text{~}}\left(g^{\text{~€}}\partial_{\text{€}}\,/\text{Œ}_{\pm 2}/\right)}{/\text{Œ}_{\pm 2}/}\end{matrix}\right)+\dfrac{\hbar^2}{2}Im\left\{\text{†}^{ab}\text{Š}_{ab\text{~}}\right\}g^{\text{~€}}\left(\begin{matrix}\dfrac{\partial_{\text{€}}\,/\text{Œ}_{\pm 1}/}{/\text{Œ}_{\pm 1}/}\\ \dfrac{\partial_{\text{€}}\,/\text{Œ}_{\pm 2}/}{/\text{Œ}_{\pm 2}/}\end{matrix}\right)\\ +\partial_{\text{~}}g^{\text{~€}}\left(\dfrac{Im\left\{\text{†}^{ab}\text{Š}_{ab\text{€}}\right\}}{4}\right)+g^{\text{~€}}\,Im\left\{\dfrac{1}{4}\text{†}^{ab}\text{Š}_{ab\text{~}}\right\}Im\left\{\dfrac{1}{4}\text{†}^{ab}\text{Š}_{ab\text{€}}\right\}\end{matrix}\right.$$

$$\left.+Re\left(g^{\text{~s}}g^{\text{€r}}\left(\begin{matrix}\dfrac{ie}{2\hbar}\left(\begin{matrix}\text{r}_{\pm\text{sr}1j}\dfrac{/\text{Œ}_{\pm j}/}{/\text{Œ}_{\pm 1}/}\\ \text{r}_{\pm\text{sr}2j}\dfrac{/\text{Œ}_{\pm j}/}{/\text{Œ}_{\pm 2}/}\end{matrix}\right)F_{\text{~€}}\\ -\dfrac{i}{4}\text{r}_{\pm\text{sr}}\text{†}^{ab}\left(\text{Š}_{ab\text{~}}\left(\begin{matrix}\dfrac{\partial_{\text{€}}\,/\text{Œ}_{\pm 1}/}{/\text{Œ}_{\pm 1}/}\\ \dfrac{\partial_{\text{€}}\,/\text{Œ}_{\pm 2}/}{/\text{Œ}_{\pm 2}/}\end{matrix}\right)+\left(\begin{matrix}\dfrac{\partial_{\text{~}}\text{Š}_{ab\text{€}}\,/\text{Œ}_{\pm 1}/}{/\text{Œ}_{\pm 1}/}\\ \dfrac{\partial_{\text{~}}\text{Š}_{ab\text{€}}\,/\text{Œ}_{\pm 2}/}{/\text{Œ}_{\pm 2}/}\end{matrix}\right)\right)\end{matrix}\right)\right)\right) \tag{B.8}$$

Moreover, by grouping the imaginary terms of eq. (B.3) as follows

$$
\begin{pmatrix}
\begin{pmatrix} \partial_{\sim}\left(\frac{i}{\hbar} g^{\sim\text{€}} \partial_{\text{€}} S_{\pm1}\right) \\ \partial_{\sim}\left(\frac{i}{\hbar} g^{\sim\text{€}} \partial_{\text{€}} S_{\pm2}\right) \end{pmatrix} + \begin{pmatrix} \frac{i}{\hbar}\partial_{\sim} S_{\pm1}\left(g^{\sim\text{€}}\partial_{\text{€}}\, \text{/Œ}_{\pm1}\text{/}\right) \\ \frac{i}{\hbar}\partial_{\sim} S_{\pm2}\left(g^{\sim\text{€}}\partial_{\text{€}}\, \text{/Œ}_{\pm2}\text{/}\right) \end{pmatrix} \\
+\begin{pmatrix}
2\left(-\frac{i}{4} Re\left\{\dagger^{\,ab}\right\}\text{Š}_{ab\sim} + \frac{ie}{\hbar} A_{\sim}\right) g^{\sim\text{€}} \begin{pmatrix} \left(\partial_{\text{€}}\, \text{/Œ}_{\pm1}\text{/}\right) \\ \left(\partial_{\text{€}}\, \text{/Œ}_{\pm2}\text{/}\right) \end{pmatrix} \\
+2 g^{\sim\text{€}} \left(\frac{1}{4} Im\left\{\dagger^{\,ab}\right\}\text{Š}_{ab\text{€}}\right) \begin{pmatrix} \left(\frac{i}{\hbar}\text{/Œ}_{\pm1}\text{/}\partial_{\sim} S_{\pm1}\right) \\ \left(\frac{i}{\hbar}\text{/Œ}_{\pm2}\text{/}\partial_{\sim} S_{\pm2}\right) \end{pmatrix} \\
+\begin{pmatrix} \text{/Œ}_{\pm1}\text{/} \\ \text{/Œ}_{\pm2}\text{/} \end{pmatrix} \partial_{\sim} g^{\sim\text{€}} \left(-\frac{i}{4}\dagger^{\,ab}\text{Š}_{ab\text{€}} + \frac{ie}{\hbar} A_{\text{€}}\right)
\end{pmatrix} \\
-g^{\sim\text{€}}\left(-\frac{1}{4} Re\left\{\dagger^{\,ab}\right\}\text{Š}_{ab\sim} + \frac{e}{\hbar} A_{\sim}\right)\left(-\frac{1}{4} Im\left\{\dagger^{\,ab}\right\}\text{Š}_{ab\text{€}}\right)\begin{pmatrix} \text{/Œ}_{\pm1}\text{/} \\ \text{/Œ}_{\pm2}\text{/} \end{pmatrix} \\
-g^{\sim\text{€}}\left(-\frac{1}{4} Im\left\{\dagger^{\,ab}\right\}\text{Š}_{ab\sim}\right)\left(-\frac{1}{4} Re\left\{\dagger^{\,ab}\right\}\text{Š}_{ab\text{€}} + \frac{e}{\hbar} A_{\text{€}}\right)\begin{pmatrix} \text{/Œ}_{\pm1}\text{/} \\ \text{/Œ}_{\pm2}\text{/} \end{pmatrix}
\end{pmatrix}
$$

$$
= -\left(\frac{ie}{2\hbar} Re\left\{ g^{\sim\text{s}} g^{\text{€r}}\, \text{r}_{\pm\text{sr}} \left( F_{\sim\text{€}} - \frac{\hbar}{2e}\dagger^{\,ab}\left(\text{Š}_{ab\sim}\begin{pmatrix} \frac{\partial_{\text{€}}\, \text{/Œ}_{\pm1}\text{/}}{\text{/Œ}_{\pm1}\text{/}} \\ \frac{\partial_{\text{€}}\, \text{/Œ}_{\pm2}\text{/}}{\text{/Œ}_{\pm2}\text{/}} \end{pmatrix} + \begin{pmatrix} \frac{\partial_{\sim}\text{Š}_{ab\text{€}}\, \text{/Œ}_{\pm1}\text{/}}{\text{/Œ}_{\pm1}\text{/}} \\ \frac{\partial_{\sim}\text{Š}_{ab\text{€}}\, \text{/Œ}_{\pm2}\text{/}}{\text{/Œ}_{\pm2}\text{/}} \end{pmatrix}\right)\right)\right\}\right)\begin{pmatrix} \text{/Œ}_{\pm1}\text{/} \\ \text{/Œ}_{\pm2}\text{/} \end{pmatrix} \qquad \text{(B.9)}
$$

the current conservation equation can finally be obtained (the derivation is not given here).

## Appendix C

It is useful to note that in the quantum case, due to the force generated by the quantum potential (that is a function of the co-ordinates $q_{\sim}$ ), generally speaking, we have that

$$
\left(\sum_{\pm i} T_{\pm i\sim}{}^{\text{€}}\right)_{,\text{€}} + T_{\sim}{}^{\text{€}}{}_{,\text{€}em} = \sum_{\pm i}\left(\text{/Œ}_{\pm i}\text{/}^{2}\, \text{T}_{\pm i\sim}{}^{\text{€}}\right)_{,\text{€}} + T_{\sim}{}^{\text{€}}{}_{,\text{€}em} \neq 0 \,. \qquad \text{(C.1)}
$$

It easy to check this point by the inspection of the quantum hydrodynamic motion equation (2.2.16) [19, 32] for a scalar uncharged particle obeying to the KGE (2.4.5) that reads

$$
\begin{aligned}
\frac{dp_{\sim}}{ds} &= \frac{d}{ds}\left(mcu_{\sim}\sqrt{1-\frac{V_{qu}}{mc^{2}}}\right) = -\frac{\text{X}}{c}\frac{\partial L}{\partial q^{\sim}} \\
&= mc\frac{\partial}{\partial q^{\sim}}\sqrt{1-\frac{V_{qu}}{mc^{2}}}
\end{aligned} \qquad \text{(C.2)}
$$

where

$$u_{\sim} = \frac{X}{c}\dot{q}_{\sim} ,\qquad \text{(C.3)}$$

that leads to

$$mc\sqrt{1-\frac{V_{qu}}{mc^2}}\,\frac{du_{\sim}}{ds} = -mcu_{\sim}\frac{d}{ds}\left(\sqrt{1-\frac{V_{qu}}{mc^2}}\right)+mc\frac{\partial}{\partial q^{\sim}}\left(\sqrt{1-\frac{V_{qu}}{mc^2}}\right)\qquad \text{(C.4)}$$

where by using the energy-impulse tensor $T_{(k)\sim}{}^{\text{€}}$ expression [19]

$$T_{(k)\sim}{}^{\text{€}} = \left(\dot{q}_{\sim}\frac{\partial L_{(k)n}}{\partial \dot{q}_{\text{€}}} - L_{(k)n}u_{\sim}{}^{\text{€}}\right) = \frac{mc^2}{X}\sqrt{1-\frac{V_{qu(k)}}{mc^2}}\left(u_{\sim}u^{\text{€}} - u_{\sim}{}^{\text{€}}\right) ,\qquad \text{(C.5)}$$

equation (C-4) more synthetically reads

$$mc\sqrt{1-\frac{V_{qu(k)}}{mc^2}}\,\frac{du_{\sim}}{ds} = -\frac{X}{c}\frac{\partial T_{(k)\sim}{}^{\text{€}}}{\partial q^{\text{€}}} .\qquad \text{(C.6)}$$

Since $T_{(k)\sim}{}^{\text{€}}$ depends by the co-ordinates, through the quantum potential, in absence of an external Hamiltonian potential there exists a local acceleration of the mass distribution being $\frac{du_{\sim}}{ds} \neq 0$.

This characteristics of the hydrodynamic quantum evolution is quite general and it confirmed also for the motion equation (2.2.17) for the generic superposition of states.
To explain this point, it is sufficient to observe that the force due to the quantum potential is self-generated by the shape of the mass distribution and, hence, there may exist a force acting on the mass density, at each point, letting it accelerating, even in absence of an external force, (e.g., the ballistic enlargement of a free Gaussian packet).

In the case of the eigenstates (that are stationary states (i.e., $\frac{du_{\sim}}{ds} = 0$)), from (C.5) it follows that

$$\frac{\partial T_{(k)\sim}{}^{\text{€}}}{\partial q^{\text{€}}} = T_{(k)\sim}{}^{\text{€}}{}_{,\text{€}} = 0\qquad \text{(C.7)}$$

The same results holds in presence of an external potential $U_{(q_i)}$ considering the overall EITD

$$T_{(k)\sim}{}^{\text{€}}{}_{tot} = T_{(k)\sim}{}^{\text{€}} + T_{ext\sim}{}^{\text{€}}\qquad \text{(C.8)}$$

where

$$T_{ext\sim}{}^{\text{€}} = U_{(q_i)}u_{\sim}^{\text{€}} ,\qquad \text{(C.9)}$$

leading to

$$mc\sqrt{1-\frac{V_{qu(k)}}{mc^2}}\,\frac{du_{\sim}}{ds}=\mathsf{T}_{(k)\sim}{}^{\text{€}}{}_{,\text{€}}+\mathsf{T}_{ext\sim}{}^{\text{€}}{}_{,\text{€}}=0\,. \tag{C.10}$$

The above property comes from the stationarity of the eigenstates generated by the fact that the quantum potential force equals the external forces point by point [23].
Moreover, since the eigenstates are the only states that survives to the decoherence, thence, it follows that

$$\begin{aligned}&\left(lim_{dec}\sum_{\pm i}\left(/\text{Œ}_{\pm i}/^{2}\,\mathsf{T}_{\pm i\sim}{}^{\text{€}}\right)_{,\text{€}\;F}\right)+T_{\sim}{}^{\text{€}}{}_{,\text{€}em}\\ &=\sum_{\pm i}\left(/\text{Œ}_{\pm i(k)}/^{2}\,\mathsf{T}_{\pm i(k)\sim}{}^{\text{€}}\right)_{,\text{€}\;F}+T_{\sim}{}^{\text{€}}{}_{,\text{€}em}=0\end{aligned} \tag{C.11}$$

## Appendix D

In quasi-Minkowskian curved space-time, with particles very far from the Planckian mass density $\frac{m_p}{l_p{}^3}=\frac{c^5}{\hbar G^2}$) , the vielbein can be approximated as following

$$e^{a}{}_{\sim}=\mathsf{u}^{a}{}_{\sim}+\frac{\mathsf{v}^{a}{}_{\sim}}{2} \tag{D.1}$$

with

$$\mathsf{v}_{\sim\text{€}}\mathsf{v}^{\sim\text{€}}<<1 \tag{D.2}$$

and, being

$$g_{\text{€}\sim}=e^{a}{}_{\sim}\mathsf{y}_{ab}e^{b}{}_{\text{€}}\,, \tag{D.3}$$

where

$$\mathsf{y}_{\text{€}\sim}=\begin{bmatrix}1&0&0&0\\0&-1&0&0\\0&0&-1&0\\0&0&0&-1\end{bmatrix}, \tag{D.4}$$

by posing

$$\text{v}_{sim}{}^{\sim}{}_{\text{€}} = \frac{\text{y}_{\sim a}\text{v}^{a}{}_{\text{€}} + \text{y}_{\text{€} a}\text{v}^{a}{}_{\sim}}{2}$$

$$= \begin{bmatrix} \text{v}^{0}{}_{0} & \text{v}^{0}{}_{1} - \text{v}^{1}{}_{0} & \text{v}^{0}{}_{2} - \text{v}^{2}{}_{0} & \text{v}^{0}{}_{3} - \text{v}^{3}{}_{0} \\ \text{v}^{0}{}_{1} - \text{v}^{1}{}_{0} & -\text{v}^{1}{}_{1} & -\left(\text{v}^{1}{}_{2} + \text{v}^{2}{}_{1}\right) & -\left(\text{v}^{1}{}_{3} + \text{v}^{3}{}_{1}\right) \\ \text{v}^{0}{}_{2} - \text{v}^{2}{}_{0} & -\left(\text{v}^{1}{}_{2} + \text{v}^{2}{}_{1}\right) & -\text{v}^{2}{}_{2} & -\left(\text{v}^{2}{}_{3} + \text{v}^{3}{}_{2}\right) \\ \text{v}^{0}{}_{3} - \text{v}^{3}{}_{0} & -\left(\text{v}^{1}{}_{3} + \text{v}^{3}{}_{1}\right) & -\left(\text{v}^{2}{}_{3} + \text{v}^{3}{}_{2}\right) & -\text{v}^{3}{}_{3} \end{bmatrix} \tag{D.5}$$

it follows that

I.

$$g_{\sim \text{€}} \cong \text{y}_{\sim \text{€}} + \frac{\text{y}_{\sim a}\text{v}^{a}{}_{\text{€}} + \text{y}_{\text{€} a}\text{v}^{a}{}_{\sim}}{2} = \begin{bmatrix} 1 & 0 & 0 & 0 \\ 0 & -1 & 0 & 0 \\ 0 & 0 & -1 & 0 \\ 0 & 0 & 0 & -1 \end{bmatrix} + \text{v}_{sim \sim \text{€}} \tag{D.6}$$

II.

$$\Gamma^{\text{s}}_{\sim \text{r}} = \frac{1}{2}\text{y}^{\text{sx}}\left(\partial_{\text{r}}\text{v}_{sim\text{x}\sim} + \partial_{\sim}\text{v}_{sim\text{xr}} - \partial_{\text{x}}\text{v}_{sim\sim\text{r}}\right) \tag{D.7}$$

III.

$$\begin{aligned} \text{Š}_{ab\sim} &= f_b^{\text{r}} e_{a\text{s}} \Gamma^{\text{s}}_{\sim\text{r}} - f_b^{\text{r}} \partial_{\sim} e_{a\text{r}} = f_b^{\text{r}} e^{b}{}_{\text{s}} g_{ba} \Gamma^{\text{s}}_{\sim\text{r}} - f_b^{\text{r}} \partial_{\sim} e_{a\text{r}} \\ &= \text{u}_{\text{s}}^{\text{r}}\left(\text{y}_{ba} + \text{v}_{sim}{}^{b}{}_{a}\right)\Gamma^{\text{s}}_{\sim\text{r}} - f_b^{\text{r}} \partial_{\sim} e_{a\text{r}} \\ &= \left(\text{y}_{ba} + \text{v}_{sim}{}^{b}{}_{a}\right)\Gamma^{\text{r}}_{\sim\text{r}} - \text{u}_b^{\text{r}} \partial_{\sim} \frac{\text{v}_{a\text{r}}}{2} \\ &= \left(\text{y}_{ba} + \text{v}_{sim}{}^{b}{}_{a}\right)\partial_{\sim} \ ln\sqrt{-g} - \partial_{\sim}\frac{\text{v}_{ab}}{2} \\ &= \left(\text{y}_{ba} + \text{v}_{sim}{}^{b}{}_{a}\right)\partial_{\sim} \ ln\sqrt{-\frac{1}{/\text{y}^{\text{st}} + \text{v}_{sim}{}^{\text{st}}/}} - \partial_{\sim}\frac{\text{v}_{ab}}{2} \\ &= \left(\text{y}_{ba} + \text{v}_{sim}{}^{b}{}_{a}\right)\partial_{\sim} \ ln\sqrt{1 + /\text{v}_{sim}{}^{\text{st}}/} - \partial_{\sim}\frac{\text{v}_{ab}}{2} \\ &= \left(\text{y}_{ba} + \text{v}_{sim}{}^{b}{}_{a}\right)\partial_{\sim} \ ln\left(1 + \frac{/\text{v}_{sim}{}^{\text{st}}/}{2}\right) - \partial_{\sim}\frac{\text{v}_{ab}}{2} \\ &= \left(\text{y}_{ba} + \text{v}_{sim}{}^{b}{}_{a}\right)\partial_{\sim} \frac{/\text{v}_{sim}{}^{\text{st}}/}{2} - \partial_{\sim}\frac{\text{v}_{ab}}{2} \\ &= \frac{1}{2}\partial_{\sim}\left(\text{y}_{ba}/\text{v}_{sim}{}^{\text{st}}/ - \text{v}_{ab}\right) \end{aligned} \tag{D.8}$$

.

where $f_b^{\text{r}}$ is the inverse vielbein.
Thence, given that

$$\dagger^{ab} = -\dagger^{ba} \qquad \text{(D.9)}$$

and that

$$\dagger^{ab}\text{y}_{ab} = 0 \qquad \text{(D.10)}$$

(2.6.6) reads

$$\left( i\hbar \text{x}^{\sim} \left( \partial_{\sim} + \frac{ie}{\hbar} A_{\sim} \right) - mc \right) \Psi = \frac{\hbar}{8} \text{x}^{\sim} \dagger^{ab} \left( \partial_{\sim} \text{v}_{ab} \right) \Psi \qquad \text{(D.11)}$$

# Appendix E

### Covariant anti-commutation rules for fermions in curved space-time

By following the postulate of physical covariance, the anti-commuting rules for the fermion field

$$i\left\{ \text{Œ}^{a}{}_{\text{€}(q',t)}, \text{Œ}^{b\dagger}{}_{\sim(q,t)} \right\} = \left\{ \text{Œ}^{a}{}_{\text{€}(q',t)}, f^{b}{}_{\sim(q,t)} \right\}$$
$$= i\hbar\,\text{u}^{(3)}(q-q')\text{u}_{ab}\text{u}_{\text{€}\sim} = i\hbar\,\text{u}^{(3)}(q-q') \frac{\partial \text{Œ}^{a}{}_{\text{€}(q',t)}}{\partial \text{Œ}^{b\sim}{}_{(q,t)}} \qquad \text{(E.1)}$$

$$\left\{ f^{a}{}_{\text{€}(q',t)}, f^{b}{}_{\sim(q,t)} \right\} = i\hbar\,\text{u}^{(3)}(q-q')\text{u}_{ab}\text{u}_{\text{€}\sim} = i\hbar\,\text{u}^{(3)}(q-q') \frac{\partial f^{a}{}_{\text{€}(q',t)}}{\partial \text{Œ}^{b\sim}{}_{(q,t)}} = 0 \qquad \text{(E.2)}$$

where $f^{a}{}_{\sim(q,t)}$ is the momentum of the field [QSFT], in curved space-time reads

$$i\left\{ \text{Œ}^{a}{}_{\text{€}(q',t)}, \text{Œ}^{b\dagger}{}_{\sim(q,t)} \right\} = \left\{ \text{Œ}^{a}{}_{\text{€}(q',t)}, f^{a}{}_{\sim(q,t)} \right\} = i\hbar\,\text{u}^{(3)}(q-q') D_{\text{Œ}^{b}{}_{\sim}} \text{Œ}^{a}{}_{\text{€}(q',t)} \qquad \text{(E.3)}$$

$$\left\{ f^{a}{}_{\text{€}(q',t)}, f^{b}{}_{\sim(q,t)} \right\} = D_{\text{Œ}^{b\sim}} f^{a}{}_{\text{€}(q',t)} \qquad \text{(E.4)}$$

where

$$D_{\left(x^{b}\right)_{\sim}} Y^{a}{}_{\text{€}(q',t)} = \frac{\partial Y^{a}{}_{\text{€}(q',t)}}{\partial \left(x^{b}\right)^{\sim}{}_{(q,t)}} - \Gamma^{m}_{\text{€}\sim(q,t)} Y^{a}{}_{m(q',t)} + \text{Š}_{\sim}{}^{a}{}_{c} Y^{c}{}_{\text{€}(q',t)} \qquad \text{(E.5)}$$

is the covariant derivative for affine and spinor connections (where $\Gamma^{m}_{\text{€}\sim}$ is the Christoffel symbol) that, for vector bi-spinor, leads to the expression

$$i\left\{Œ^{a}{}_{€(q',t)}, Œ^{b\dagger}{}_{\sim(q,t)}\right\} = \left\{Œ^{a}{}_{€(q',t)}, f^{b}{}_{\sim(q,t)}\right\}$$

$$= i\hbar u^{(3)}(q-q')\begin{pmatrix} \dfrac{\partial Œ^{a}{}_{€(q',t)}}{\partial Œ^{b\sim}{}_{(q,t)}} - \Gamma^{m}_{€\sim(q',t)} Œ^{a}{}_{m(q',t)} \\ + Š_{€}{}^{a}{}_{c} Œ_{€}{}^{c} \end{pmatrix}$$

$$= i\hbar u^{(3)}(q-q')\left( g_{€\sim} - Œ^{ar}{}_{(q,t)} \frac{\partial g_{r€(q',t)}}{\partial Œ^{b\sim}{}_{(q,t)}} - \Gamma^{m}_{€\sim(q',t)} Œ^{a}{}_{m(q',t)} + Š_{€}{}^{a}{}_{c} Œ_{€}{}^{c} \right) \qquad \text{(E.6)}$$

where $Œ^{a}{}_{€} = \begin{pmatrix} Œ^{1}{}_{€} \\ Œ^{2}{}_{€} \\ Œ^{3}{}_{€} \\ Œ^{4}{}_{€} \end{pmatrix} = \Psi_{€}$ .

In the case of the Dirac spinor $\Psi = Œ^{a} = \begin{pmatrix} Œ^{1} \\ Œ^{2} \\ Œ^{3} \\ Œ^{4} \end{pmatrix}$, (E.3-6) reads

$$i\left\{Œ^{a}{}_{(q',t)}, Œ^{b\dagger}{}_{(q,t)}\right\} = \left\{Œ^{a}{}_{(q',t)}, f^{b}{}_{(q,t)}\right\} = i\hbar u^{(3)}(q-q')\frac{\partial Œ^{a}{}_{(q',t)}}{\partial Œ^{b}{}_{(q,t)}} \qquad \text{(E.7)}$$

$$\left\{f^{a}{}_{€(q',t)}, f^{b}{}_{\sim(q,t)}\right\} = i\hbar u^{(3)}(q-q')\frac{\partial f^{a}{}_{(q',t)}}{\partial Œ^{b}{}_{(q,t)}} \qquad \text{(E.8)}$$

where, in weak gravity, the momentum of the field $f^{a}$ reads

$$f^{a} = \qquad \text{(E.9)}$$

## Appendix F

As shown by Landau and Lifsits [30] in the limit of Newtonian gravity

$$g_{\sim€} \cong y_{\sim€} + v_{\sim€}\,, \qquad \text{(F.1)}$$

where

$$v_{\sim€} v^{\sim€} << 1, \qquad \text{(F.2)}$$

it holds

$$g_{00} = 1 + \frac{2\{}{c^{2}} \qquad \text{(F.3)}$$

where $\{$ is the Newtonian potential that reads

$$\frac{\partial}{\partial q^{r}}\frac{\partial \{}{\partial q^{r}} = c^{2} R_{0}^{\ 0} .$$ (F.4)

where $R_{0}^{\ 0}$ can be easily calculated by the GE (4.1.21) [30].
Moreover, since from (4.1.6) it follows that

$$V_{sim\sim€} \cong \begin{bmatrix} \frac{2\{}{c^{2}} & 0 & 0 & 0 \\ 0 & 0 & 0 & 0 \\ 0 & 0 & 0 & 0 \\ 0 & 0 & 0 & 0 \end{bmatrix}$$ (F.5)

we obtain

$$\mathcal{L}_{I} = \frac{\hbar}{8c^{2}}\left(\overline{\Psi} X^{\sim} \dagger^{00}\left[\partial_{\sim}\{\right]\Psi\right) = \dagger^{00} J^{\sim}\partial_{\sim}\{ = 0 .$$ (F.6)

It is noteworthy that the non-zero contribution to the gravitational perturbation

$$\mathcal{L}_{I} = \frac{\hbar}{8}\dagger^{ab} J^{\sim}\left[\partial_{\sim} V_{ab}\right],$$ (F.7)

due to the antisymmetry of $\dagger^{ab}$, comes from the out-diagonal terms of the metric tensor [30] that become relevant in high (non-Newtonian) gravity fields.